\documentclass[11pt,english]{article}
\usepackage[T1]{fontenc}
\usepackage[latin9]{inputenc}
\usepackage{color}
\usepackage{babel}
\usepackage{cprotect}
\usepackage{float}
\usepackage{booktabs}
\usepackage{mathrsfs}
\usepackage{mathtools}
\usepackage{url}
\usepackage{amsmath}
\usepackage{amsthm}
\usepackage{amssymb}
\usepackage{graphicx}
\usepackage{geometry}
\usepackage{setspace}
\usepackage[authoryear]{natbib}
\usepackage[pdfusetitle,
 bookmarks=true,bookmarksnumbered=false,bookmarksopen=false,
 breaklinks=true,pdfborder={0 0 0},pdfborderstyle={},backref=false,colorlinks=true]
 {hyperref}
\hypersetup{
 linkcolor=blue, citecolor=blue}

\makeatletter

\providecommand{\tabularnewline}{\\}

\theoremstyle{plain}
\newtheorem{assumption}{\protect\assumptionname}
\theoremstyle{plain}
\newtheorem{thm}{\protect\theoremname}
\theoremstyle{plain}
\newtheorem{cor}{\protect\corollaryname}
\theoremstyle{plain}
\newtheorem{lyxalgorithm}{\protect\algorithmname}

\allowdisplaybreaks

\usepackage{bbm}

\usepackage{amsmath}

\DeclareFontFamily{U}{mathx}{}
\DeclareFontShape{U}{mathx}{m}{n}{<-> mathx10}{}
\DeclareSymbolFont{mathx}{U}{mathx}{m}{n}
\DeclareMathAccent{\widehat}{0}{mathx}{"70}
\DeclareMathAccent{\widecheck}{0}{mathx}{"71}

\date{}

\newtheoremstyle{remboldstyle}
  {}{}{}{}{\bfseries}{.}{.5em}{{\thmname{#1 }}{\thmnumber{#2}}{\thmnote{ (#3)}}}
\theoremstyle{remboldstyle}
\newtheorem{rembold}{Remark}

\usepackage{appendix}
\appendixtitleon

\usepackage{scalerel,stackengine}
\stackMath
\newcommand\reallywidecheck[1]{%
\savestack{\tmpbox}{\stretchto{%
  \scaleto{%
    \scalerel*[\widthof{\ensuremath{#1}}]{\kern-.6pt\bigwedge\kern-.6pt}%
    {\rule[-\textheight/2]{1ex}{\textheight}}
  }{\textheight}%
}{0.5ex}}%
\stackon[1pt]{#1}{\scalebox{-1}{\tmpbox}}%
}
\date{}
\allowdisplaybreaks



\newtheoremstyle{ntnboldstyle}
  {}{}{}{}{\bfseries}{.}{.5em}{{\thmname{#1}}{\thmnote{(#2)}}}
\theoremstyle{ntnboldstyle}

\makeatletter
\def\@seccntformat#1{\@ifundefined{#1@cntformat}%
   {\csname the#1\endcsname\quad}  
   {\csname #1@cntformat\endcsname}
}
\let\oldappendix\appendix 
\renewcommand\appendix{%
    \oldappendix
    \newcommand{\section@cntformat}{\appendixname~\thesection\quad}
}
\makeatother

\usepackage{natbib}
\usepackage{hyperref} 
\defcitealias{vuong2017counterfactual}{VX}
\defcitealias{feng2019estimation}{FVX}
\defcitealias{Ma2023}{MMY}

\makeatother

\providecommand{\algorithmname}{Algorithm}
\providecommand{\assumptionname}{Assumption}
\providecommand{\corollaryname}{Corollary}
\providecommand{\theoremname}{Theorem}

\begin{document}
\title{\textbf{Inference on the Distribution of Individual Treatment Effects
in Nonseparable Triangular Models}\let\thefootnote\relax\footnotetext{Date: \today}\textbf{\thanks{We acknowledge the financial support from the National Natural Science
Foundation of China under grant 72394392 (Ma), the Natural Sciences
and Engineering Research Council of Canada (NSERC) under grant RGPIN-2024-04877
and the Social Sciences and Humanities Research Council of Canada
(SSHRC) under grant 435-2025-0755 (Marmer), and the Japan Society
for the Promotion of Science KAKENHI under grant 25K05032 (Yu).}}}
\author{Jun Ma\thanks{School of Economics, Renmin University of China, P.R. China. Email:
jun.ma@ruc.edu.cn}\and Vadim Marmer\thanks{Vancouver School of Economics, University of British Columbia, Canada.
Email: vadim.marmer@ubc.ca}\and Zhengfei Yu\thanks{Faculty of Humanities and Social Sciences, University of Tsukuba,
Japan. Email: yu.zhengfei.gn@u.tsukuba.ac.jp}}
\maketitle
\begin{abstract}
In this paper, we develop inference methods for the distribution of
heterogeneous individual treatment effects (ITEs) in the nonseparable
triangular model with a binary endogenous treatment and a binary instrument
of \citet{vuong2017counterfactual} and \citet*{feng2019estimation}.
We focus on the estimation of the cumulative distribution function
(CDF) of the ITE, which can be used to address a wide range of practically
important questions such as inference on the proportion of individuals
with positive ITEs, the quantiles of the distribution of ITEs, and
the interquartile range as a measure of the spread of the ITEs, as
well as comparison of the ITE distributions across sub-populations.
Moreover, our CDF-based approach can deliver more precise results
than density-based approach previously considered in the literature.
We establish weak convergence to tight Gaussian processes for the
empirical CDF and quantile function computed from nonparametric ITE
estimates of \citet*{feng2019estimation}. Using those results, we
develop bootstrap-based nonparametric inferential methods, including
uniform confidence bands for the CDF and quantile function of the
ITE distribution. \\

\noindent\textbf{Keywords: }Distribution of individual treatment
effects, nonparametric triangular models, two-step nonparametric estimation,
bootstrap, uniform confidence bands

\noindent\textbf{JEL classification:} C12, C14, C31, C36
\end{abstract}

\section{Introduction}

Heterogeneity of individual treatment effects (ITEs), including scenarios
with endogenous treatment, has received substantial attention in the
literature. When ITEs are heterogeneous, the econometrician is often
interested in the properties of their distribution, e.g., the CDF
and quantile function, as they contain important policy-relevant information
beyond average treatment effects. Recently, using a triangular model
with binary endogenous treatment, \citet[VX, hereafter]{vuong2017counterfactual}
and \citet*[FVX, hereafter]{feng2019estimation} established nonparametric
identification of heterogeneous ITEs and proposed their nonparametric
estimation. The estimated ITEs (also referred to as pseudo ITEs) can
be used further to estimate the distribution of the ITEs.

In this paper, we develop the asymptotic theory of the empirical CDF
and quantile function of the nonparametrically estimated (pseudo)
ITEs, which has been lacking in the literature so far. Such results
are nontrivial because of the multi-step nonparametric estimation
procedure required for their construction. We further use the results
to develop easy-to-implement nonparametric bootstrap methods for inference
on the CDF and quantile function of the ITE distribution. Our methods
can be used, e.g., for inference on the proportion of the population
with positive or negative ITEs and the dispersion of ITEs as measured
by the interquartile range (IQR). Moreover, our procedure can be used
to compare the ITE distributions between different sub-populations.
E.g., one can use our results to test whether the distribution of
the ITEs in one sub-population stochastically dominates that for another
sub-population.

Suppose that the econometrician observes data on an outcome variable,
a binary endogenous treatment, a binary instrument, and exogenous
covariates. We assume that the outcome variable and the endogenous
treatment are generated from the nonseparable nonparametric triangular
model of \citetalias{vuong2017counterfactual}  that satisfies the
rank invariance assumption. We further assume that the econometrician
uses the nonparametric method of \citetalias{feng2019estimation}
to construct pseudo ITEs as the estimates of the true ITEs for each
individual. In the next step, the econometrician uses the pseudo ITEs
to construct the empirical CDF or quantile function as the estimates
of the true ITE CDF or quantile function, respectively. The second
step can be performed for the entire sample or in sub-groups determined
by chosen values of discretely distributed exogenous covariates. E.g.,
the econometrician can perform the second step by gender, education
levels, income quartiles, etc., as well as intersections of such groups.

The first contribution of the paper is to show that the properly scaled
difference between the empirical CDF of the pseudo ITEs and the CDF
of the true ITEs weakly converges to a tight Gaussian process, with
a similar result holding for the empirical quantile function of the
pseudo ITEs. Importantly, we show that due to the two-step estimation,
the asymptotic variances of the empirical CDF and quantile function
of pseudo ITEs are ``inflated'' relative to their infeasible counterparts
based on true unobserved ITEs.

For our second contribution, we use the weak convergence results to
develop bootstrap inference methods for the CDF and quantile function
of the distribution of the ITE. Both pointwise confidence intervals
and uniform confidence bands (UCBs) are considered, as the pointwise
confidence interval is useful, e.g., for inference on the percentage
of the population with positive ITEs and the IQR, while the UCB is
useful for inference on the entire CDF or quantile function and comparing
the distributions of the ITEs between different sub-populations.\footnote{A UCB is a collection of random intervals that cover the unknown curve
of interest simultaneously over a range of values with a pre-specified
confidence level.} Our method for constructing confidence intervals for the percentages
has the desirable range-preserving property: the bootstrap percentile
confidence intervals are always sub-intervals of $\left[0,1\right]$.\footnote{See, e.g., \citet[Section 13.7]{Efron1994}.}

Our proposed inference methods exhibit excellent finite-sample performance
in Monte Carlo simulations. We further demonstrate their practical
value by revisiting a well-known empirical application: the effect
of participation in 401(k) retirement programs on personal savings,
see, e.g., \citet*{Chernozhukov2006} and \citetalias{feng2019estimation},
where our methods can be used to conduct valid inference on important
distributional features such as the proportion of individuals with
positive ITEs and stochastic dominance relationships between the distributions
of ITEs in different subpopulations. In the case of 401(k) programs,
our method reveals rich features of the ITE distributions. For instance,
the 95\% confidence interval for the proportion of households with
a positive ITE is $\left[0.851,0.919\right]$, suggesting that program
participation increased savings for the majority of households, though
a nontrivial minority experienced negative effects. Moreover, for
young individuals (with age in the first quartile), the 95\% confidence
interval is $\left[0.706,0.884\right]$, suggesting that up to 29.4\%
of young individuals may experience negative ITEs. The median ITE
has a 95\% confidence interval of $\left[6.96,9.74\right]$ thousand
dollars, indicating a significantly positive central tendency of the
treatment effect distribution. The 95\% confidence interval for the
IQR, $\left[16.68,23.38\right]$, underscores substantial heterogeneity
in the ITEs. A subgroup analysis reveals that as income or age increases,
the ITE distribution shifts to the right, with both the median and
the quartiles moving upward, and the spread of the distribution widening.
The UCBs of the quantile functions further indicate that, across all
quantiles between the $0.2$ and $0.9$ levels, the ITE is consistently
larger for higher-income groups than for lower-income groups.

Our paper contributes to the growing literature on causal inference
methods that emphasize heterogeneous treatment effects (see, e.g.,
\citealp{angrist2004treatment,Heckman1997,Heckman2006} among others).
The \citetalias{vuong2017counterfactual}  model we employ belongs
to a broad class of triangular models widely used for causal inference.\footnote{See, e.g., \citet{abrevaya2021estimation,Chesher2003,Chesher2005,d2015identification,Imbens2009,Jun2011,Newey1999,Torgovitsky2015,Vytlacil2007},
among others.} \citetalias{vuong2017counterfactual}  showed the identification
of the ``counterfactual mappings'', which can be used to obtain
the counterfactual outcome for each individual. \citetalias{feng2019estimation} 
proposed convenient extremum estimators for the counterfactual mappings
and established their asymptotic properties. Using estimated/pseudo
ITEs, \citetalias{feng2019estimation}  also proposed a kernel estimator
for the probability density function (PDF) of the ITE distribution.
The asymptotic theory of the density estimator was further developed
in \citet*[MMY, hereafter]{Ma2023}. \citetalias{Ma2023}  showed
that this estimator converges at the optimal rate (\citealp{Stone1982}),
established its asymptotic normality, and proposed a bootstrap-based
UCB for inference on the density function of the ITE distribution.
Our paper continues this line of research by developing corresponding
inference methods for the CDF and quantile function of the ITE distribution.\footnote{Like MMY, our paper also contributes to the literature of multi-step
nonparametric estimation using nonparametrically generated variables.
See, e.g., \citet{Ma2019} and \citet{Mammen2012} among others.} Combined with the results in \citetalias{Ma2023}, the econometrician
can use our results to characterize the commonly used distributional
features for the ITE. The methods for inference on the proportion
of positive/negative ITEs, the median, the IQR and also the stochastic
order relation between ITE distributions cannot be derived from the
results on PDF estimation and inference in \citetalias{Ma2023}. E.g,
when comparing two distributions, first-order stochastic dominance
is evident when one quantile function lies entirely above the other,
even though their PDFs may still intersect.

While our results are complementary to \citetalias{feng2019estimation} 
and \citetalias{Ma2023}, their derivation employs different techniques
from those used in \citetalias{Ma2023}. The main difference is that
the density estimator in \citetalias{feng2019estimation} and \citetalias{Ma2023}
is a differentiable function of the pseudo ITEs. \citetalias{Ma2023}
utilizes this fact and \textit{U}-process theory to establish its
properties. On the other hand, the empirical CDF estimator we focus
on here is non-differentiable, and we use the approach of \citet{van2007empirical}
instead. One should also note that the CDF-based approach developed
here is tuning-parameter-free, unlike the PDF-based approach in \citetalias{feng2019estimation} 
and \citetalias{Ma2023}.\footnote{See \citet{liu2022sample} and \citet{liu2024tuning} among others
for recent examples of tuning-free methods in the causal inference
literature.}

A related strand of literature is concerned with quantile treatment
effects (QTEs). When the treatment is endogenous, QTEs are often estimated
using the local quantile treatment effect (LQTE) model (\citealp{Abadie2002,Froelich2013})
or the instrumental variable quantile regression (IVQR) model (\citealp*{Chernozhukov2005,chernozhukov2006instrumental}).
Unlike the LQTE model, the approach of \citetalias{vuong2017counterfactual} 
and \citetalias{feng2019estimation}  allows for the identification
and estimation of ITEs for the entire population rather than just
for compliers. This is possible due to somewhat stronger assumptions
of \citetalias{vuong2017counterfactual}, such as the rank invariance
condition enabling the identification of ITEs. Nevertheless, we believe
that the ability to estimate effects for a broader population can
be important in practice.\footnote{Neither LQTE nor IVQR can identify the ITE distribution without the
rank invariance condition. An alternative strand of the literature
avoids the rank invariance assumption and employs a copula-based approach
to derive sharp bounds on the ITE distribution, typically in the context
of randomized experiments or under selection-on-observables assumptions
(see, e.g., \citealp{fan2009partial,fan2010sharp,fan2012confidence,firpo2019partial}
among others).} Moreover, the approach of \citetalias{feng2019estimation}  is computationally
attractive as it only involves a one-dimensional optimization problem.

The rest of the paper is organized as follows. Section \ref{sec:The-model}
reviews the model and the identification and estimation of ITEs as
proposed in \citetalias{vuong2017counterfactual}  and \citetalias{feng2019estimation}.
Section \ref{sec:Asymptotic-properties} shows the asymptotic normality
and weak convergence results for the empirical CDF and quantiles of
the pseudo ITEs. Section \ref{sec:Bootstrap-inference} describes
the construction of bootstrap percentile confidence intervals and
bootstrap UCBs for the ITE CDF and quantiles. Section \ref{sec:Extensions}
presents extensions of the methods proposed in the preceding section,
including inference on the ITE distributions of broader subgroups
and the differences of ITE quantiles of subgroups. Section \ref{sec:Monte-Carlo-simulations}
provides numerical evidence that shows the validity of the asymptotic
theory of Section \ref{sec:Asymptotic-properties} and evaluates the
finite sample performances of the inference methods proposed in Section
\ref{sec:Bootstrap-inference}. Section \ref{sec:Empirical-application}
revisits the empirical application in \citetalias{feng2019estimation},
which assesses the effect of participation in the 401(k) retirement
program on savings. Proofs of all main results are presented in an
online appendix.\cprotect\footnote{The appendix is available at \url{https://ruc-econ.github.io/ITE_CDF_app_V3.pdf}.}

\textbf{Notation}. We use ``$a\coloneqq b$'' to denote ``$a$
is defined by $b$'', and ``$a\eqqcolon b$'' is understood as
``$b$ is defined by $a$''. The closed interval $\left[a-b,a+b\right]$
is denoted as $a\pm b$ . Let $\mathrm{sgn}\left(u\right)\coloneqq2\times\mathbbm{1}\left(u>0\right)-1$
denote the left continuous sign function, where $\mathbbm{1}\left(\cdot\right)$
denotes the indicator function. For $a\in\mathbb{R}$, let $\left\lceil a\right\rceil \coloneqq\min\left\{ z\in\mathbb{Z}:z\geq a\right\} $
be the smallest integer greater than or equal to $a$. Let $a^{\top}$
denote the transpose of $a$. For a positive integer $T$, $\left[T\right]\coloneqq\left\{ 1,...,T\right\} $.
Let $\mathscr{S}_{V}$ denote the support of the distribution of a
random vector $V$, and let $\mathscr{S}_{V\mid W=w}$ denote the
support of the conditional distribution of $V$ given $W=w$. The
conditional CDF and PDF of the distribution of $V$ given $W=w$ are
denoted as $F_{V\mid W}\left(\cdot\mid w\right)$ and $f_{V\mid W}\left(\cdot\mid w\right)$,
respectively. Convergence in distribution in the general sense (\citealp[Chapter 18.2]{van2000asymptotic})
is denoted as ``$\rightsquigarrow$''. Let $\ell^{\infty}\left[a,b\right]$
denote the set of bounded real-valued functions on the closed interval
$\left[a,b\right]$. For any $f\in\ell^{\infty}\left[a,b\right]$,
let $\left\Vert f\right\Vert _{\left[a,b\right]}\coloneqq\sup_{t\in\left[a,b\right]}\left|f\left(t\right)\right|$
denote the sup-norm of $f$ on $\left[a,b\right]$. Let $C\left[a,b\right]$
denote the set of continuous functions on $\left[a,b\right]$. Let
$D\left[a,b\right]$ denote the set of c�dl�g functions on $\left[a,b\right]$
(i.e., for all $f\in D\left[a,b\right]$, $f$ is right continuous
at each point in $\left[a,b\right)$ and has a left limit at each
point in $\left(a,b\right]$). All the three spaces are endowed with
the sup-norm metric. Let $\mathit{BL}_{1}\left(\mathbb{D}\right)$
be the collection of real valued functions defined on a Banach space
$\mathbb{D}$ (endowed with a norm $\left\Vert \cdot\right\Vert $)
that satisfy the following condition: $h\in\mathit{BL}_{1}\left(\mathbb{D}\right)$
if and only if $\left|h\left(x\right)-h\left(y\right)\right|\leq\left\Vert x-y\right\Vert $
for all $x,y\in\mathbb{D}$ and $\sup_{x\in\mathbb{D}}\left|h\left(x\right)\right|\leq1$.

\section{Model and estimation of ITEs\label{sec:The-model}}

For completeness, in Section \ref{subsec:The-triangular-model}, we
review the model setup and assumptions of \citetalias{vuong2017counterfactual} 
and \citetalias{feng2019estimation}. Similarly, in Section \ref{subsec:The-ITEs},
we review the definition of ITEs, the additional assumption imposed
by \citetalias{Ma2023}, and the estimation method of \citetalias{feng2019estimation}.
The main objects of interest, the ITE CDF and quantile function as
well as their estimators are defined in Section \ref{subsec:Empirical-CDF-and-Q}.

\subsection{Triangular model\label{subsec:The-triangular-model}}

Let $Y$ be a continuously distributed outcome variable and let $D$
be an endogenous binary treatment variable. The model assumes that
$Y$ and $D$ are determined by the following outcome and selection
equations:
\begin{eqnarray}
Y & = & g\left(D,X,\epsilon\right)\label{eq:outcome equation}\\
D & = & \mathbbm{1}\left(\eta\leq s\left(Z,X\right)\right).\label{eq:selection equation}
\end{eqnarray}
In the outcome equation (\ref{eq:outcome equation}), $X$ is a vector
of observed explanatory variables (covariates), $\epsilon$ is the
unobserved scalar-valued disturbance, and $g$ is an unknown function.
The right hand side of (\ref{eq:outcome equation}) is of a completely
nonseparable form.\footnote{The outcome model (\ref{eq:outcome equation}) does not assume additive
or weak separability (see, e.g., \citealp{Vytlacil2007}). See Section
2.2 of \citetalias{vuong2017counterfactual}  and \citet{abrevaya2021estimation}
for examples of nonseparable specifications.} The selection equation (\ref{eq:selection equation}) has the form
of a latent index model, where $Z$ is a binary instrument (or instrumental
variable, IV) excluded from the outcome equation, $\eta$ is the unobserved
scalar-valued cost of the treatment to the individual, $s$ is an
unknown function, and $s\left(Z,X\right)$ is understood as the benefit
from the treatment. The treatment is taken up if the net utility from
taking up the treatment is positive.

Let $Y\left(d,x\right)\coloneqq g\left(d,x,\epsilon\right)$ and $D\left(z,x\right)\coloneqq\mathbbm{1}\left(\eta\leq s\left(z,x\right)\right)$
denote the potential outcome and treatment, and $\mathsf{co}_{x}$
denote the ``complier'' event ``$X=x$ and $D\left(0,x\right)<D\left(1,x\right)$''.
Lastly, let $\mathscr{S}_{Y\left(d,x\right)\mid\mathsf{co}_{x}}$
and $f_{Y\left(d,x\right)\mid\mathsf{co}_{x}}$ denote the support
and Lebesgue density of the conditional distribution of $Y\left(d,x\right)$
given $\mathsf{co}_{x}$. The assumptions on the data generating process
(DGP) from \citetalias{vuong2017counterfactual}  and \citetalias{feng2019estimation}
are summarized as follows.
\begin{assumption}[DGP]
\label{assu: DGP1}(a) For all $\left(d,x\right)\in\mathscr{S}_{\left(D,X\right)}$,
$g\left(d,x,\cdot\right)$ is continuously differentiable and strictly
increasing. (b) $Z$ is independent of $\left(\epsilon,\eta\right)$
conditionally on $X$. (c) For all $x\in\mathscr{S}_{X}$, $s\left(0,x\right)<s\left(1,x\right)$
and $\Pr\left[D=1\mid Z=1,X=x\right]>\Pr\left[D=1\mid Z=0,X=x\right]$.
(d) For all $x\in\mathscr{S}_{X}$, the conditional distribution of
$\left(\epsilon,\eta\right)$ given $X=x$ is absolutely continuous
with respect to the Lebesgue measure, has a compact support, and its
PDF is continuous and bounded. (e) $\mathscr{S}_{\left(D,X\right)}$
and $\mathscr{S}_{\left(Z,X\right)}$ are both $\left\{ 0,1\right\} \times\mathscr{S}_{X}$.
(f) For all $\left(d,x\right)\in\mathscr{S}_{\left(D,X\right)}$,
$\mathscr{S}_{Y\left(d,x\right)\mid\mathsf{co}_{x}}=\mathscr{S}_{Y\left(d,x\right)\mid X=x}$.
(g) For all $\left(d,x\right)\in\mathscr{S}_{\left(D,X\right)}$,
$f_{Y\left(d,x\right)\mid\mathsf{co}_{x}}$ is bounded away from zero.
(h) For all $x\in\mathscr{S}_{X}$ and $d\in\left\{ 0,1\right\} $,
the conditional distribution of $Y\left(d,x\right)$ has the support
$\mathscr{S}_{Y\left(d,x\right)\mid X=x}=\left[\underline{y}_{dx},\overline{y}_{dx}\right]$
with known boundaries $-\infty<\underline{y}_{dx}<\overline{y}_{dx}<+\infty$.
(i) $X$ is discretely distributed and $\mathscr{S}_{X}$ is finite.
\end{assumption}
The monotonicity of $g\left(d,x,\cdot\right)$ in Part (a) imposes
rank invariance on the potential outcomes. Part (b) is the IV exogeneity
assumption and Part (c) is the IV relevance assumption. Given the
assumption in Part (b) and equations (\ref{eq:outcome equation})--(\ref{eq:selection equation}),
$Z$ is independent of $\left(Y\left(1,x\right),Y\left(0,x\right),D\left(1,x\right),D\left(0,x\right)\right)$
conditionally on $X=x$. Part (c) and equation (\ref{eq:selection equation}),
imply the monotonicity assumption of potential treatments: $D\left(0,x\right)\leq D\left(1,x\right)$.\footnote{See, e.g., \citet{Vytlacil2002}. Note also that the independence
and monotonicity assumptions jointly have testable implications \citep[see, e.g., ][]{Kitagawa2015}.} Parts (d,e) are mild regularity conditions. The support condition
in Part (f) is crucial for the identification result of Lemma 1 of
\citetalias{vuong2017counterfactual}  and is related to the effectiveness
of the IV.\footnote{See Section 2.1 of \citetalias{vuong2017counterfactual}. In particular,
Part (f) is satisfied if the conditional distribution of $\left(\epsilon,\eta\right)$
given $X=x$ has a rectangular support for all $x\in\mathscr{S}_{X}$.} Parts (a,c,d) together with equations (\ref{eq:outcome equation})--(\ref{eq:selection equation})
ensure that the conditional distribution of $Y\left(d,x\right)$ given
$\mathsf{co}_{x}$ is absolutely continuous with respect to the Lebesgue
measure, and thus the existence of a continuous and bounded Lebesgue
density $f_{Y\left(d,x\right)\mid\mathsf{co}_{x}}$ is guaranteed.
Given the conditions of Parts (a,d), $\mathscr{S}_{Y\left(d,x\right)\mid X=x}$
is a compact interval. Moreover, Lemma 1 of \citetalias{vuong2017counterfactual} 
shows that $\mathscr{S}_{Y\left(d,x\right)\mid X=x}=\mathscr{S}_{Y\mid D=d,X=x}$
and, therefore, the end points $\underline{y}_{dx}$ and $\overline{y}_{dx}$
of $\mathscr{S}_{Y\left(d,x\right)\mid X=x}$ are identifiable and
estimable.  Part (h) assumes that $\underline{y}_{dx}$ and $\overline{y}_{dx}$
are known,  however, in practice, $\underline{y}_{dx}$ and $\overline{y}_{dx}$
can be estimated by the minimum and the maximum of the observed outcomes,
respectively.\footnote{As discussed in \citetalias{feng2019estimation}, Parts (g,h,i) can
be relaxed at the cost of technical complications. See Section 3 therein.}

\subsection{ITEs and their estimation\label{subsec:The-ITEs}}

The ITE is defined as 
\begin{equation}
\varDelta\coloneqq g\left(1,X,\epsilon\right)-g\left(0,X,\epsilon\right).\label{eq:Delta through Y}
\end{equation}
Note that $\varDelta$ is random conditionally on $X$ due to the
unobserved $\epsilon$, i.e., the treatment effects vary among individuals
with the same observed characteristics. Since the disturbances $\epsilon$
and $\eta$ are allowed to be correlated conditionally on $X$, whether
or not individuals select into treatment can be correlated with the
gain from treatment.\footnote{The property is referred to as ``essential heterogeneity'' in the
causal inference literature. See, e.g., \citet{Heckman2006}.}

Let $\varDelta_{x}\left(e\right)\coloneqq g\left(1,x,e\right)-g\left(0,x,e\right)$.
As discussed in \citetalias{Ma2023}, the assumptions imposed in the
preceding section alone are insufficient to ensure that the conditional
distribution of $\varDelta$ given $X=x$ is absolutely continuous
with respect to the Lebesgue measure. Therefore, as in \citetalias{Ma2023},
we introduce the following seemingly minimal assumption which guarantees
that the conditional distribution of $\varDelta$ given $X=x$ has
a continuous PDF denoted as $f_{\varDelta\mid X}\left(\cdot\mid x\right)$.
Let $\left(\underline{\epsilon}_{x},\overline{\epsilon}_{x}\right)$
be the end points of $\mathscr{S}_{\epsilon\mid X=x}$; that is, $\underline{\epsilon}_{x}<\overline{\epsilon}_{x}$
and $\mathscr{S}_{\epsilon\mid X=x}=\left[\underline{\epsilon}_{x},\overline{\epsilon}_{x}\right]$.
\begin{assumption}[Existence and continuity of the conditional PDF of ITE]
\label{assu: DGP4}(a) There is a partition of $\left[\underline{\epsilon}_{x},\overline{\epsilon}_{x}\right]$,
$\underline{\epsilon}_{x}=\epsilon_{x,0}<\epsilon_{x,1}<\cdots<\epsilon_{x,m}=\overline{\epsilon}_{x}$
with $\left[\underline{\epsilon}_{x},\overline{\epsilon}_{x}\right]=\bigcup_{j=1}^{m}\left[\epsilon_{x,j-1},\epsilon_{x,j}\right]$,
such that $\varDelta_{x}$ is piecewise monotone: for all $j=1,...,m$,
the restriction $\varDelta_{x,j}$ of $\varDelta_{x}$ on $\left[\epsilon_{x,j-1},\epsilon_{x,j}\right]$,
is strictly monotone. (b) The images of $\left(\epsilon_{x,j-1},\epsilon_{x,j}\right)$
under the mapping $\varDelta_{x,j}$ for $j=1,...,m$ are all the
same.
\end{assumption}
Note that the knowledge of the partition introduced in Assumption
\ref{assu: DGP4} is not required for the implementation of our methods.

Denote $d'\coloneqq1-d$, and let $g^{-1}\left(d',x,\cdot\right)$
be the inverse function of $g\left(d',x,\cdot\right)$. For $y\in\mathscr{S}_{Y\left(d',x\right)\mid X=x}$,
define the corresponding counterfactual mapping $\phi_{dx}\left(y\right)\coloneqq g\left(d,x,g^{-1}\left(d',x,y\right)\right)$,
i.e., $\phi_{dx}\left(y\right)$ is the counterfactual outcome if
the observed treatment status $d'$ were $d$. Using the counterfactual
mappings, we can write the ITE as 
\begin{equation}
\varDelta=D\left(Y-\phi_{0X}\left(Y\right)\right)+\left(1-D\right)\left(\phi_{1X}\left(Y\right)-Y\right).\label{eq:Delta}
\end{equation}
Lemma 1 of \citetalias{vuong2017counterfactual}  provides a constructive
nonparametric identification result for the counterfactual mappings.
This result and (\ref{eq:Delta}) establish the identification of
the distribution of $\varDelta$.

Next, we review the estimation procedure of \citetalias{feng2019estimation}.
Lemma 1 of \citetalias{feng2019estimation} shows that $\phi_{dx}\left(y\right)$
is the unique minimizer of the strictly convex function $\varUpsilon_{dx}\left(\cdot,y\right)$,
where
\begin{multline}
\varUpsilon_{dx}\left(t,y\right)\coloneqq\left(\mathrm{E}\left[\mathbbm{1}\left(D=d\right)\left|Y-t\right|\mid Z=d,X=x\right]-\mathrm{E}\left[\mathbbm{1}\left(D=d'\right)\mathrm{sgn}\left(Y-y\right)\mid Z=d,X=x\right]\cdot t\right)\\
-\left(\mathrm{E}\left[\mathbbm{1}\left(D=d\right)\left|Y-t\right|\mid Z=d',X=x\right]-\mathrm{E}\left[\mathbbm{1}\left(D=d'\right)\mathrm{sgn}\left(Y-y\right)\mid Z=d',X=x\right]\cdot t\right).\label{eq:Q_dx definition}
\end{multline}
The fact that $\phi_{dx}\left(y\right)$ uniquely minimizes $\varUpsilon_{dx}\left(\cdot,y\right)$
motivates using an extremum estimator for its estimation.

Since estimation is performed for each given value of $x\in\mathscr{S}_{X}$,
we make the following assumption, which allows us to treat the sample
size $n_{x}$ of a sub-sample with the covariate values being $x$
as non-random. It is a simplification that does not affect the properties
of the estimation and inference procedures.
\begin{assumption}[Sampling]
\label{assu:sampling}Data $\left\{ W_{i}\coloneqq\left(Y_{i},D_{i},Z_{i}\right)^{\top}\right\} _{i=1}^{n_{x}}$
are i.i.d. observations generated from the model defined by equations
(\ref{eq:outcome equation})--(\ref{eq:selection equation}) and
Assumptions \ref{assu: DGP1} and \ref{assu: DGP4}, with the covariate
values set to $x\in\mathscr{S}_{X}$.
\end{assumption}
Let $\widehat{\varUpsilon}_{dx}^{\left(-i\right)}\left(t,y\right)$
denote the leave-$i$-out sample analogue of $\varUpsilon_{dx}\left(t,y\right)$:
\begin{multline}
\widehat{\varUpsilon}_{dx}^{\left(-i\right)}\left(t,y\right)\coloneqq\frac{\sum_{j\in\left[n_{x}\right]\setminus\left\{ i\right\} }\left\{ \mathbbm{1}\left(D_{j}=d,Z_{j}=d\right)\left|Y_{j}-t\right|-\mathbbm{1}\left(D_{j}=d',Z_{j}=d\right)\mathrm{sgn}\left(Y_{j}-y\right)t\right\} }{\sum_{j\in\left[n_{x}\right]\setminus\left\{ i\right\} }\mathbbm{1}\left(Z_{j}=d\right)}\\
-\frac{\sum_{j\in\left[n_{x}\right]\setminus\left\{ i\right\} }\left\{ \mathbbm{1}\left(D_{j}=d,Z_{j}=d'\right)\left|Y_{j}-t\right|-\mathbbm{1}\left(D_{j}=d',Z_{j}=d'\right)\mathrm{sgn}\left(Y_{j}-y\right)t\right\} }{\sum_{j\in\left[n_{x}\right]\setminus\left\{ i\right\} }\mathbbm{1}\left(Z_{j}=d'\right)}.\label{eq:Q_hat definition}
\end{multline}
The leave-$i$-out nonparametric estimator of $\phi_{dx}\left(y\right),d\in\left\{ 0,1\right\} $,
can be constructed as
\begin{equation}
\widehat{\phi}_{dx}^{\left(-i\right)}\left(y\right)\coloneqq\underset{t\in\left[\underline{y}_{dx},\overline{y}_{dx}\right]}{\arg\min}\widehat{\varUpsilon}_{dx}^{\left(-i\right)}\left(t,y\right).\label{eq:phi_hat definition}
\end{equation}
One can now estimate the ITEs by replacing $\phi_{dx}(y)$ in (\ref{eq:Delta})
with its leave-$i$-out nonparametric estimator $\widehat{\phi}_{dx}^{\left(-i\right)}\left(y\right)$:
\begin{equation}
\widehat{\varDelta}_{i}=D_{i}\left(Y_{i}-\widehat{\phi}_{0x}^{\left(-i\right)}\left(Y_{i}\right)\right)+\left(1-D_{i}\right)\left(\widehat{\phi}_{1x}^{\left(-i\right)}\left(Y_{i}\right)-Y_{i}\right),\,i=1,...,n_{x}.\label{eq:pseudo ITE definition}
\end{equation}
Using these estimated/pseudo ITEs, one can estimate various features
of the distribution of $\varDelta$.

\subsection{\label{subsec:Empirical-CDF-and-Q}Empirical CDF and quantile function
of pseudo ITEs}

We estimate the conditional CDF $F_{\varDelta\mid X}\left(\cdot\mid x\right)$
given $X=x$ of ITEs using the empirical CDF of the pseudo ITEs $\left\{ \widehat{\varDelta}_{i}\right\} _{i=1}^{n_{x}}$:
\begin{equation}
\widehat{F}_{\varDelta\mid X}\left(v\mid x\right)\coloneqq\frac{1}{n_{x}}\sum_{i=1}^{n_{x}}\mathbbm{1}\left(\widehat{\varDelta}_{i}\leq v\right),\,v\in\mathbb{R}.\label{eq:ITE CDF estimator}
\end{equation}
Related quantities of practical interest are, e.g., the proportion
$F_{\varDelta\mid X}\left(0\mid x\right)$ of population with positive
ITEs or the proportion $1-F_{\varDelta\mid X}\left(0\mid x\right)$
of population with negative ITEs.

For $\tau\in\left(0,1\right)$, the $\tau$-th quantile of the ITE
distribution conditional on $X=x$ is defined as $Q_{\varDelta\mid X}\left(\tau\mid x\right)\coloneqq\inf\left\{ y\in\mathbb{R}:F_{\varDelta\mid X}\left(y\mid x\right)\geq\tau\right\} $.
We estimate $Q_{\varDelta\mid X}\left(\tau\mid x\right)$ using the
corresponding empirical quantile of the pseudo ITEs $\left\{ \widehat{\varDelta}_{i}\right\} _{i=1}^{n_{x}}$:
\begin{equation}
\widehat{Q}_{\varDelta\mid X}\left(\tau\mid x\right)\coloneqq\inf\left\{ y\in\mathbb{R}:\widehat{F}_{\varDelta\mid X}\left(y\mid x\right)\geq\tau\right\} .\label{eq: ITE Quantile}
\end{equation}
The econometrician may be interested in the conditional median $Q_{\varDelta\mid X}\left(0.5\mid x\right)$
as a measure of centrality of the ITE distribution or the conditional
population IQR
\begin{equation}
\mathit{IR}_{\varDelta\mid X=x}\coloneqq Q_{\varDelta\mid X}\left(0.75\mid x\right)-Q_{\varDelta\mid X}\left(0.25\mid x\right)\label{eq:interquartile range}
\end{equation}
as a measure of dispersion.

\section{Asymptotic properties\label{sec:Asymptotic-properties}}

Section \ref{subsec:Asymptotic-normality-CDF} presents the asymptotic
theory for the ITE CDF estimator (\ref{eq:ITE CDF estimator}) and
discusses the key steps in the proof. Section \ref{subsec:Asymptotic-normality-quantiles}
presents the asymptotic theory for the quantile estimator (\ref{eq:ITE CDF estimator}).

\subsection{Asymptotic Gaussianity of the empirical CDF\label{subsec:Asymptotic-normality-CDF}}

Let $\text{\ensuremath{\left[\underline{v}_{x},\overline{v}_{x}\right]}}$
be any inner closed sub-interval of $\mathscr{S}_{\varDelta\mid X=x}$.
Denote
\begin{equation}
S_{F}\left(v\mid x\right)\coloneqq\sqrt{n_{x}}\left(\widehat{F}_{\varDelta\mid X}\left(v\mid x\right)-F_{\varDelta\mid X}\left(v\mid x\right)\right),\,v\in\text{\ensuremath{\left[\underline{v}_{x},\overline{v}_{x}\right]}}.\label{eq:S_F definition}
\end{equation}
Our first result is that the process $S_{F}\left(\cdot\mid x\right)$,
as a map from the underlying probability space into $\ell^{\infty}\left[\underline{v}_{x},\overline{v}_{x}\right]$,
converges in distribution to a tight Gaussian process. The asymptotic
normality of $S_{F}\left(v\mid x\right)$ for any fixed $v\in\left[\underline{v}_{x},\overline{v}_{x}\right]$
immediately follows from this result. 

Before we discuss the key steps in the proof of the convergence in
distribution result for $S_{F}\left(\cdot\mid x\right)$, we introduce
the following notations. Let
\begin{eqnarray*}
p_{z\mid x} & \coloneqq & \mathrm{Pr}\left[Z=z\mid X=x\right],\\
\pi_{x}\left(Z_{i}\right) & \coloneqq & \frac{\mathbbm{1}\left(Z_{i}=0\right)}{p_{0\mid x}}-\frac{\mathbbm{1}\left(Z_{i}=1\right)}{p_{1\mid x}},\\
H_{x}\left(e\right) & \coloneqq & \frac{1}{n_{x}}\sum_{i=1}^{n_{x}}\left\{ \mathbbm{1}\left(\epsilon_{i}\leq e\right)-F_{\epsilon\mid X}\left(e\mid x\right)\right\} \pi_{x}\left(Z_{i}\right).
\end{eqnarray*}
By \citet[Theorem 8.19]{Kosorok2007} and \citet[Corollary 9.32(v)]{Kosorok2007},
we have
\begin{equation}
\sqrt{n_{x}}\cdot H_{x}(\cdot)\rightsquigarrow\mathbb{H}_{x}(\cdot)\coloneqq\sqrt{p_{1\mid x}^{-1}+p_{0\mid x}^{-1}}\cdot\mathbb{B}_{0}\left(F_{\epsilon\mid X}\left(\cdot\mid x\right)\right)\textrm{ in }\text{\ensuremath{\ell^{\infty}\left[\underline{\epsilon}_{x},\overline{\epsilon}_{x}\right]}},\label{eq:H_weak_convergence}
\end{equation}
where $\left\{ \mathbb{B}_{0}\left(t\right):t\in\left[0,1\right]\right\} $
is a standard Brownian bridge, whose sample path is continuous almost
surely. Therefore, $\mathbb{H}_{x}$ concentrates on $C\left[\underline{\epsilon}_{x},\overline{\epsilon}_{x}\right]\subseteq\ell^{\infty}\left[\underline{\epsilon}_{x},\overline{\epsilon}_{x}\right]$
(i.e., $\Pr\left[\mathbb{H}_{x}\in C\left[\underline{\epsilon}_{x},\overline{\epsilon}_{x}\right]\right]=1$)
and $\mathbb{H}_{x}$ is a tight random element in $\ell^{\infty}\left[\underline{\epsilon}_{x},\overline{\epsilon}_{x}\right]$
(i.e., for every $\varepsilon>0$, there exists a compact set $K\subseteq\ell^{\infty}\left[\underline{\epsilon}_{x},\overline{\epsilon}_{x}\right]$
such that $\mathrm{Pr}\left[\mathbb{H}_{x}\notin K\right]\leq\varepsilon$). 

The following notations are used to define an intermediate surrogate
for $\widehat{F}_{\varDelta\mid X}\left(v\mid x\right)$. Let
\begin{eqnarray*}
\zeta_{dx}\left(y\right) & \coloneqq & f_{Y\left(d,x\right)\mid\mathsf{co}_{x}}\left(y\right)\left(\Pr\left[D=d\mid Z=1,X=x\right]-\Pr\left[D=d\mid Z=0,X=x\right]\right),\\
\varsigma_{dx}\left(e\right) & \coloneqq & \left(-1\right)^{d'}\zeta_{dx}\left(g\left(d,x,e\right)\right).
\end{eqnarray*}
 Then, let
\[
\widecheck{F}{}_{\varDelta\mid X}\left(v\mid x\right)\coloneqq\frac{1}{n_{x}}\sum_{i=1}^{n_{x}}\sum_{d\in\left\{ 0,1\right\} }\mathbbm{1}\left(\varDelta_{i}+\frac{H_{x}\left(\epsilon_{i}\right)}{\varsigma_{dx}\left(\epsilon_{i}\right)}\leq v\right)\mathbbm{1}\left(D_{i}=d'\right)
\]
be the intermediate surrogate of $\widehat{F}_{\varDelta\mid X}\left(v\mid x\right)$.
In the appendix, using the Bahadur-type representation result given
by Lemma 2 in \citetalias{Ma2023}, we show that
\begin{equation}
\widehat{F}_{\varDelta\mid X}\left(v\mid x\right)-\widecheck{F}{}_{\varDelta\mid X}\left(v\mid x\right)=o_{p}\left(n_{x}^{-1/2}\right),\label{eq:F_hat - F_check rate}
\end{equation}
uniformly in $v\in\text{\ensuremath{\left[\underline{v}_{x},\overline{v}_{x}\right]}}$.

Let
\[
\widetilde{F}_{\varDelta\mid X}\left(v\mid x\right)\coloneqq\frac{1}{n_{x}}\sum_{i=1}^{n_{x}}\mathbbm{1}\left(\varDelta_{i}\leq v\right),\,v\in\mathbb{R},
\]
be the infeasible estimator using the true ITEs. Define the operator
$\varPsi_{dx}:\ell^{\infty}\left[\underline{\epsilon}_{x},\overline{\epsilon}_{x}\right]\rightarrow\ell^{\infty}\left[\underline{v}_{x},\overline{v}_{x}\right]$
by 
\begin{equation}
\varPsi_{dx}h\left(v\right)\coloneqq\mathrm{E}\left[\mathbbm{1}\left(h\left(\epsilon\right)\leq v\right)\mathbbm{1}\left(D=d'\right)\mid X=x\right],\,h\in\ell^{\infty}\left[\underline{\epsilon}_{x},\overline{\epsilon}_{x}\right].\label{eq:operator definition}
\end{equation}
Then, in the appendix, we show that 
\begin{equation}
\widecheck{F}{}_{\varDelta\mid X}\left(v\mid x\right)-\widetilde{F}_{\varDelta\mid X}\left(v\mid x\right)-\sum_{d\in\left\{ 0,1\right\} }\left\{ \varPsi_{dx}\left(\varDelta_{x}+\frac{H_{x}}{\varsigma_{dx}}\right)-\varPsi_{dx}\varDelta_{x}\right\} \left(v\right)=o_{p}\left(n_{x}^{-1/2}\right),\label{eq:F_check - Psi_difference}
\end{equation}
uniformly in $v\in\left[\underline{v}_{x},\overline{v}_{x}\right]$.
Note that (\ref{eq:H_weak_convergence}) and the continuous mapping
theorem (CMT, see, e.g., \citealp[Theorem 7.7]{Kosorok2007}) imply
$\left\Vert H_{x}\right\Vert _{\left[\underline{\epsilon}_{x},\overline{\epsilon}_{x}\right]}\rightarrow_{p}0$.
Also, it is clear that all sample paths of $H_{x}$ reside in the
space $D\left[\underline{\epsilon}_{x},\overline{\epsilon}_{x}\right]$.
To establish the result in (\ref{eq:F_check - Psi_difference}), since
the function class
\[
\left\{ e\mapsto\mathbbm{1}\left(\varDelta_{x}\left(e\right)+\frac{h\left(e\right)}{\varsigma_{dx}\left(e\right)}\leq v\right):\left(v,h\right)\in\left[\underline{v}_{x},\overline{v}_{x}\right]\times D\left[\underline{\epsilon}_{x},\overline{\epsilon}_{x}\right]\right\} 
\]
does not satisfy the bounded complexity (Donsker) condition, we follow
the arguments of \citet{van2007empirical}, which make use of (\ref{eq:H_weak_convergence})
and also the fact that the limit $\mathbb{H}_{x}$ concentrates on
the much smaller separable Banach space $C\left[\underline{\epsilon}_{x},\overline{\epsilon}_{x}\right]$
. Now by using (\ref{eq:F_hat - F_check rate}) and (\ref{eq:F_check - Psi_difference}),
we obtain the following approximation for $S_{F}\left(v\mid x\right)$:
\begin{eqnarray}
S_{F}\left(v\mid x\right) & = & \sqrt{n_{x}}\left(\widetilde{F}_{\varDelta\mid X}\left(v\mid x\right)-F_{\varDelta\mid X}\left(v\mid x\right)\right)+\sqrt{n_{x}}\cdot\sum_{d\in\left\{ 0,1\right\} }\left\{ \varPsi_{dx}\left(\varDelta_{x}+\frac{H_{x}}{\varsigma_{dx}}\right)-\varPsi_{dx}\varDelta_{x}\right\} \left(v\right)\nonumber \\
 &  & +o_{p}\left(n_{x}^{-1/2}\right),\label{eq:S_F approximation}
\end{eqnarray}
uniformly in $v\in\left[\underline{v}_{x},\overline{v}_{x}\right]$.

Let $\left\{ \mathbb{B}_{1}\left(t\right):t\in\left[0,1\right]\right\} $
be a standard Brownian bridge and define the Gaussian process
\[
\mathbb{F}_{1}\left(v\mid x\right)\coloneqq\mathbb{B}_{1}\left(F_{\varDelta\mid X}\left(v\mid x\right)\right),\,v\in\left[\underline{v}_{x},\overline{v}_{x}\right].
\]
Since $\mathbb{B}_{1}$ has continuous sample paths almost surely,
under the model assumptions, $\mathbb{F}_{1}\left(\cdot\mid x\right)$
concentrates on $C\left[\underline{v}_{x},\overline{v}_{x}\right]$.
By the functional central limit theorem (see, e.g., \citealp[Theorem 19.3]{van2000asymptotic}),
\begin{equation}
\sqrt{n_{x}}\left(\widetilde{F}_{\varDelta\mid X}\left(\cdot\mid x\right)-F_{\varDelta\mid X}\left(\cdot\mid x\right)\right)\rightsquigarrow\mathbb{F}_{1}\left(\cdot\mid x\right)\textrm{ in }\ell^{\infty}\text{\ensuremath{\left[\underline{v}_{x},\overline{v}_{x}\right]}}.\label{eq:F_til weak convergence}
\end{equation}

In the appendix, we show that $\varPsi_{dx}$ is Hadamard differentiable
(see, e.g., \citealp[Section 20.2]{van2000asymptotic} for the definition)
at $\varDelta_{x}$ with derivative denoted by $\psi_{dx}$. By the
functional delta method (see, e.g., \citealp[Theorem 20.8]{van2000asymptotic}),
we have
\begin{equation}
\sqrt{n_{x}}\sum_{d\in\left\{ 0,1\right\} }\left\{ \varPsi_{dx}\left(\varDelta_{x}+\frac{H_{x}}{\varsigma_{dx}}\right)-\varPsi_{dx}\varDelta_{x}\right\} \left(v\right)=\sum_{d\in\left\{ 0,1\right\} }\psi_{dx}\left(\frac{\sqrt{n_{x}}\cdot H_{x}}{\varsigma_{dx}}\right)\left(v\right)+o_{p}\left(1\right),\label{eq:functional delta method result}
\end{equation}
uniformly in $v\in\left[\underline{v}_{x},\overline{v}_{x}\right]$.
We can show that the leading term on the right hand side of (\ref{eq:functional delta method result})
is uncorrelated with the first term on the right hand side of (\ref{eq:S_F approximation}).
Before characterizing its limiting distribution, we introduce the
following notations. Let 
\[
f_{\left(\epsilon,D\right)\mid X}\left(e,d\mid x\right)\coloneqq f_{\epsilon\mid\left(D,X\right)}\left(e\mid d,x\right)\mathrm{Pr}\left[D=d\mid X=x\right]
\]
denote the conditional density of $\left(\epsilon,D\right)$ given
$X=x$, and also let
\begin{eqnarray}
\rho_{dx,j}\left(v\right) & \coloneqq & f_{\left(\epsilon,D\right)\mid X}\left(\varDelta_{x,j}^{-1}\left(v\right),d\mid x\right)\left(\varDelta_{x,j}^{-1}\right)'\left(v\right),\nonumber \\
\omega_{x,j}\left(v\right) & \coloneqq & -\sum_{d\in\left\{ 0,1\right\} }\frac{\left|\rho_{d'x,j}\left(v\right)\right|}{\varsigma_{dx}\left(\varDelta_{x,j}^{-1}\left(v\right)\right)}.\label{eq:Omega definition}
\end{eqnarray}
Let $\left\{ \mathbb{B}_{2}\left(t\right):t\in\left[0,1\right]\right\} $
be a standard Brownian bridge that is independent of $\left\{ \mathbb{B}_{1}\left(t\right):t\in\left[0,1\right]\right\} $.
Define the Gaussian process
\[
\mathbb{F}_{2}\left(v\mid x\right)\coloneqq\sqrt{p_{1\mid x}^{-1}+p_{0\mid x}^{-1}}\left\{ \sum_{j=1}^{m}\omega_{x,j}\left(v\right)\mathbb{B}_{2}\left(F_{\epsilon\mid X}\left(\varDelta_{x,j}^{-1}\left(v\right)\mid x\right)\right)\right\} ,\,v\in\left[\underline{v}_{x},\overline{v}_{x}\right].
\]
It is clear that under the model assumptions, $\mathbb{F}_{2}\left(\cdot\mid x\right)$
also concentrates on $C\left[\underline{v}_{x},\overline{v}_{x}\right]$.
Then we can show that the leading term on the right hand side of (\ref{eq:functional delta method result})
also converges in distribution: 
\begin{equation}
\sum_{d\in\left\{ 0,1\right\} }\psi_{dx}\left(\frac{\sqrt{n_{x}}\cdot H_{x}}{\varsigma_{dx}}\right)\rightsquigarrow\mathbb{F}_{2}\left(\cdot\mid x\right)\textrm{ in }\ell^{\infty}\text{\ensuremath{\left[\underline{v}_{x},\overline{v}_{x}\right]}}.\label{eq:psi weak convergence}
\end{equation}

Now it follows from (\ref{eq:S_F approximation}), (\ref{eq:F_til weak convergence}),
(\ref{eq:functional delta method result}), and (\ref{eq:psi weak convergence})
that $S_{F}\left(\cdot\mid x\right)$ converges in distribution to
a tight Gaussian process in $\ell^{\infty}\text{\ensuremath{\left[\underline{v}_{x},\overline{v}_{x}\right]}}$.
We present it as the first main result of this paper in the following
theorem.
\begin{thm}
\label{thm:weak_convergence}Suppose that Assumptions \ref{assu: DGP1},
\ref{assu: DGP4} and \ref{assu:sampling} hold. We have: (i) $S_{F}\left(\cdot\mid x\right)\rightsquigarrow\mathbb{F}\left(\cdot\mid x\right)$
in $\ell^{\infty}\left[\underline{v}_{x},\overline{v}_{x}\right]$,
as $n_{x}\uparrow\infty$, where $\mathbb{F}\left(\cdot\mid x\right)\coloneqq\mathbb{F}_{1}\left(\cdot\mid x\right)+\mathbb{F}_{2}\left(\cdot\mid x\right)$;
(ii) For any $v\in\left[\underline{v}_{x},\overline{v}_{x}\right]$,
we have $S_{F}\left(v\mid x\right)\rightsquigarrow\mathbb{F}\left(v\mid x\right)$,
where $\mathbb{F}\left(v\mid x\right)\sim\mathrm{N}\left(0,V_{F}\left(v\mid x\right)\right)$,
$V_{F}\left(v\mid x\right)\coloneqq V_{1}\left(v\mid x\right)+V_{2}\left(v\mid x\right)$
and 
\begin{eqnarray*}
V_{1}\left(v\mid x\right) & \coloneqq & F_{\varDelta\mid X}\left(v\mid x\right)\left(1-F_{\varDelta\mid X}\left(v\mid x\right)\right),\\
V_{2}\left(v\mid x\right) & \coloneqq & \mathrm{E}\left[\left\{ \sum_{j=1}^{m}\omega_{x,j}\left(v\right)\left\{ \mathbbm{1}\left(\epsilon\leq\varDelta_{x,j}^{-1}\left(v\right)\right)-F_{\epsilon\mid X}\left(\varDelta_{x,j}^{-1}\left(v\right)\mid x\right)\right\} \right\} ^{2}\mid X=x\right]\left(p_{1\mid x}^{-1}+p_{0\mid x}^{-1}\right).
\end{eqnarray*}
\end{thm}
\begin{rembold}\label{rmk: CDF estimator}Part (ii) shows that while
the empirical CDF using pseudo ITEs is still $\sqrt{n_{x}}$-consistent,
estimation of ITEs can have non-negligible contribution to the asymptotic
variance. $V_{1}\left(v\mid x\right)$ is the variance of the asymptotic
distribution of $\sqrt{n_{x}}\left(\widetilde{F}_{\varDelta\mid X}\left(v\mid x\right)-F_{\varDelta\mid X}\left(v\mid x\right)\right)$.
By using arguments similar to those in Remark 3 of \citetalias{Ma2023},
we can show that $V_{2}\left(v\mid x\right)>0$ under our assumptions.
Therefore, the asymptotic variance of $\widehat{F}_{\varDelta\mid X}\left(v\mid x\right)$
is always larger than that of the infeasible estimator $\widetilde{F}_{\varDelta\mid X}\left(v\mid x\right)$.
Given some consistent estimator of $V_{F}\left(v\mid x\right)$, we
can easily construct an asymptotically valid confidence interval for
$F_{\varDelta\mid X}\left(v\mid x\right)$. However, it is clear that
plug-in estimation of $V_{2}\left(v\mid x\right)$ is infeasible,
since it requires knowledge about the partition in Assumption \ref{assu: DGP4}
and also depends on several infinite-dimensional nuisance parameters
that are hard to estimate. E.g., estimation of $\varsigma_{dx}$ requires
using tuning parameters and nonparametric estimation of $\varDelta_{x,j}^{-1}$
is also complicated, since $\varDelta_{x,j}$ depends on the unknown
outcome equation. In Section \ref{sec:Bootstrap-inference}, we propose
constructing bootstrap percentile confidence intervals to circumvent
this problem and show that nonparametric bootstrap approximation to
the asymptotic distribution of $\mathbb{F}\left(v\mid x\right)$ is
asymptotically valid.\end{rembold}

\begin{rembold}\label{rmk: CDF confidence band}By the CMT, $\left\Vert S_{F}\left(\cdot\mid x\right)\right\Vert _{\left[\underline{v}_{x},\overline{v}_{x}\right]}\rightsquigarrow\left\Vert \mathbb{F}\left(\cdot\mid x\right)\right\Vert _{\left[\underline{v}_{x},\overline{v}_{x}\right]}$.
Since $\mathbb{F}\left(\cdot\mid x\right)$ concentrates on the separable
Banach space $C\left[\underline{v}_{x},\overline{v}_{x}\right]$,
the CDF of $\left\Vert \mathbb{F}\left(\cdot\mid x\right)\right\Vert _{\left[\underline{v}_{x},\overline{v}_{x}\right]}$
is continuous everywhere on $\mathbb{R}$ (see, e.g., \citealp[Exercise 2.4.4]{gine2016mathematical}).
Let $1-\alpha$ be the desired coverage probability for some $\alpha\in\left(0,1\right)$.
If the $\left(1-\alpha\right)$-th quantile of $\left\Vert \mathbb{F}\left(\cdot\mid x\right)\right\Vert _{\left[\underline{v}_{x},\overline{v}_{x}\right]}$
is known or can be consistently estimated by some estimator $\tilde{s}_{1-\alpha}$,
we can easily construct a UCB for the conditional CDF $F_{\varDelta\mid X}\left(\cdot\mid x\right)$
on $\left[\underline{v}_{x},\overline{v}_{x}\right]$.\footnote{If $\widetilde{s}_{1-\alpha}$ is a consistent estimator for the $\left(1-\alpha\right)$-th
quantile of $\left\Vert \mathbb{F}\left(\cdot\mid x\right)\right\Vert _{\left[\underline{v}_{x},\overline{v}_{x}\right]}$,
it follows from Slutsky's theorem and \citet[Lemma 21.1(ii)]{van2000asymptotic}
that the probability of the event $\left\Vert S_{F}\left(\cdot\mid x\right)\right\Vert _{\left[\underline{v}_{x},\overline{v}_{x}\right]}\leq\widetilde{s}_{1-\alpha}$
converges to $1-\alpha$. This result immediately implies that $\left\{ \widehat{F}_{\varDelta\mid X}\left(v\mid x\right)\pm\widetilde{s}_{1-\alpha}/\sqrt{n_{x}}:v\in\left[\underline{v}_{x},\overline{v}_{x}\right]\right\} $
is an asymptotically valid UCB. } However, due to the presence of the $\mathbb{F}_{2}$ term, whose
distribution depends on the unknown partition in Assumption \ref{assu: DGP4}
and also several other unknown infinite-dimensional nuisance parameters,
the distribution of $\left\Vert \mathbb{F}\left(\cdot\mid x\right)\right\Vert _{\left[\underline{v}_{x},\overline{v}_{x}\right]}$
cannot be tabulated or easily approximated by simulations. In Section
\ref{sec:Bootstrap-inference}, we show that the nonparametric bootstrap
estimator for the distribution of $\left\Vert \mathbb{F}\left(\cdot\mid x\right)\right\Vert _{\left[\underline{v}_{x},\overline{v}_{x}\right]}$
is consistent, relatively to the Kolmogorov-Smirnov distance.\footnote{The Kolmogorov-Smirnov distance between the probability distributions
of two random vectors is defined to be the sup-norm of $F-G$, where
$F$ and $G$ are their CDFs.}\end{rembold}

\subsection{Asymptotic Gaussianity of the empirical quantiles\label{subsec:Asymptotic-normality-quantiles}}

\begin{figure}[t]
\caption{Numerical example: $\tilde{V}_{1}$ versus $\tilde{V}_{2}$}

\centering{}\label{numerical example}\includegraphics[scale=0.4]{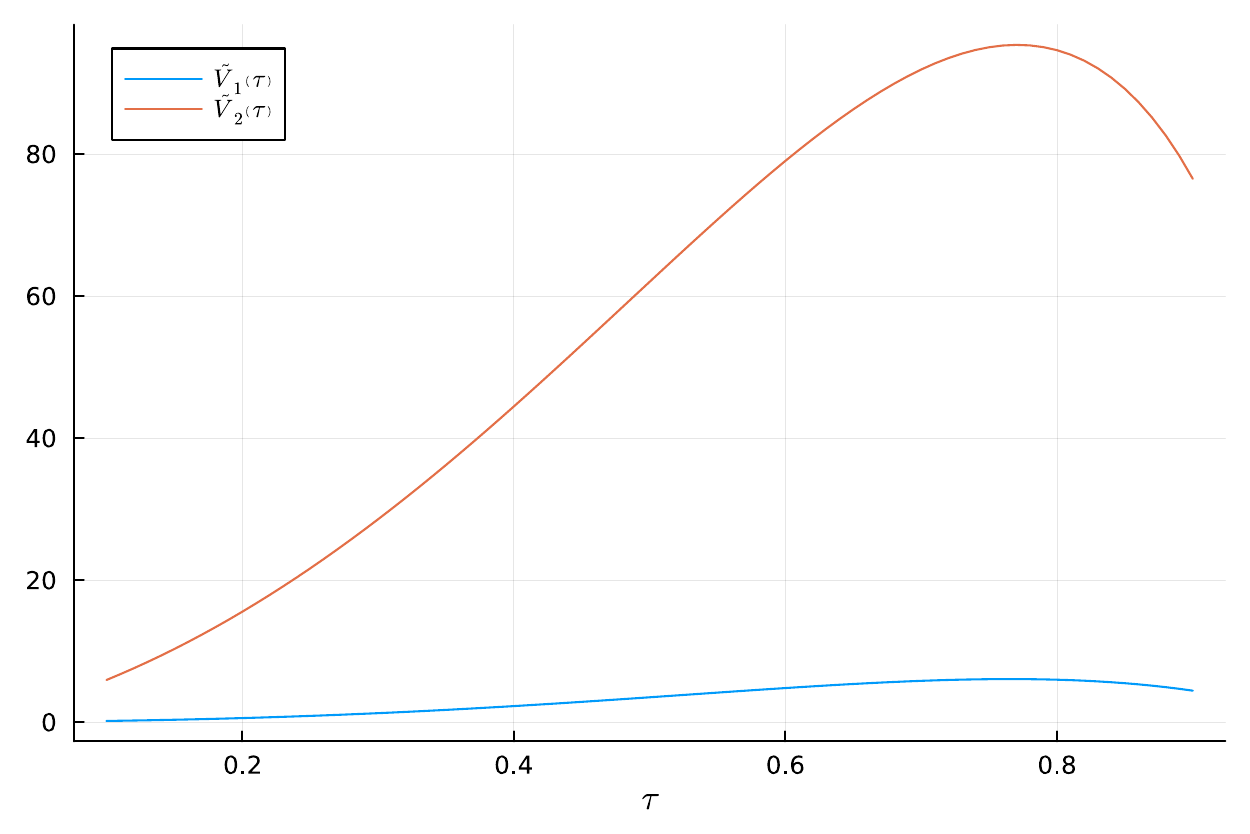}
\end{figure}

The estimator $\widehat{Q}_{\varDelta\mid X}\left(\cdot\mid x\right)$
of the ITE quantile function defined in (\ref{eq: ITE Quantile})
is a left continuous step function on $\left(0,1\right)$: for $\tau\in\left(0,1\right)$,
\begin{eqnarray*}
\widehat{Q}_{\varDelta\mid X}\left(\tau\mid x\right) & = & \sum_{j=1}^{n_{x}}\mathbbm{1}\left(\tau\in\left(\frac{j-1}{n_{x}},\frac{j}{n_{x}}\right]\right)\widehat{\varDelta}_{\left\langle j\right\rangle }\\
 & = & \widehat{\varDelta}_{\left\langle \left\lceil \tau n_{x}\right\rceil \right\rangle },
\end{eqnarray*}
where $\widehat{\varDelta}_{\left\langle 1\right\rangle }\leq\cdots\leq\widehat{\varDelta}_{\left\langle n_{x}\right\rangle }$
are the order statistics corresponding to the pseudo ITEs. Then, we
can show that the quantile estimator also has an asymptotically normal
distribution. This result is presented in the following corollary
to Theorem \ref{thm:weak_convergence}.
\begin{cor}
\label{cor:quantile}Suppose that Assumptions \ref{assu: DGP1}, \ref{assu: DGP4}
and \ref{assu:sampling} hold. (i) Let $0<\underline{\tau}<\overline{\tau}<1$.
We have
\[
S_{Q}\left(\cdot\mid x\right)\coloneqq\sqrt{n_{x}}\left(\widehat{Q}_{\varDelta\mid X}\left(\cdot\mid x\right)-Q_{\varDelta\mid X}\left(\cdot\mid x\right)\right)\rightsquigarrow\mathbb{Q}\left(\cdot\mid x\right)\textrm{ in }\text{\ensuremath{\ell^{\infty}\left[\underline{\tau},\overline{\tau}\right]}},
\]
where $\mathbb{Q}\left(\cdot\mid x\right)\coloneqq\mathbb{Q}_{1}\left(\cdot\mid x\right)+\mathbb{Q}_{2}\left(\cdot\mid x\right)$
and
\[
\mathbb{Q}_{j}\left(\tau\mid x\right)\coloneqq-\frac{\mathbb{F}_{j}\left(Q_{\varDelta\mid X}\left(\tau\mid x\right)\mid x\right)}{f_{\varDelta\mid X}\left(Q_{\varDelta\mid X}\left(\tau\mid x\right)\mid x\right)},\,\tau\in\left[\underline{\tau},\overline{\tau}\right],\,j=1,2;
\]
(ii) For any fixed $\tau\in\left[\underline{\tau},\overline{\tau}\right]$,
$S_{Q}\left(\tau\mid x\right)\rightsquigarrow\mathbb{Q}\left(\tau\mid x\right)$,
where $\mathbb{Q}\left(\tau\mid x\right)\sim\mathrm{N}\left(0,V_{Q}\left(\tau\mid x\right)\right)$,
$V_{Q}\left(\tau\mid x\right)\coloneqq\tilde{V}_{1}\left(\tau\mid x\right)+\tilde{V}_{2}\left(\tau\mid x\right)$
and 
\[
\tilde{V}_{j}\left(\tau\mid x\right)\coloneqq\frac{V_{j}\left(Q_{\varDelta\mid X}\left(\tau\mid x\right)\mid x\right)}{\left\{ f_{\varDelta\mid X}\left(Q_{\varDelta\mid X}\left(\tau\mid x\right)\mid x\right)\right\} ^{2}},\,j=1,2.
\]
\end{cor}
\begin{rembold}\label{rmk: numerical example}We now give a numerical
example. We consider the DGP for the Monte Carlo simulations in Section
\ref{sec:Monte-Carlo-simulations} and present numerical calculations
to illustrate the effect of estimation of the ITEs. Figure \ref{numerical example}
shows the contrast between the two variance components across $\tau\in\left[0.1,0.9\right]$.
It suggests that the contribution $\tilde{V}_{2}\left(\tau\right)$
from the ITE estimation errors to the asymptotic variance can be substantial
and much larger than the asymptotic variance $\tilde{V}_{1}\left(\tau\right)$
of the infeasible estimator.\end{rembold}

\begin{rembold}\label{rmk: quantile estimator}By the CMT, we have
$S_{Q}\left(\tau\mid x\right)\rightsquigarrow\mathbb{Q}\left(\tau\mid x\right)$
and $\left\Vert S_{Q}\left(\cdot\mid x\right)\right\Vert _{\left[\underline{\tau},\overline{\tau}\right]}\rightsquigarrow\left\Vert \mathbb{Q}\left(\cdot\mid x\right)\right\Vert _{\left[\underline{\tau},\overline{\tau}\right]}$.
Asymptotically valid confidence intervals and UCBs for the ITE quantiles
can be constructed by using consistent estimators of the distributions
of $\mathbb{Q}\left(\tau\mid x\right)$ and $\left\Vert \mathbb{Q}\left(\cdot\mid x\right)\right\Vert _{\left[\underline{\tau},\overline{\tau}\right]}$.
Similarly, the asymptotic variance of $\mathbb{Q}\left(\tau\mid x\right)$
and the distribution of $\left\Vert \mathbb{Q}\left(\cdot\mid x\right)\right\Vert _{\left[\underline{\tau},\overline{\tau}\right]}$
depend on infinite-dimensional nuisance parameters that are hard to
estimate (e.g., nonparametric estimation of $f_{\varDelta\mid X}$
requires using tuning parameters). In Section \ref{sec:Bootstrap-inference},
we show that nonparametric bootstrap approximation to these distributions
is asymptotically valid and this result implies that bootstrap percentile
confidence intervals and UCBs using bootstrap critical values are
asymptotically valid.\end{rembold}

\begin{rembold}Let $\widehat{\mathit{IR}}_{\varDelta\mid X=x}$ be
the ``plug-in'' estimator (i.e., the difference of $\widehat{Q}_{\varDelta\mid X}\left(0.75\mid x\right)$
and $\widehat{Q}_{\varDelta\mid X}\left(0.25\mid x\right)$). Since
$f\mapsto f\left(0.75\right)-f\left(0.25\right)$ as a map from $\ell^{\infty}\left[\underline{\tau},\overline{\tau}\right]$
into $\mathbb{R}$ is clearly continuous, by the CMT, we have
\[
\sqrt{n_{x}}\left(\widehat{\mathit{IR}}_{\varDelta\mid X=x}-\mathit{IR}_{\varDelta\mid X=x}\right)=S_{Q}\left(0.75\mid x\right)-S_{Q}\left(0.25\mid x\right)\rightsquigarrow\mathbb{Q}\left(0.75\mid x\right)-\mathbb{Q}\left(0.25\mid x\right).
\]
By using estimators of the quantiles of the Gaussian random variable
$\mathbb{Q}\left(0.75\mid x\right)-\mathbb{Q}\left(0.25\mid x\right)$,
we can construct confidence intervals for $\mathit{IR}_{\varDelta\mid X=x}$.
Results in the next section show that we can consistently estimate
the quantiles of $\mathbb{Q}\left(0.75\mid x\right)-\mathbb{Q}\left(0.25\mid x\right)$
by using nonparametric bootstrap.\end{rembold}

\section{Bootstrap inference\label{sec:Bootstrap-inference}}

It has been discussed in Remarks \ref{rmk: CDF estimator}, \ref{rmk: CDF confidence band}
and \ref{rmk: quantile estimator} that bootstrapping seems to be
a feasible approach to estimate the asymptotic distributions. In Section
\ref{subsec:Constructing-confidence-intervals-bands}, we discuss
the construction and the algorithms of the bootstrap-based confidence
intervals and UCBs. Section \ref{subsec:Asymptotic-validity-bootstrap}
is devoted to the presentation of the results showing the asymptotic
validity of the inference methods proposed in Section \ref{subsec:Constructing-confidence-intervals-bands}.

\subsection{Constructing bootstrap confidence intervals and UCBs\label{subsec:Constructing-confidence-intervals-bands}}

A nonparametric bootstrap sample $\left\{ W_{i}^{\dagger}\coloneqq\left(Y_{i}^{\dagger},D_{i}^{\dagger},Z_{i}^{\dagger}\right)^{\top}\right\} _{i=1}^{n_{x}}$
consists of $n_{x}$ independent draws from the original sample $\left\{ W_{i}\right\} _{i=1}^{n_{x}}$
with replacement. Let $\widehat{\varUpsilon}_{dx}^{\left(-i\right)\dagger}\left(t,y\right)$
denote the bootstrap analogue of $\widehat{\varUpsilon}_{dx}^{\left(-i\right)}\left(t,y\right)$,
i.e., $\widehat{\varUpsilon}_{dx}^{\left(-i\right)\dagger}\left(t,y\right)$
is given by the right hand side of (\ref{eq:Q_hat definition}) with
$\left\{ W_{j}\right\} _{j\in\left[n_{x}\right]\setminus\left\{ i\right\} }$
replaced by $\left\{ W_{j}^{\dagger}\right\} _{j\in\left[n_{x}\right]\setminus\left\{ i\right\} }$.
Let $\widehat{\phi}_{dx}^{\left(-i\right)\dagger}\left(y\right)$
be the bootstrap analogue of $\widehat{\phi}_{dx}^{\left(-i\right)}\left(y\right)$
defined by
\[
\widehat{\phi}_{dx}^{\left(-i\right)\dagger}\left(y\right)\coloneqq\underset{t\in\left[\underline{y}_{dx},\overline{y}_{dx}\right]}{\arg\min}\widehat{\varUpsilon}_{dx}^{\left(-i\right)\dagger}\left(t,y\right).
\]
Similarly, we construct the bootstrap analogues 
\[
\widehat{\varDelta}_{i}^{\dagger}\coloneqq D_{i}^{\dagger}\left(Y_{i}^{\dagger}-\widehat{\phi}_{0x}^{\left(-i\right)\dagger}\left(Y_{i}^{\dagger}\right)\right)+\left(1-D_{i}^{\dagger}\right)\left(\widehat{\phi}_{1x}^{\left(-i\right)\dagger}\left(Y_{i}^{\dagger}\right)-Y_{i}^{\dagger}\right).
\]
 and 
\begin{equation}
\widehat{F}_{\varDelta\mid X}^{\dagger}\left(v\mid x\right)\coloneqq\frac{1}{n_{x}}\sum_{i=1}^{n_{x}}\mathbbm{1}\left(\widehat{\varDelta}_{i}^{\dagger}\leq v\right),\,v\in\mathbb{R}.\label{eq:F_hat_dag definition}
\end{equation}

Let $v$ be an interior point of $\mathscr{S}_{\varDelta\mid X=x}$.
Let $\mathrm{Pr}_{\dagger}\left[\cdot\right]$ denote the conditional
probability given the original sample. Now we construct the (asymptotically
valid) bootstrap confidence interval for $F_{\varDelta\mid X}\left(v\mid x\right)$.
For $p\in\left(0,1\right)$, let 
\begin{equation}
s_{F,p}\left(v\mid x\right)\coloneqq\inf\left\{ u\in\mathbb{R}:\mathrm{Pr}_{\dagger}\left[\widehat{F}_{\varDelta\mid X}^{\dagger}\left(v\mid x\right)\leq u\right]\geq p\right\} \label{eq:s_F_pt definition}
\end{equation}
be the $p$-th quantile of the resampling distribution of $\widehat{F}_{\varDelta\mid X}^{\dagger}\left(v\mid x\right)$
(i.e., the conditional distribution of $\widehat{F}_{\varDelta\mid X}^{\dagger}\left(v\mid x\right)$
given the original data). Note that the resampling distribution of
$\widehat{F}_{\varDelta\mid X}^{\dagger}\left(v\mid x\right)$ can
be easily simulated. The bootstrap percentile confidence interval
with nominal coverage probability $1-\alpha$ for $F_{\varDelta\mid X}\left(v\mid x\right)$
is given by $\left[s_{F,\alpha/2}\left(v\mid x\right),s_{F,1-\alpha/2}\left(v\mid x\right)\right]$.
The following algorithm summarizes the procedure that uses simulations
to calculate the confidence interval $\left[s_{F,\alpha/2}\left(v\mid x\right),s_{F,1-\alpha/2}\left(v\mid x\right)\right]$.
Let $B$ denote the number of bootstrap replications. 
\begin{lyxalgorithm}[Bootstrap percentile confidence interval for cumulative probabilities]
\label{alg:confidence interval for CDF}Step 1: In each of the replications
$r\in\left[B\right]$, independently draw $\left\{ W_{i}^{\dagger\left(r\right)}\right\} _{i=1}^{n_{x}}$
with replacement from the original sample. Step 2: For all $r\in\left[B\right]$,
compute the pseudo ITEs $\left\{ \widehat{\varDelta}_{i}^{\dagger\left(r\right)}\right\} _{i=1}^{n_{x}}$
by applying (\ref{eq:Q_hat definition}), (\ref{eq:phi_hat definition}),
and (\ref{eq:pseudo ITE definition}) to the bootstrap sample in the
$r$-th replication. Step 3: Compute $\widehat{F}_{\varDelta\mid X}^{\dagger\left(r\right)}\left(v\mid x\right)$
using the formula (\ref{eq:F_hat_dag definition}) with $\widehat{\varDelta}_{i}^{\dagger}$
replaced by $\widehat{\varDelta}_{i}^{\dagger\left(r\right)}$, for
all $r\in\left[B\right]$. Step 4: Order $\left\{ \widehat{F}_{\varDelta\mid X}^{\dagger\left(r\right)}\left(v\mid x\right)\right\} _{r=1}^{B}$
and compute the corresponding order statistics $F_{\left\langle 1\right\rangle }^{\dagger}\leq\cdots\leq F_{\left\langle B\right\rangle }^{\dagger}$.
Step 5: Return the confidence interval $\left[F_{\left\langle \left\lceil B\times\left(\alpha/2\right)\right\rceil \right\rangle }^{\dagger},F_{\left\langle \left\lceil B\times\left(1-\alpha/2\right)\right\rceil \right\rangle }^{\dagger}\right]$
for $F_{\varDelta\mid X}\left(v\mid x\right)$.
\end{lyxalgorithm}
For any $\tau\in\left(0,1\right)$, it is also straightforward to
construct a bootstrap confidence interval for the $\tau$-th quantile
$Q_{\varDelta\mid X}\left(\tau\mid x\right)$ by adapting the preceding
algorithm. For $\tau\in\left(0,1\right)$, denote
\begin{eqnarray}
\widehat{Q}_{\varDelta\mid X}^{\dagger}\left(\tau\mid x\right) & \coloneqq & \inf\left\{ y\in\mathbb{R}:\widehat{F}_{\varDelta\mid X}^{\dagger}\left(y\mid x\right)\geq\tau\right\} \nonumber \\
 & = & \widehat{\varDelta}_{\left\langle \left\lceil \tau n_{x}\right\rceil \right\rangle }^{\dagger},\label{eq:Q_dag definition}
\end{eqnarray}
where $\widehat{\varDelta}_{\left\langle 1\right\rangle }^{\dagger}\leq\cdots\leq\widehat{\varDelta}_{\left\langle n_{x}\right\rangle }^{\dagger}$
are the order statistics corresponding to the pseudo ITEs from the
bootstrap sample. Let $\widehat{\mathit{IR}}_{\varDelta\mid X=x}^{\dagger}\coloneqq\widehat{Q}_{\varDelta\mid X}^{\dagger}\left(0.75\mid x\right)-\widehat{Q}_{\varDelta\mid X}^{\dagger}\left(0.25\mid x\right)$
be the bootstrap analogue of the estimated IQR. For $p\in\left(0,1\right)$,
let
\begin{eqnarray*}
s_{Q,p}\left(\tau\mid x\right) & \coloneqq & \inf\left\{ u\in\mathbb{R}:\mathrm{Pr}_{\dagger}\left[\widehat{Q}_{\varDelta\mid X}^{\dagger}\left(\tau\mid x\right)\leq u\right]\geq p\right\} \,\textrm{and}\\
s_{\mathit{IR},p} & \coloneqq & \inf\left\{ u\in\mathbb{R}:\mathrm{Pr}_{\dagger}\left[\widehat{\mathit{IR}}_{\varDelta\mid X=x}^{\dagger}\leq u\right]\geq p\right\} 
\end{eqnarray*}
be the $p$-th quantiles of the resampling distributions of $\widehat{Q}_{\varDelta\mid X}^{\dagger}\left(\tau\mid x\right)$
and $\widehat{\mathit{IR}}_{\varDelta\mid X=x}^{\dagger}$. Similarly,
these resampling distributions can be simulated. The bootstrap percentile
confidence intervals for the quantile and the IQR are given by $\left[s_{Q,\alpha/2}\left(\tau\mid x\right),s_{Q,1-\alpha/2}\left(\tau\mid x\right)\right]$
and $\left[s_{\mathit{IR},\alpha/2},s_{\mathit{IR},1-\alpha/2}\right]$.
The following algorithm summarizes the simulation procedure for calculating
these confidence intervals.
\begin{lyxalgorithm}[Bootstrap percentile confidence intervals for the quantiles]
\label{alg:confidence interval for quantile}Steps 1-2: Same as those
in Algorithm \ref{alg:confidence interval for CDF}. Step 3: Order
$\left\{ \widehat{\varDelta}_{i}^{\dagger\left(r\right)}\right\} _{i=1}^{n_{x}}$
to get the corresponding order statistics $\widehat{\varDelta}_{\left\langle 1\right\rangle }^{\dagger\left(r\right)}\leq\cdots\leq\widehat{\varDelta}_{\left\langle n_{x}\right\rangle }^{\dagger\left(r\right)}$,
for all $r\in\left[B\right]$. Step 4: Compute $\widehat{Q}_{\varDelta\mid X}^{\dagger\left(r\right)}\left(\tau\mid x\right)$
and $\widehat{Q}_{\varDelta\mid X}^{\dagger\left(r\right)}\left(0.75\mid x\right)-\widehat{Q}_{\varDelta\mid X}^{\dagger\left(r\right)}\left(0.25\mid x\right)$
using the formula (\ref{eq:Q_dag definition}) with $\widehat{\varDelta}_{\left\langle j\right\rangle }^{\dagger}$
replaced by $\widehat{\varDelta}_{\left\langle j\right\rangle }^{\dagger\left(r\right)}$
for all $r\in\left[B\right]$. Step 5: Order $\left\{ \widehat{Q}_{\varDelta\mid X}^{\dagger\left(r\right)}\left(\tau\mid x\right)\right\} _{r=1}^{B}$
and $\left\{ \widehat{Q}_{\varDelta\mid X}^{\dagger\left(r\right)}\left(0.75\mid x\right)-\widehat{Q}_{\varDelta\mid X}^{\dagger\left(r\right)}\left(0.25\mid x\right)\right\} _{r=1}^{B}$,
and compute the corresponding order statistics $Q_{\left\langle 1\right\rangle }^{\dagger}\leq\cdots\leq Q_{\left\langle B\right\rangle }^{\dagger}$
and $\mathit{IR}_{\left\langle 1\right\rangle }^{\dagger}\leq\cdots\leq\mathit{IR}_{\left\langle B\right\rangle }^{\dagger}$.
Step 6: Return the confidence interval $\left[Q_{\left\langle \left\lceil B\times\left(\alpha/2\right)\right\rceil \right\rangle }^{\dagger},Q_{\left\langle \left\lceil B\times\left(1-\alpha/2\right)\right\rceil \right\rangle }^{\dagger}\right]$
for the quantile and the confidence interval $\left[\mathit{IR}_{\left\langle \left\lceil B\times\left(\alpha/2\right)\right\rceil \right\rangle }^{\dagger},\mathit{IR}_{\left\langle \left\lceil B\times\left(1-\alpha/2\right)\right\rceil \right\rangle }^{\dagger}\right]$
for the IQR.
\end{lyxalgorithm}
Next, we consider constructing bootstrap UCBs for the CDF over any
inner closed sub-interval $\left[\underline{v}_{x},\overline{v}_{x}\right]$
of $\mathscr{S}_{\varDelta\mid X=x}$. Denote 
\begin{equation}
S_{F}^{\dagger}\left(v\mid x\right)\coloneqq\sqrt{n_{x}}\left(\widehat{F}_{\varDelta\mid X}^{\dagger}\left(v\mid x\right)-\widehat{F}_{\varDelta\mid X}\left(v\mid x\right)\right).\label{eq:S_F_dag definition}
\end{equation}
For $p\in\left(0,1\right)$, let 
\begin{equation}
s_{F,p}^{\mathsf{unif}}\coloneqq\inf\left\{ u\in\mathbb{R}:\mathrm{Pr}_{\dagger}\left[\left\Vert S_{F}^{\dagger}\left(\cdot\mid x\right)\right\Vert _{\left[\underline{v}_{x},\overline{v}_{x}\right]}\leq u\right]\geq p\right\} \label{eq:s_F_unif definition}
\end{equation}
be the $p$-th quantile of the resampling distribution of $\left\Vert S_{F}^{\dagger}\left(\cdot\mid x\right)\right\Vert _{\left[\underline{v}_{x},\overline{v}_{x}\right]}$.
Then, we construct the UCB with the nominal coverage probability $1-\alpha$
from the following continuum 
\begin{equation}
\mathit{CB}_{F}\left(v\mid x\right)\coloneqq\widehat{F}_{\varDelta\mid X}\left(v\mid x\right)\pm\frac{s_{F,1-\alpha}^{\mathsf{unif}}}{\sqrt{n_{x}}},\,v\in\left[\underline{v}_{x},\overline{v}_{x}\right],\label{eq:CB}
\end{equation}
of random intervals using the critical value $s_{F,1-\alpha}^{\mathsf{unif}}$.
The following discretization algorithm summarizes the simulation procedure
for computing the bootstrap UCB $\left\{ \mathit{CB}_{F}\left(v\mid x\right):v\in\left[\underline{v}_{x},\overline{v}_{x}\right]\right\} $
for the ITE CDF. Let $T$ be a large positive integer and let $\mathcal{V}_{x}\coloneqq\left\{ v_{x}^{\left(1\right)},...,v_{x}^{\left(T\right)}\right\} $
be equally spaced grid points in $\left[\underline{v}_{x},\overline{v}_{x}\right]$.
\begin{lyxalgorithm}[Bootstrap UCB for the CDF]
\label{alg:confidence bands for CDF}Steps 1-2: Same as those in
Algorithm \ref{alg:confidence interval for CDF}. Step 3: Compute
$\widehat{F}_{\varDelta\mid X}^{\dagger\left(r\right)}\left(v\mid x\right)$
for $\left\{ r,v\right\} \in\left[B\right]\times\mathcal{V}_{x}$
and compute $\widehat{F}_{\varDelta\mid X}\left(v\mid x\right)$ for
$v\in\mathcal{V}_{x}$. Step 4: Compute and order
\[
\left\{ \max_{v\in\mathcal{V}_{x}}\left|\widehat{F}_{\varDelta\mid X}^{\dagger\left(r\right)}\left(v\mid x\right)-\widehat{F}_{\varDelta\mid X}\left(v\mid x\right)\right|\right\} _{r=1}^{B}
\]
to get the corresponding order statistics $s_{F,\left\langle 1\right\rangle }^{\dagger}\leq\cdots\leq s_{F,\left\langle B\right\rangle }^{\dagger}$
and the critical value $s_{F,\left\langle \left\lceil B\left(1-\alpha\right)\right\rceil \right\rangle }^{\dagger}$.
Step 5: Return the UCB $\left\{ \widehat{F}_{\varDelta\mid X}\left(v\mid x\right)\pm s_{F,\left\langle \left\lceil B\left(1-\alpha\right)\right\rceil \right\rangle }^{\dagger}\right\} _{v\in\mathcal{V}_{x}}$.
\end{lyxalgorithm}
Similarly, we can also construct bootstrap UCBs for the ITE quantile
function over the range $\left[\underline{\tau},\overline{\tau}\right]$
for any $0<\underline{\tau}<\overline{\tau}<1$. Let
\begin{equation}
S_{Q}^{\dagger}\left(\tau\mid x\right)\coloneqq\sqrt{n_{x}}\left(\widehat{Q}_{\varDelta\mid X}^{\dagger}\left(\tau\mid x\right)-\widehat{Q}_{\varDelta\mid X}\left(\tau\mid x\right)\right).\label{eq:S_Q_dag definition}
\end{equation}
The bootstrap UCB with the nominal coverage probability $1-\alpha$
is given by the continuum of intervals
\begin{equation}
\mathit{CB}_{Q}\left(\tau\mid x\right)\coloneqq\widehat{Q}_{\varDelta\mid X}\left(\tau\mid x\right)\pm\frac{s_{Q,1-\alpha}^{\mathsf{unif}}}{\sqrt{n_{x}}},\,\tau\in\left[\underline{\tau},\overline{\tau}\right],\label{eq:CB_Q}
\end{equation}
where $s_{Q,1-\alpha}^{\mathsf{unif}}$ is the $\left(1-\alpha\right)$-th
quantile of the resampling distribution of $\left\Vert S_{Q}^{\dagger}\left(\cdot\mid x\right)\right\Vert _{\left[\underline{\tau},\overline{\tau}\right]}$.
We summarize the procedure for computing $\left\{ \mathit{CB}_{Q}\left(\tau\mid x\right):\tau\in\left[\underline{\tau},\overline{\tau}\right]\right\} $
in the following algorithm. Let $T$ be a large positive integer and
let $\mathcal{T}\coloneqq\left\{ \tau^{\left(1\right)},...,\tau^{\left(T\right)}\right\} $
be equally spaced grid points in $\left[\underline{\tau},\overline{\tau}\right]$.
\begin{lyxalgorithm}[Bootstrap UCB for the quantile function]
\label{alg:confidence band quantile}Steps 1-3: Same as those in
Algorithm \ref{alg:confidence interval for quantile}. Step 4: Compute
$\widehat{Q}_{\varDelta\mid X}^{\dagger\left(r\right)}\left(\tau\mid x\right)$
for $\left\{ r,\tau\right\} \in\left[B\right]\times\mathcal{T}$ and
compute $\widehat{Q}_{\varDelta\mid X}\left(\tau\mid x\right)$ for
$\tau\in\mathcal{T}$. Step 5: Compute
\[
\left\{ \max_{\tau\in\mathcal{T}}\left|\widehat{Q}_{\varDelta\mid X}^{\dagger\left(r\right)}\left(\tau\mid x\right)-\widehat{Q}_{\varDelta\mid X}\left(\tau\mid x\right)\right|\right\} _{r=1}^{B}
\]
 and order them to get the corresponding order statistics $s_{Q,\left\langle 1\right\rangle }^{\dagger}\leq\cdots\leq s_{Q,\left\langle B\right\rangle }^{\dagger}$
and the critical value $s_{Q,\left\langle \left\lceil B\left(1-\alpha\right)\right\rceil \right\rangle }^{\dagger}$.
Step 6: Return the UCB $\left\{ \widehat{Q}_{\varDelta\mid X}\left(\tau\mid x\right)\pm s_{Q,\left\langle \left\lceil B\left(1-\alpha\right)\right\rceil \right\rangle }^{\dagger}\right\} _{\tau\in\mathcal{T}}$.
\end{lyxalgorithm}
Next, we consider variable-width UCBs that are based on studentized
statistics. One of the advantages of variable-width UCBs is that they
adjust to local variability and are narrower where the function is
estimated more precisely, i.e., the estimator has a smaller pointwise
variance. We follow the approach of \citet{Chernozhukov2018} to construct
a variable-width UCB. Recall that $s_{Q,p}\left(\tau\mid x\right)$
is defined to be the $p$-th quantile of the resampling distribution
of $\widehat{Q}_{\varDelta\mid X}^{\dagger}\left(\tau\mid x\right)$.
Then it is clear that $\sqrt{n_{x}}\left(s_{Q,p}\left(\tau\mid x\right)-\widehat{Q}_{\varDelta\mid X}\left(\tau\mid x\right)\right)$
is the $p$-th quantile of the resampling distribution of $S_{Q}^{\dagger}\left(\tau\mid x\right)$.
In the proof of Corollary \ref{cor:bootstrap_quantile}, we show that
$\sqrt{n_{x}}\left(s_{Q,p}\left(\tau\mid x\right)-\widehat{Q}_{\varDelta\mid X}\left(\tau\mid x\right)\right)$
consistently estimates the $p$-th quantile of $\mathbb{Q}\left(\tau\mid x\right)\sim\mathrm{N}\left(0,V_{Q}\left(\tau\mid x\right)\right)$.
Therefore, a consistent estimator of $V_{Q}\left(\tau\mid x\right)$
is given by 
\[
n_{x}\left(\frac{s_{Q,0.75}\left(\tau\mid x\right)-s_{Q,0.25}\left(\tau\mid x\right)}{z_{0.75}-z_{0.25}}\right)^{2},
\]
where $z_{p}$ denotes the $p$-th quantile of $\mathrm{N}\left(0,1\right)$
and $s_{Q,0.75}\left(\tau\mid x\right)-s_{Q,0.25}\left(\tau\mid x\right)$
is the IQR of the resampling distribution of $\widehat{Q}_{\varDelta\mid X}^{\dagger}\left(\tau\mid x\right)$.
Let 
\[
\tilde{s}_{Q,p}^{\mathsf{unif}}\coloneqq\inf\left\{ u\in\mathbb{R}:\mathrm{Pr}_{\dagger}\left[\underset{\tau\in\left[\underline{\tau},\overline{\tau}\right]}{\sup}\frac{\left|\widehat{Q}_{\varDelta\mid X}^{\dagger}\left(\tau\mid x\right)-\widehat{Q}_{\varDelta\mid X}\left(\tau\mid x\right)\right|}{\left(s_{Q,0.75}\left(\tau\mid x\right)-s_{Q,0.25}\left(\tau\mid x\right)\right)/\left(z_{0.75}-z_{0.25}\right)}\leq u\right]\geq p\right\} 
\]
be the quantile of the resampling distribution of the supremum of
the studentized version of $\left|S_{Q}^{\dagger}\left(\cdot\mid x\right)\right|$.
A variable-width UCB is given by the continuum $\left\{ \widetilde{\mathit{CB}}_{Q}\left(\tau\mid x\right):\tau\in\left[\underline{\tau},\overline{\tau}\right]\right\} $
of intervals, where
\begin{equation}
\widetilde{\mathit{CB}}_{Q}\left(\tau\mid x\right)\coloneqq\widehat{Q}_{\varDelta\mid X}\left(\tau\mid x\right)\pm\tilde{s}_{Q,1-\alpha}^{\mathsf{unif}}\left(\frac{s_{Q,0.75}\left(\tau\mid x\right)-s_{Q,0.25}\left(\tau\mid x\right)}{z_{0.75}-z_{0.25}}\right),\,\tau\in\left[\underline{\tau},\overline{\tau}\right].\label{eq:CB_Q variable width}
\end{equation}

A procedure to calculate the variable-width UCB consists of steps
that are adaptations of those in Algorithms \ref{alg:confidence interval for quantile}
and \ref{alg:confidence band quantile}. We summarize the procedure
in the following algorithm.
\begin{lyxalgorithm}[Variable-width bootstrap UCB for the quantile function]
\label{alg:variable width quantile}Step 1-4: Same as those in Algorithms
\ref{alg:confidence bands for CDF}. Step 5: Compute the order statistics
$Q_{\left\langle 1\right\rangle }^{\dagger}\left(\tau\mid x\right)\leq\cdots\leq Q_{\left\langle B\right\rangle }^{\dagger}\left(\tau\mid x\right)$
corresponding to $\left\{ \widehat{Q}_{\varDelta\mid X}^{\dagger\left(r\right)}\left(\tau\mid x\right)\right\} _{r\in\left[B\right]}$
for all $\tau\in\mathcal{T}$. Step 6: compute
\[
\left\{ \underset{\tau\in\mathcal{T}}{\max}\frac{\left|\widehat{Q}_{\varDelta\mid X}^{\dagger\left(r\right)}\left(\tau\mid x\right)-\widehat{Q}_{\varDelta\mid X}\left(\tau\mid x\right)\right|}{\left(Q_{\left\langle \left\lceil B\times0.75\right\rceil \right\rangle }^{\dagger}\left(\tau\mid x\right)-Q_{\left\langle \left\lceil B\times0.25\right\rceil \right\rangle }^{\dagger}\left(\tau\mid x\right)\right)/\left(z_{0.75}-z_{0.25}\right)}\right\} _{r=1}^{B}
\]
and get the corresponding statistics $\tilde{s}_{Q,\left\langle 1\right\rangle }^{\dagger}\leq\cdots\leq\tilde{s}_{Q,\left\langle B\right\rangle }^{\dagger}$
and the critical value $\tilde{s}_{Q,\left\langle \left\lceil B\left(1-\alpha\right)\right\rceil \right\rangle }^{\dagger}$.
Step 7: Return the variable-width UCB
\[
\left\{ \widehat{Q}_{\varDelta\mid X}\left(\tau\mid x\right)\pm\tilde{s}_{Q,\left\langle \left\lceil B\left(1-\alpha\right)\right\rceil \right\rangle }^{\dagger}\left(\frac{Q_{\left\langle \left\lceil B\times0.75\right\rceil \right\rangle }^{\dagger}\left(\tau\mid x\right)-Q_{\left\langle \left\lceil B\times0.25\right\rceil \right\rangle }^{\dagger}\left(\tau\mid x\right)}{z_{0.75}-z_{0.25}}\right)\right\} _{\tau\in\mathcal{T}}.
\]
\end{lyxalgorithm}
A variable-width UCB for the CDF can be defined analogously. The procedure
for computation is similar to Algorithm \ref{alg:variable width quantile}.
We omit the details for simplicity.

Now it remains to show the asymptotic validity of these inference
methods. We will show that the validity results essentially follow
from bootstrap analogues of Theorem \ref{thm:weak_convergence} and
Corollary \ref{cor:quantile}.

\subsection{Asymptotic validity\label{subsec:Asymptotic-validity-bootstrap}}

Let $\mathrm{E}_{\dagger}\left[\cdot\right]$ denote the conditional
expectation given the original sample. Suppose that $\mathbb{W}_{n_{x}}$
is a map (from the underlying probability space) into some Banach
space $\mathbb{D}$. $\mathbb{W}_{n_{x}}$ depends on the bootstrap
sample, and let $\mathbb{W}$ be a tight random element in $\mathbb{D}$,
we use ``$\mathbb{W}_{n_{x}}\rightsquigarrow_{\dagger}\mathbb{W}$
in $\mathbb{D}$'' to denote convergence in distribution conditional
on the original data: ``$\mathbb{W}_{n_{x}}\rightsquigarrow_{\dagger}\mathbb{W}$
in $\mathbb{D}$'' is understood as
\[
\underset{h\in\mathit{BL}_{1}\left(\mathbb{D}\right)}{\sup}\left|\mathrm{E}_{\dagger}\left[h\left(\mathbb{W}_{n_{x}}\right)\right]-\mathrm{E}\left[h\left(\mathbb{W}\right)\right]\right|\rightarrow_{p}0,
\]
as $n_{x}\uparrow\infty$ (see \citealp[Chapter 23.2.1]{van2000asymptotic}).
The following result shows that for any inner closed sub-interval
$\text{\ensuremath{\left[\underline{v}_{x},\overline{v}_{x}\right]}}$
of $\mathscr{S}_{\varDelta\mid X=x}$, the bootstrap analogue $S_{F}^{\dagger}\left(\cdot\mid x\right)$
of $S_{F}\left(\cdot\mid x\right)$, defined by (\ref{eq:S_F_dag definition}),
as a map from the underlying probability space into $\ell^{\infty}\left[\underline{v}_{x},\overline{v}_{x}\right]$
converges in distribution to the same limiting random element $\mathbb{F}\left(\cdot\mid x\right)$.
It can be viewed as a bootstrap analogue of Theorem \ref{thm:weak_convergence}(i).
\begin{thm}
\label{thm:bootstrap_weak_convergence}Suppose that Assumptions \ref{assu: DGP1},
\ref{assu: DGP4} and \ref{assu:sampling} hold. We have $S_{F}^{\dagger}\left(\cdot\mid x\right)\rightsquigarrow_{\dagger}\mathbb{F}\left(\cdot\mid x\right)$
in $\ell^{\infty}\left[\underline{v}_{x},\overline{v}_{x}\right]$.
\end{thm}
\begin{rembold}Since both $f\mapsto f\left(v\right)$ and $f\mapsto\left\Vert f\right\Vert _{\left[\underline{v}_{x},\overline{v}_{x}\right]}$
as maps from $\ell^{\infty}\left[\underline{v}_{x},\overline{v}_{x}\right]$
to $\mathbb{R}$ are Lipschitz continuous, by the bootstrap analogue
of the CMT (see, e.g., \citealp[Proposition 10.7]{Kosorok2007}),
we have $S_{F}^{\dagger}\left(v\mid x\right)\rightsquigarrow_{\dagger}\mathbb{F}\left(v\mid x\right)$
and $\left\Vert S_{F}^{\dagger}\left(\cdot\mid x\right)\right\Vert _{\left[\underline{v}_{x},\overline{v}_{x}\right]}\rightsquigarrow_{\dagger}\left\Vert \mathbb{F}\left(\cdot\mid x\right)\right\Vert _{\left[\underline{v}_{x},\overline{v}_{x}\right]}$
in $\mathbb{R}$. For fixed $v\in\left[\underline{v}_{x},\overline{v}_{x}\right]$,
\begin{equation}
\underset{u\in\mathbb{R}}{\sup}\left|\mathrm{Pr}_{\dagger}\left[S_{F}^{\dagger}\left(v\mid x\right)\leq u\right]-\Pr\left[\mathbb{F}\left(v\mid x\right)\leq u\right]\right|\rightarrow_{p}0\label{eq:S_F bootstrap convergence}
\end{equation}
follows from $S_{F}^{\dagger}\left(v\mid x\right)\rightsquigarrow_{\dagger}\mathbb{F}\left(v\mid x\right)$,
the subsequence lemma (see, e.g., \citealp[Theorem 18.6]{davidson1994stochastic})
and \citet[Lemma 10.12]{Kosorok2007}. And similarly, we have
\begin{equation}
\underset{u\in\mathbb{R}}{\sup}\left|\mathrm{Pr}_{\dagger}\left[\left\Vert S_{F}^{\dagger}\left(\cdot\mid x\right)\right\Vert _{\left[\underline{v}_{x},\overline{v}_{x}\right]}\leq u\right]-\Pr\left[\left\Vert \mathbb{F}\left(\cdot\mid x\right)\right\Vert _{\left[\underline{v}_{x},\overline{v}_{x}\right]}\leq u\right]\right|\rightarrow_{p}0.\label{eq:S_F_sup bootstrap convergence}
\end{equation}
(\ref{eq:S_F bootstrap convergence}) and (\ref{eq:S_F_sup bootstrap convergence})
show that the resampling distributions of $S_{F}^{\dagger}\left(v\mid x\right)$
and $\left\Vert S_{F}^{\dagger}\left(\cdot\mid x\right)\right\Vert _{\left[\underline{v}_{x},\overline{v}_{x}\right]}$
consistently estimate the distributions of $\mathbb{F}\left(v\mid x\right)$
and $\left\Vert \mathbb{F}\left(\cdot\mid x\right)\right\Vert _{\left[\underline{v}_{x},\overline{v}_{x}\right]}$,
relatively to the Kolmogorov-Smirnov distance.\end{rembold}

The asymptotic validity of the confidence interval $\left[s_{F,\alpha/2}\left(v\mid x\right),s_{F,1-\alpha/2}\left(v\mid x\right)\right]$
for $F_{\varDelta\mid X}\left(v\mid x\right)$ and the UCB $\left\{ \mathit{CB}_{F}\left(v\mid x\right):v\in\left[\underline{v}_{x},\overline{v}_{x}\right]\right\} $
for $F_{\varDelta\mid X}\left(v\mid x\right)$ over $v\in\left[\underline{v}_{x},\overline{v}_{x}\right]$
essentially follows from the stochastic convergence results (\ref{eq:S_F bootstrap convergence})
and (\ref{eq:S_F_sup bootstrap convergence}) stated in the preceding
remark and also the fact that the Kolmogorov-Smirnov distance between
the distribution of $S_{F}\left(v\mid x\right)$ (or $\left\Vert S_{F}\left(\cdot\mid x\right)\right\Vert _{\left[\underline{v}_{x},\overline{v}_{x}\right]}$)
and the distribution of $\mathbb{F}\left(v\mid x\right)$ (or $\left\Vert \mathbb{F}\left(\cdot\mid x\right)\right\Vert _{\left[\underline{v}_{x},\overline{v}_{x}\right]}$)
converges to zero, which follows from \citet[Lemma 2.11]{van2000asymptotic}
and the continuity of the CDF of $\left\Vert \mathbb{F}\left(\cdot\mid x\right)\right\Vert _{\left[\underline{v}_{x},\overline{v}_{x}\right]}$.
We present the asymptotic validity results in the following corollary.
For simplicity, we give the result for the constant-width UCB only.
The validity of the variable-width UCB follows from similar arguments.

\begin{cor}
\label{cor:consistency_of_CDF_bootstrap_interval}Under Assumptions
\ref{assu: DGP1}, \ref{assu: DGP4} and \ref{assu:sampling}, we
have: (i) for all $v\in\left[\underline{v}_{x},\overline{v}_{x}\right]$,
as $n_{x}\uparrow\infty$,
\begin{equation}
\Pr\left[F_{\varDelta\mid X}\left(v\mid x\right)\in\left[s_{F,\alpha/2}\left(v\mid x\right),s_{F,1-\alpha/2}\left(v\mid x\right)\right]\right]\rightarrow1-\alpha;\label{eq:F pointwise interval validity}
\end{equation}
(ii) as $n_{x}\uparrow\infty$,
\begin{equation}
\Pr\left[F_{\varDelta\mid X}\left(v\mid x\right)\in\mathit{CB}_{F}\left(v\mid x\right),\,\forall v\in\left[\underline{v}_{x},\overline{v}_{x}\right]\right]\rightarrow1-\alpha.\label{eq:F uniform band validity}
\end{equation}
\end{cor}
Similarly, we can show a bootstrap analogue of Corollary \ref{cor:quantile}(i).
By using this result and similar arguments as those used in the proof
of Corollary \ref{cor:consistency_of_CDF_bootstrap_interval}, we
can show the asymptotic validity of the bootstrap percentile confidence
intervals $\left[s_{Q,\alpha/2}\left(\tau\mid x\right),s_{Q,1-\alpha/2}\left(\tau\mid x\right)\right]$
and $\left[s_{\mathit{IR},\alpha/2},s_{\mathit{IR},1-\alpha/2}\right]$
for the quantile $Q_{\varDelta\mid X}\left(\tau\mid x\right)$ and
the IQR defined by (\ref{eq:interquartile range}), and also the UCB
$\left\{ \mathit{CB}_{Q}\left(\tau\mid x\right):\tau\in\left[\underline{\tau},\overline{\tau}\right]\right\} $
for $Q_{\varDelta\mid X}\left(\tau\mid x\right)$ over $\tau\in\left[\underline{\tau},\overline{\tau}\right]$.
These results are summarized in the following corollary.
\begin{cor}
\label{cor:bootstrap_quantile}Under Assumptions \ref{assu: DGP1},
\ref{assu: DGP4} and \ref{assu:sampling}, we have: (i) $S_{Q}^{\dagger}\left(\cdot\mid x\right)\rightsquigarrow_{\dagger}\mathbb{Q}\left(\cdot\mid x\right)$
in $\ell^{\infty}\left[\underline{\tau},\overline{\tau}\right]$;
(ii) for each $\tau\in\left(0,1\right)$, as $n_{x}\uparrow\infty$,
\[
\Pr\left[Q_{\varDelta\mid X}\left(\tau\mid x\right)\in\left[s_{Q,\alpha/2}\left(\tau\mid x\right),s_{Q,1-\alpha/2}\left(\tau\mid x\right)\right]\right]\rightarrow1-\alpha;
\]
(iii) as $n_{x}\uparrow\infty$,
\[
\Pr\left[\mathit{IR}_{\varDelta\mid X=x}\in\left[s_{\mathit{IR},\alpha/2},s_{\mathit{IR},1-\alpha/2}\right]\right]\rightarrow1-\alpha;
\]
(iv) as $n_{x}\uparrow\infty$,
\[
\Pr\left[Q_{\varDelta\mid X}\left(\tau\mid x\right)\in\mathit{CB}_{Q}\left(\tau\mid x\right),\,\forall\tau\in\left[\underline{\tau},\overline{\tau}\right]\right]\rightarrow1-\alpha.
\]
\end{cor}

\section{Extensions\label{sec:Extensions}}

This section is devoted to the presentation of several useful extensions
to the results and algorithms given in the preceding section. Section
\ref{subsec:Conditioning-on-sub-vectors} considers inference on the
ITE distribution conditional on a sub-vector of the covariate vector
$X$.

In many empirical applications, the econometrician is interested in
analyzing and comparing heterogeneous treatment effects in subgroups
corresponding to different covariate values. Let $x_{1}$ and $x_{2}$
be two different values in $\mathscr{S}_{X}$. It would be of interest
to compare the two ITE distributions ``$\varDelta$ given $X=x_{1}$''
versus ``$\varDelta$ given $X=x_{2}$''. To this end, being interested
in comparing central tendencies (or dispersions), one can employ the
estimation and inference methods proposed in the preceding section
and compare the confidence intervals for $Q_{\varDelta\mid X}\left(0.5\mid x_{1}\right)$
and $Q_{\varDelta\mid X}\left(0.5\mid x_{2}\right)$ (or those for
$\mathit{IR}_{\varDelta\mid X=x_{1}}$ and $\mathit{IR}_{\varDelta\mid X=x_{2}}$).
Another more transparent approach is to construct confidence intervals
for the differences $Q_{\varDelta\mid X}\left(0.5\mid x_{1}\right)-Q_{\varDelta\mid X}\left(0.5\mid x_{2}\right)$
or $\mathit{IR}_{\varDelta\mid X=x_{1}}-\mathit{IR}_{\varDelta\mid X=x_{2}}$.
One may be also interested in making judgement about equality of the
entire ITE distributions, rather than comparing certain summary measures.
This can be facilitated by computing and comparing the UCBs of $Q_{\varDelta\mid X}\left(\cdot\mid x_{1}\right)$
and $Q_{\varDelta\mid X}\left(\cdot\mid x_{2}\right)$. Similarly,
one can also refer to an estimate and a UCB of the quantile difference
function $Q_{\varDelta\mid X}\left(\cdot\mid x_{1}\right)-Q_{\varDelta\mid X}\left(\cdot\mid x_{2}\right)$.
E.g., a constant quantile difference function suggests that the two
ITE distributions are the same up to a location shift and a monotonic
quantile difference function suggests that one ITE distribution is
more dispersed than the other. In Section \ref{subsec:comparison},
we present results and algorithms related to the problem of inference
on quantile differences.

\subsection{Conditioning on sub-vectors of the covariates\label{subsec:Conditioning-on-sub-vectors}}

Suppose that $\tilde{X}$ is a sub-vector of $X$ and let $\tilde{X}_{i}$
denote the corresponding sub-vector of $X_{i}$. Let $A$ be a subset
of $\mathscr{S}_{\tilde{X}}$. Let $F_{\varDelta\mid\tilde{X}}\left(v\mid A\right)\coloneqq\mathrm{Pr}\left[\varDelta\leq v\mid\tilde{X}\in A\right]$
be the conditional CDF of $\varDelta$ given $\tilde{X}\in A$. For
$\tau\in\left(0,1\right)$, let $Q_{\varDelta\mid\tilde{X}}\left(\tau\mid A\right)\coloneqq\inf\left\{ y\in\mathbb{R}:F_{\varDelta\mid\tilde{X}}\left(y\mid A\right)\geq\tau\right\} $
denote the $\tau$-th quantile. Note that $A$ can be taken to be
$\mathscr{S}_{\tilde{X}}$ such that $F_{\varDelta\mid\tilde{X}}\left(\cdot\mid A\right)$
equals the unconditional CDF $F_{\varDelta}$. Similarly, let
\[
\mathit{IR}_{\varDelta\mid\tilde{X}\in A}\coloneqq Q_{\varDelta\mid\tilde{X}}\left(0.75\mid A\right)-Q_{\varDelta\mid\tilde{X}}\left(0.25\mid A\right)
\]
be the IQR of the conditional distribution of $\varDelta$ given $\tilde{X}\in A$.
We consider the problem of estimation and inference for $F_{\varDelta\mid\tilde{X}}\left(v\mid A\right)$,
$Q_{\varDelta\mid\tilde{X}}\left(\tau\mid A\right)$ and $\mathit{IR}_{\varDelta\mid\tilde{X}\in A}$.

Our sample consists of i.i.d. observations $\left\{ W_{i}\right\} _{i=1}^{n_{A}}$
with observed covariates $X_{i}$ satisfying $\tilde{X}_{i}\in A$,
where we redefine $W_{i}$ as $W_{i}\coloneqq\left(Y_{i},D_{i},Z_{i},X_{i}^{\top}\right)^{\top}$
collecting the observed variables from the $i$-th individual for
notational convenience. Under this sampling assumption, the probability
masses of $X$ are given by $\left\{ \mathrm{Pr}\left[X=x\mid\tilde{X}\in A\right]:x\in\mathscr{S}_{X\mid\tilde{X}\in A}\right\} $,
where $\mathscr{S}_{X\mid\tilde{X}\in A}$ denotes the conditional
support of $X$ given $\tilde{X}\in A$. For each $x\in\mathscr{S}_{X\mid\tilde{X}\in A}$,
we redefine $\widehat{\varUpsilon}_{dx}^{\left(-i\right)}\left(t,y\right)$
as 
\begin{multline}
\widehat{\varUpsilon}_{dx}^{\left(-i\right)}\left(t,y\right)\coloneqq\\
\frac{\sum_{j\in\left[n_{A}\right]\setminus\left\{ i\right\} }\left\{ \mathbbm{1}\left(D_{j}=d,Z_{j}=d,X_{j}=x\right)\left|Y_{j}-t\right|-\mathbbm{1}\left(D_{j}=d',Z_{j}=d,X_{j}=x\right)\mathrm{sgn}\left(Y_{j}-y\right)t\right\} }{\sum_{j\in\left[n_{A}\right]\setminus\left\{ i\right\} }\mathbbm{1}\left(Z_{j}=d,X_{j}=x\right)}\\
-\frac{\sum_{j\in\left[n_{A}\right]\setminus\left\{ i\right\} }\left\{ \mathbbm{1}\left(D_{j}=d,Z_{j}=d',X_{j}=x\right)\left|Y_{j}-t\right|-\mathbbm{1}\left(D_{j}=d',Z_{j}=d',X_{j}=x\right)\mathrm{sgn}\left(Y_{j}-y\right)t\right\} }{\sum_{j\in\left[n_{A}\right]\setminus\left\{ i\right\} }\mathbbm{1}\left(Z_{j}=d',X_{j}=x\right)},\\
\label{eq:Upsilon_hat subvector}
\end{multline}
i.e., the leave-$i$-out sample analogue of the right hand side of
(\ref{eq:Q_dx definition}) using $\left\{ W_{i}\right\} _{i=1}^{n_{A}}$
as the sample. The leave-$i$-out nonparametric estimator $\widehat{\phi}_{dx}^{\left(-i\right)}\left(y\right)$
of $\phi_{dx}\left(y\right)$ can be defined similarly as $\widehat{\phi}_{dx}^{\left(-i\right)}\left(y\right)\coloneqq\textrm{argmin}_{t\in\left[\underline{y}_{dx},\overline{y}_{dx}\right]}\widehat{\varUpsilon}_{dx}^{\left(-i\right)}\left(t,y\right)$.
We redefine $\widehat{\varDelta}_{i}$ as the pseudo ITE
\begin{equation}
\widehat{\varDelta}_{i}\coloneqq D_{i}\left(Y_{i}-\widehat{\phi}_{0X_{i}}^{\left(-i\right)}\left(Y_{i}\right)\right)+\left(1-D_{i}\right)\left(\widehat{\phi}_{1X_{i}}^{\left(-i\right)}\left(Y_{i}\right)-Y_{i}\right),\label{eq:pseudo all}
\end{equation}
for the $i$-th individual in the sample. 

Let
\begin{equation}
\widehat{F}_{\varDelta\mid\tilde{X}}\left(v\mid A\right)\coloneqq\frac{1}{n_{A}}\sum_{i=1}^{n_{A}}\mathbbm{1}\left(\widehat{\varDelta}_{i}\leq v\right)\label{eq:F_hat_A}
\end{equation}
be the nonparametric estimator of $F_{\varDelta\mid\tilde{X}}\left(v\mid A\right)$
using the pseudo ITEs defined by (\ref{eq:pseudo all}). For each
$\tau\in\left(0,1\right)$, let 
\begin{eqnarray}
\widehat{Q}_{\varDelta\mid\tilde{X}}\left(\tau\mid A\right) & \coloneqq & \inf\left\{ y\in\mathbb{R}:\widehat{F}_{\varDelta\mid\tilde{X}}\left(y\mid A\right)\geq\tau\right\} \nonumber \\
 & = & \widehat{\varDelta}_{\left\langle \left\lceil \tau n_{A}\right\rceil \right\rangle }\label{eq:Q_hat_A definition}
\end{eqnarray}
be the estimated quantile, where $\widehat{\varDelta}_{\left\langle 1\right\rangle }\leq\cdots\leq\widehat{\varDelta}_{\left\langle n_{A}\right\rangle }$
are the order statistics corresponding to $\left\{ \widehat{\varDelta}_{i}\right\} _{i=1}^{n_{A}}$.
Similarly, we let $\widehat{\mathit{IR}}_{\varDelta\mid\tilde{X}\in A}\coloneqq\widehat{Q}_{\varDelta\mid\tilde{X}}\left(0.75\mid A\right)-\widehat{Q}_{\varDelta\mid\tilde{X}}\left(0.25\mid A\right)$
be the estimator of $\mathit{IR}_{\varDelta\mid\tilde{X}\in A}$.
Let $\left[\underline{v}_{A},\overline{v}_{A}\right]$ be an inner
closed sub-interval of $\mathscr{S}_{\varDelta\mid\tilde{X}\in A}$.
Let
\begin{equation}
S_{F}\left(v\mid A\right)\coloneqq\sqrt{n_{A}}\left(\widehat{F}_{\varDelta\mid\tilde{X}}\left(v\mid A\right)-F_{\varDelta\mid\tilde{X}}\left(v\mid A\right)\right),\,v\in\left[\underline{v}_{A},\overline{v}_{A}\right],\label{eq:S_F_A definition}
\end{equation}
and let $S_{Q}\left(\tau\mid A\right)$ be defined analogously. By
using the same arguments as those in the proof of Theorem \ref{thm:weak_convergence}(i),
we can show that $S_{F}\left(\cdot\mid A\right)$ converges in distribution
to a tight Gaussian process in $\ell^{\infty}\left[\underline{v}_{A},\overline{v}_{A}\right]$.
An analogous result can be established for $S_{Q}\left(\cdot\mid A\right)$
that takes values in $\ell^{\infty}\left[\underline{\tau},\overline{\tau}\right]$.

A nonparametric bootstrap sample $\left\{ W_{i}^{\dagger}\right\} _{i=1}^{n_{A}}$
is obtained by independently drawing $n_{A}$ observations from the
original sample $\left\{ W_{i}\right\} _{i=1}^{n_{A}}$ and let $Y_{i}^{\dagger},D_{i}^{\dagger},Z_{i}^{\dagger}$
and $X_{i}^{\dagger}$ be the corresponding components of the vector
$W_{i}^{\dagger}$. By replacing $\left\{ W_{j}\right\} _{j\in\left[n_{A}\right]\setminus\left\{ i\right\} }$
on the right hand side of (\ref{eq:Upsilon_hat subvector}) with $\left\{ W_{j}^{\dagger}\right\} _{j\in\left[n_{A}\right]\setminus\left\{ i\right\} }$,
we get the bootstrap analogue $\widehat{\varUpsilon}_{dx}^{\left(-i\right)\dagger}\left(t,y\right)$
of $\widehat{\varUpsilon}_{dx}^{\left(-i\right)}\left(t,y\right)$.
Let $\widehat{\phi}_{dx}^{\left(-i\right)\dagger}\left(y\right)\coloneqq\textrm{argmin}_{t\in\left[\underline{y}_{dx},\overline{y}_{dx}\right]}\widehat{\varUpsilon}_{dx}^{\left(-i\right)\dagger}\left(t,y\right)$
be the bootstrap analogue of $\widehat{\phi}_{dx}^{\left(-i\right)}\left(y\right)$
and by using this counterfactual mapping estimator from the bootstrap
sample and replacing $\left(Y_{i},D_{i},X_{i}\right)$ and $\left(\widehat{\phi}_{0X_{i}}^{\left(-i\right)},\widehat{\phi}_{1X_{i}}^{\left(-i\right)}\right)$
on the right hand side of (\ref{eq:pseudo all}) with their bootstrap
analogues, we construct the pseudo ITEs $\left\{ \widehat{\varDelta}_{i}^{\dagger}\right\} _{i=1}^{n_{A}}$
from the bootstrap sample. Let
\begin{eqnarray}
\widehat{F}_{\varDelta\mid\tilde{X}}^{\dagger}\left(v\mid A\right) & \coloneqq & \frac{1}{n_{A}}\sum_{i=1}^{n_{A}}\mathbbm{1}\left(\widehat{\varDelta}_{i}^{\dagger}\leq v\right)\nonumber \\
\widehat{Q}_{\varDelta\mid\tilde{X}}^{\dagger}\left(\tau\mid A\right) & \coloneqq & \inf\left\{ y\in\mathbb{R}:\widehat{F}_{\varDelta\mid\tilde{X}}^{\dagger}\left(y\mid A\right)\geq\tau\right\} \nonumber \\
\widehat{\mathit{IR}}_{\varDelta\mid\tilde{X}\in A}^{\dagger} & \coloneqq & \widehat{Q}_{\varDelta\mid\tilde{X}}^{\dagger}\left(0.75\mid A\right)-\widehat{Q}_{\varDelta\mid\tilde{X}}^{\dagger}\left(0.25\mid A\right)\label{eq:F_dag_A definition}
\end{eqnarray}
be bootstrap analogues of $\widehat{F}_{\varDelta\mid\tilde{X}}\left(v\mid A\right)$,
$\widehat{Q}_{\varDelta\mid\tilde{X}}\left(\tau\mid A\right)$ and
$\widehat{\mathit{IR}}_{\varDelta\mid\tilde{X}\in A}$. Note that
we have $\widehat{Q}_{\varDelta\mid\tilde{X}}^{\dagger}\left(\tau\mid A\right)=\widehat{\varDelta}_{\left\langle \left\lceil \tau n_{A}\right\rceil \right\rangle }^{\dagger}$,
where $\widehat{\varDelta}_{\left\langle 1\right\rangle }^{\dagger}\leq\cdots\leq\widehat{\varDelta}_{\left\langle n_{A}\right\rangle }^{\dagger}$
are the order statistics corresponding to $\left\{ \widehat{\varDelta}_{i}^{\dagger}\right\} _{i=1}^{n_{A}}$.
Bootstrap percentile confidence intervals for $F_{\varDelta\mid\tilde{X}}\left(v\mid A\right)$,
$Q_{\varDelta\mid\tilde{X}}\left(\tau\mid A\right)$ and $\mathit{IR}_{\varDelta\mid\tilde{X}\in A}$
can be defined by using the $\left(\alpha/2\right)$-th and the $\left(1-\alpha/2\right)$-th
quantiles of the resampling distributions of $\widehat{F}_{\varDelta\mid\tilde{X}}^{\dagger}\left(v\mid A\right)$,
$\widehat{Q}_{\varDelta\mid\tilde{X}}^{\dagger}\left(\tau\mid A\right)$
and $\widehat{\mathit{IR}}_{\varDelta\mid\tilde{X}\in A}^{\dagger}$
as the end points. 

The end points of these bootstrap confidence intervals can be easily
estimated by Monte Carlo simulations. It is straightforward to adapt
Algorithms \ref{alg:confidence interval for CDF} and \ref{alg:confidence interval for quantile}
to obtain bootstrap percentile confidence intervals. In the first
two steps, in the $r$-th bootstrap replication, we independently
draw a bootstrap sample $\left\{ W_{i}^{\dagger\left(r\right)}\right\} _{i=1}^{n_{A}}$
and compute the pseudo ITEs $\left\{ \widehat{\varDelta}_{i}^{\dagger\left(r\right)}\right\} _{i=1}^{n_{A}}$
using the procedure described in the preceding paragraph. Then by
using the formulae given by (\ref{eq:F_dag_A definition}) with $\left\{ \widehat{\varDelta}_{i}^{\dagger}\right\} _{i=1}^{n_{A}}$
replaced by $\left\{ \widehat{\varDelta}_{i}^{\dagger\left(r\right)}\right\} _{i=1}^{n_{A}}$,
we can easily compute $\widehat{F}_{\varDelta\mid\tilde{X}}^{\dagger\left(r\right)}\left(v\mid A\right)$
and $\widehat{Q}_{\varDelta\mid\tilde{X}}^{\dagger\left(r\right)}\left(\tau\mid A\right)=\widehat{\varDelta}_{\left\langle \left\lceil \tau n_{A}\right\rceil \right\rangle }^{\dagger\left(r\right)}$,
where $\widehat{\varDelta}_{\left\langle 1\right\rangle }^{\dagger\left(r\right)}\leq\cdots\leq\widehat{\varDelta}_{\left\langle n_{A}\right\rangle }^{\dagger\left(r\right)}$
are the order statistics corresponding to $\left\{ \widehat{\varDelta}_{i}^{\dagger\left(r\right)}\right\} _{i=1}^{n_{A}}$.
The rest of the steps are identical to those in Algorithms \ref{alg:confidence interval for CDF}
and \ref{alg:confidence interval for quantile}.

The UCBs (\ref{eq:CB}) and (\ref{eq:CB_Q}) constructed in Section
\ref{subsec:Constructing-confidence-intervals-bands} can also be
easily extended. A bootstrap UCB for $F_{\varDelta\mid\tilde{X}}\left(v\mid A\right)$
over $v\in\left[\underline{v}_{A},\overline{v}_{A}\right]$ with nominal
coverage probability $1-\alpha$ centers around $\widehat{F}_{\varDelta\mid\tilde{X}}\left(v\mid A\right)$
and has radius given by the $\left(1-\alpha\right)$-th quantile of
the resampling distribution of $\left\Vert \widehat{F}_{\varDelta\mid\tilde{X}}^{\dagger}\left(\cdot\mid A\right)-\widehat{F}_{\varDelta\mid\tilde{X}}\left(\cdot\mid A\right)\right\Vert _{\left[\underline{v}_{A},\overline{v}_{A}\right]}$.
A bootstrap UCB for $Q_{\varDelta\mid\tilde{X}}\left(\tau\mid A\right)$
over $\tau\in\left[\underline{\tau},\overline{\tau}\right]$ can be
constructed analogously. A straightforward adaptation leads to the
construction of a variable-width bootstrap UCB for $Q_{\varDelta\mid\tilde{X}}\left(\cdot\mid A\right)$
similar to (\ref{eq:CB_Q variable width}).

We again easily adapt Algorithms \ref{alg:confidence bands for CDF}
and \ref{alg:confidence band quantile}. The first two or three steps
are the same as those in the algorithms for computing the bootstrap
percentile confidence intervals. Then, we compute $\widehat{F}_{\varDelta\mid\tilde{X}}^{\dagger\left(r\right)}\left(v\mid A\right)-\widehat{F}_{\varDelta\mid\tilde{X}}\left(v\mid A\right)$
for $\left(r,v\right)\in\left[B\right]\times\mathcal{V}_{A}$, where
$\mathcal{V}_{A}\coloneqq\left\{ v_{A}^{\left(1\right)},...,v_{A}^{\left(T\right)}\right\} $
are equally spaced grid points in $\left[\underline{v}_{A},\overline{v}_{A}\right]$
and $\widehat{Q}_{\varDelta\mid\tilde{X}}^{\dagger\left(r\right)}\left(\tau\mid A\right)-\widehat{Q}_{\varDelta\mid\tilde{X}}\left(\tau\mid A\right)$
for $\left(r,\tau\right)\in\left[B\right]\times\mathcal{T}$. The
simulated critical values are given by the $\left(1-\alpha\right)$-th
empirical quantiles of 
\[
\left\{ \max_{v\in\mathcal{V}_{A}}\left|\widehat{F}_{\varDelta\mid\tilde{X}}^{\dagger\left(r\right)}\left(v\mid A\right)-\widehat{F}_{\varDelta\mid\tilde{X}}\left(v\mid A\right)\right|\right\} _{r=1}^{B}\textrm{ and }\left\{ \max_{\tau\in\mathcal{T}}\left|\widehat{Q}_{\varDelta\mid\tilde{X}}^{\dagger\left(r\right)}\left(\tau\mid A\right)-\widehat{Q}_{\varDelta\mid\tilde{X}}\left(\tau\mid A\right)\right|\right\} _{r=1}^{B},
\]
 respectively. As those in Algorithms \ref{alg:confidence bands for CDF}
and \ref{alg:confidence band quantile}, the UCBs are collections
of intervals centered around $\left\{ \widehat{F}_{\varDelta\mid\tilde{X}}\left(v\mid A\right)\right\} _{v\in\mathcal{V}_{A}}$
and $\left\{ \widehat{Q}_{\varDelta\mid\tilde{X}}\left(\tau\mid A\right)\right\} _{\tau\in\mathcal{T}}$
with radii given by these critical values. The variable-width counterparts
can be computed analogously.

Let $S_{F}^{\dagger}\left(v\mid A\right)$ be the bootstrap analogue
of (\ref{eq:S_F_A definition}) defined analogously to (\ref{eq:S_F_dag definition}).
Similarly, let $S_{Q}^{\dagger}\left(\tau\mid A\right)$ denote the
bootstrap analogue of $S_{Q}\left(\tau\mid A\right)$. To justify
the validity of the inference methods just proposed, we can use the
same arguments as those in the proofs of Theorem \ref{thm:bootstrap_weak_convergence}
and Corollary \ref{cor:bootstrap_quantile}(i) to show that $S_{F}^{\dagger}\left(\cdot\mid A\right)$
and $S_{Q}^{\dagger}\left(\cdot\mid A\right)$ converge in distribution
conditionally on the original data to the same limits as those of
$S_{F}\left(\cdot\mid A\right)$ and $S_{Q}\left(\cdot\mid A\right)$.
The asymptotic validity follows from these results and arguments in
the proofs of Corollaries \ref{cor:consistency_of_CDF_bootstrap_interval}
and \ref{cor:bootstrap_quantile}.

\subsection{Comparison of ITE distributions\label{subsec:comparison}}

Let $A_{0}$ and $A_{1}$ be two disjoint subsets of $\mathscr{S}_{\tilde{X}}$
respectively. We consider the problem of comparing the ITE distributions
conditional on $\tilde{X}\in A_{0}$ and $\tilde{X}\in A_{1}$ respectively.
Let $\delta\left(\tau\right)\coloneqq Q_{\varDelta\mid\tilde{X}}\left(\tau\mid A_{1}\right)-Q_{\varDelta\mid\tilde{X}}\left(\tau\mid A_{0}\right)$
for $\tau\in\left[\underline{\tau},\overline{\tau}\right]$ denote
the difference of the $\tau$-th quantiles. In empirical applications,
it may be interesting to learn about $\delta\left(\tau\right)$. E.g.,
we can conclude which subgroup of individuals tend to have a larger
median effect by constructing a confidence interval for $\delta\left(0.5\right)$
and drawing inference on the sign of $\delta\left(0.5\right)$. Similarly,
the difference of dispersions of ITE distributions can be measured
by $\mathit{IR}_{\varDelta\mid\tilde{X}\in A_{1}}-\mathit{IR}_{\varDelta\mid\tilde{X}\in A_{0}}=\delta\left(0.75\right)-\delta\left(0.25\right)$
and knowledge about the sign of this quantity is useful in determining
which subgroup of individuals tend to have more dispersed ITEs.

Our sample is the union of two independent samples $\left\{ W_{0,i}\right\} _{i=1}^{n_{0}}$
and $\left\{ W_{1,i}\right\} _{i=1}^{n_{1}}$. Let $n\coloneqq n_{0}+n_{1}$
be the sample size. Let $\widehat{\delta}\left(\tau\right)\coloneqq\widehat{Q}_{\varDelta\mid\tilde{X}}\left(\tau\mid A_{1}\right)-\widehat{Q}_{\varDelta\mid\tilde{X}}\left(\tau\mid A_{0}\right)$
be the estimator of $\delta\left(\tau\right)$ based on (\ref{eq:Q_hat_A definition})
defined in the preceding subsection. Under the additional assumption
that the limits of $n_{0}/n$ and $n_{1}/n$ as $n_{0},n_{1}\uparrow\infty$
exist, we can show that $\sqrt{n}\left(\widehat{\delta}-\delta\right)$
converges in distribution in $\ell^{\infty}\left[\underline{\tau},\overline{\tau}\right]$
to the sum of two independent tight Gaussian processes. Let $\widehat{\delta}^{\dagger}\left(\tau\right)\coloneqq\widehat{Q}_{\varDelta\mid\tilde{X}}^{\dagger}\left(\tau\mid A_{1}\right)-\widehat{Q}_{\varDelta\mid\tilde{X}}^{\dagger}\left(\tau\mid A_{0}\right)$
denote the bootstrap analogue of $\widehat{\delta}\left(\tau\right)$
constructed from bootstrap samples $\left\{ W_{0,i}^{\dagger}\right\} _{i=1}^{n_{0}}$
and $\left\{ W_{1,i}^{\dagger}\right\} _{i=1}^{n_{1}}$ of $\left\{ W_{0,i}\right\} _{i=1}^{n_{0}}$
and $\left\{ W_{1,i}\right\} _{i=1}^{n_{1}}$. We can show that $\sqrt{n}\left(\widehat{\delta}^{\dagger}-\widehat{\delta}\right)$
converges in distribution conditionally on the original data to the
same limiting tight Gaussian process. The asymptotic validity of all
inference methods follow from these results. Bootstrap percentile
confidence intervals for $\delta\left(\tau\right)$ (or $\delta\left(0.75\right)-\delta\left(0.25\right)$)
can be defined by using the $\left(\alpha/2\right)$-th and $\left(1-\alpha/2\right)$-th
quantiles of the resampling distribution of $\widehat{\delta}^{\dagger}\left(\tau\right)$
(or $\widehat{\delta}^{\dagger}\left(0.75\right)-\widehat{\delta}^{\dagger}\left(0.25\right)$)
as the end points. We summarize the procedure for computing these
confidence intervals in the following algorithm.
\begin{lyxalgorithm}[Bootstrap percentile confidence intervals for quantile differences]
\label{alg:difference quantiles}Step 1: In each of the replications
$r\in\left[B\right]$, independently draw $\left\{ W_{0,i}^{\dagger\left(r\right)}\right\} _{i=1}^{n_{0}}$
and $\left\{ W_{1,i}^{\dagger\left(r\right)}\right\} _{i=1}^{n_{1}}$
with replacement from $\left\{ W_{0,i}\right\} _{i=1}^{n_{0}}$ and
$\left\{ W_{1,i}\right\} _{i=1}^{n_{1}}$. Step 2: For all $r\in\left[B\right]$,
compute the pseudo ITEs $\left\{ \widehat{\varDelta}_{0,i}^{\dagger\left(r\right)}\right\} _{i=1}^{n_{0}}$
and $\left\{ \widehat{\varDelta}_{1,i}^{\dagger\left(r\right)}\right\} _{i=1}^{n_{1}}$.
Step 3: Order the pseudo ITEs to get the order statistics $\widehat{\varDelta}_{0,\left\langle 1\right\rangle }^{\dagger\left(r\right)}\leq\cdots\leq\widehat{\varDelta}_{0,\left\langle n_{0}\right\rangle }^{\dagger\left(r\right)}$
and $\widehat{\varDelta}_{1,\left\langle 1\right\rangle }^{\dagger\left(r\right)}\leq\cdots\leq\widehat{\varDelta}_{1,\left\langle n_{1}\right\rangle }^{\dagger\left(r\right)}$
for all $r\in\left[B\right]$. Step 4: Compute $\widehat{\delta}^{\dagger\left(r\right)}\left(\tau\right)\coloneqq\widehat{Q}_{\varDelta\mid\tilde{X}}^{\dagger\left(r\right)}\left(\tau\mid A_{1}\right)-\widehat{Q}_{\varDelta\mid\tilde{X}}^{\dagger\left(r\right)}\left(\tau\mid A_{0}\right)$
and $\widehat{\delta}^{\dagger\left(r\right)}\left(0.75\right)-\widehat{\delta}^{\dagger\left(r\right)}\left(0.25\right)$
for all $r\in\left[B\right]$. Step 5: Order $\left\{ \widehat{\delta}^{\dagger\left(r\right)}\left(\tau\right)\right\} _{r=1}^{B}$
and $\left\{ \widehat{\delta}^{\dagger\left(r\right)}\left(0.75\right)-\widehat{\delta}^{\dagger\left(r\right)}\left(0.25\right)\right\} _{r=1}^{B}$
to get the order statistics $\delta_{\left\langle 1\right\rangle }^{\dagger}\leq\cdots\leq\delta_{\left\langle B\right\rangle }^{\dagger}$
and $\tilde{\delta}_{\left\langle 1\right\rangle }^{\dagger}\leq\cdots\leq\tilde{\delta}_{\left\langle B\right\rangle }^{\dagger}$.
Step 6: Return the confidence intervals $\left[\delta_{\left\langle \left\lceil B\times\left(\alpha/2\right)\right\rceil \right\rangle }^{\dagger},\delta_{\left\langle \left\lceil B\times\left(1-\alpha/2\right)\right\rceil \right\rangle }^{\dagger}\right]$
and $\left[\tilde{\delta}_{\left\langle \left\lceil B\times\left(\alpha/2\right)\right\rceil \right\rangle }^{\dagger},\tilde{\delta}_{\left\langle \left\lceil B\times\left(1-\alpha/2\right)\right\rceil \right\rangle }^{\dagger}\right]$
for $\delta\left(\tau\right)$ and $\mathit{IR}_{\varDelta\mid\tilde{X}\in A_{1}}-\mathit{IR}_{\varDelta\mid\tilde{X}\in A_{0}}$,
respectively.
\end{lyxalgorithm}
In applications, one may also be interested in comparing the entire
ITE distributions of two subgroups. To this end, one can use a UCB
for $\delta\left(\tau\right)$ over $\tau\in\left[\underline{\tau},\overline{\tau}\right]$
with $\underline{\tau}$ and $\overline{\tau}$ chosen to be close
to 0 and 1 (e.g., $\left[\underline{\tau},\overline{\tau}\right]=\left[0.1,0.9\right]$).
It is straightforward to extend the method proposed in Section \ref{subsec:Constructing-confidence-intervals-bands}.
The desired UCB with nominal coverage probability $1-\alpha$ centers
around $\widehat{\delta}\left(\tau\right)$ and has radius given by
the $\left(1-\alpha\right)$-th quantile of the resampling distribution
of $\left\Vert \widehat{\delta}^{\dagger}-\widehat{\delta}\right\Vert _{\left[\underline{\tau},\overline{\tau}\right]}$.
We summarize the procedure for this UCB in the following algorithm.
\begin{lyxalgorithm}[Bootstrap UCB for quantile differences]
\label{alg:difference quantile functions}Steps 1-3: Same as those
in Algorithm \ref{alg:difference quantiles}. Step 4: Compute $\widehat{\delta}^{\dagger\left(r\right)}\left(\tau\right)$
for $\left(r,\tau\right)\in\left[B\right]\times\mathcal{T}$ and compute
$\widehat{\delta}\left(\tau\right)$ for $\tau\in\mathcal{T}$. Step
5: Compute $\left\{ \max_{\tau\in\mathcal{T}}\left|\widehat{\delta}^{\dagger\left(r\right)}\left(\tau\right)-\widehat{\delta}\left(\tau\right)\right|\right\} _{r=1}^{B}$
and order them to get the corresponding order statistics $s_{\delta,\left\langle 1\right\rangle }^{\dagger}\leq\cdots\leq s_{\delta,\left\langle B\right\rangle }^{\dagger}$
and the critical value $s_{\delta,\left\langle \left\lceil B\left(1-\alpha\right)\right\rceil \right\rangle }^{\dagger}$.
Step 6: Return the UCB $\left\{ \widehat{\delta}\left(\tau\right)\pm s_{\delta,\left\langle \left\lceil B\left(1-\alpha\right)\right\rceil \right\rangle }^{\dagger}\right\} _{\tau\in\mathcal{T}}$.
\end{lyxalgorithm}
A variable-width UCB for the quantile difference function can be constructed
by following the approach of \citet{Chernozhukov2018} and using the
calculations in Algorithms \ref{alg:difference quantiles} and \ref{alg:difference quantile functions}.
The following algorithm summarizes the procedure.
\begin{lyxalgorithm}[Variable-width bootstrap UCB for quantile differences]
\label{alg:variable width quantile difference}Steps 1-3: Same as
those in Algorithm \ref{alg:difference quantile functions}. Step
4: Compute the order statistics $\delta_{\left\langle 1\right\rangle }^{\dagger}\left(\tau\right)\leq\cdots\leq\delta_{\left\langle B\right\rangle }^{\dagger}\left(\tau\right)$
corresponding to $\left\{ \widehat{\delta}^{\dagger\left(r\right)}\left(\tau\right)\right\} _{r=1}^{B}$
for all $\tau\in\mathcal{T}$. Step 5: Compute
\[
\left\{ \max_{\tau\in\mathcal{T}}\frac{\left|\widehat{\delta}^{\dagger\left(r\right)}\left(\tau\right)-\widehat{\delta}\left(\tau\right)\right|}{\left(\delta_{\left\langle \left\lceil B\times0.75\right\rceil \right\rangle }^{\dagger}\left(\tau\right)-\delta_{\left\langle \left\lceil B\times0.25\right\rceil \right\rangle }^{\dagger}\left(\tau\right)\right)/\left(z_{0.75}-z_{0.25}\right)}\right\} _{r=1}^{B}
\]
and get the order statistics $\tilde{s}_{\delta,\left\langle 1\right\rangle }^{\dagger}\leq\cdots\leq\tilde{s}_{\delta,\left\langle B\right\rangle }^{\dagger}$
and the critical value $\tilde{s}_{\delta,\left\langle \left\lceil B\left(1-\alpha\right)\right\rceil \right\rangle }^{\dagger}$.
Step 6: Return the variable-width UCB
\[
\left\{ \widehat{\delta}\left(\tau\right)\pm\tilde{s}_{\delta,\left\langle \left\lceil B\left(1-\alpha\right)\right\rceil \right\rangle }^{\dagger}\left(\frac{\delta_{\left\langle \left\lceil B\times0.75\right\rceil \right\rangle }^{\dagger}\left(\tau\right)-\delta_{\left\langle \left\lceil B\times0.25\right\rceil \right\rangle }^{\dagger}\left(\tau\right)}{z_{0.75}-z_{0.25}}\right)\right\} _{\tau\in\mathcal{T}}.
\]
\end{lyxalgorithm}
We can use the UCB constructed by Algorithm \ref{alg:difference quantile functions}
or Algorithm \ref{alg:variable width quantile difference} to test
the equality of the two ITE distributions. The null hypothesis in
this case is ``$\mathrm{H}_{0}^{\mathsf{a}}$: $\delta\left(\tau\right)=0$,
for all $\tau\in\left[\underline{\tau},\overline{\tau}\right]$''
and the alternative hypothesis is ``$\mathrm{H}_{1}^{\mathsf{a}}$:
$\delta\left(\tau\right)\neq0$ for some unknown $\tau\in\left[\underline{\tau},\overline{\tau}\right]$''.
We do not reject $\mathrm{H}_{0}^{\mathsf{a}}$ if the zero function
$\left[\underline{\tau},\overline{\tau}\right]\ni\tau\mapsto0$ is
covered by the confidence band (i.e., $\sup_{\tau\in\mathcal{T}}\left|\widehat{\delta}\left(\tau\right)\right|\leq s_{\delta,\left\langle \left\lceil B\left(1-\alpha\right)\right\rceil \right\rangle }^{\dagger}$)
and reject $\mathrm{H}_{0}^{\mathsf{a}}$ otherwise ($\sup_{\tau\in\mathcal{T}}\left|\widehat{\delta}\left(\tau\right)\right|>s_{\delta,\left\langle \left\lceil B\left(1-\alpha\right)\right\rceil \right\rangle }^{\dagger}$).
Note that the asymptotic validity of the UCB immediately implies the
asymptotic validity of the test. 

In empirical applications, it can be interesting to learn whether
the conditional ITE distribution given $\tilde{X}\in A_{0}$ is the
same as the conditional distribution given $\tilde{X}\in A_{1}$ up
to a location shift (i.e., $\delta:\left[\underline{\tau},\overline{\tau}\right]\rightarrow\mathbb{R}$
is some unknown constant function) or there is also difference in
dispersions. This testing ``equality up to a location shift'' problem
is a generalization of equality testing. Let $\gamma\left(\tau\right)\coloneqq\delta\left(\tau\right)-\left(\int_{\underline{\tau}}^{\overline{\tau}}\delta\left(t\right)\mathrm{d}t\right)/\left(\overline{\tau}-\underline{\tau}\right)$
for $\tau\in\left[\underline{\tau},\overline{\tau}\right]$. The problem
can be formulated as testing the null hypothesis ``$\mathrm{H}_{0}^{\mathsf{b}}$:
$\gamma\left(\tau\right)=0$, for all $\tau\in\left[\underline{\tau},\overline{\tau}\right]$''
against the alternative hypothesis ``$\mathrm{H}_{1}^{\mathsf{b}}$:
$\gamma\left(\tau\right)\neq0$, for some unknown $\tau\in\left[\underline{\tau},\overline{\tau}\right]$''.
Let $\widehat{\gamma}\left(\tau\right)\coloneqq\widehat{\delta}\left(\tau\right)-\left(\int_{\underline{\tau}}^{\overline{\tau}}\widehat{\delta}\left(t\right)\mathrm{d}t\right)/\left(\overline{\tau}-\underline{\tau}\right)$
be the estimator of $\gamma\left(\tau\right)$. The bootstrap analogue
$\widehat{\gamma}^{\dagger}\left(\tau\right)$ of $\widehat{\gamma}\left(\tau\right)$
is defined analogously.\footnote{It follows from the continuity of the map $f\mapsto f-\left(\int_{\underline{\tau}}^{\overline{\tau}}f\left(t\right)\mathrm{d}t\right)/\left(\overline{\tau}-\underline{\tau}\right)$
and CMT that $\sqrt{n}\left(\widehat{\gamma}-\gamma\right)$ (or $\sqrt{n}\left(\widehat{\gamma}^{\dagger}-\widehat{\gamma}\right)$)
converges in distribution (conditionally on the original data) to
a tight Gaussian process.} Similarly, an asymptotically valid test of equality up to a location
shift can be based on using an asymptotically valid UCB for $\gamma\left(\tau\right)$
over $\tau\in\left[\underline{\tau},\overline{\tau}\right]$, whose
construction is a straightforward extension of the UCB for $\delta\left(\tau\right)$
over $\tau\in\left[\underline{\tau},\overline{\tau}\right]$. For
practical computation, we can easily adapt Algorithm \ref{alg:difference quantile functions}
or Algorithm \ref{alg:variable width quantile difference}. Steps
1-3 are the same as those in Algorithm \ref{alg:difference quantiles}.
Then, we compute $\left\{ \left|\widehat{\gamma}^{\dagger\left(r\right)}\left(\tau\right)-\widehat{\gamma}\left(\tau\right)\right|\right\} _{\left(r,\tau\right)\in\left[B\right]\times\mathcal{T}}$
and order $\left\{ \max_{\tau\in\mathcal{T}}\left|\widehat{\gamma}^{\dagger\left(r\right)}\left(\tau\right)-\widehat{\gamma}\left(\tau\right)\right|\right\} _{r=1}^{B}$
to get the order statistics $s_{\gamma,\left\langle 1\right\rangle }^{\dagger}\leq\cdots\leq s_{\gamma,\left\langle B\right\rangle }^{\dagger}$
and the critical value $s_{\gamma,\left\langle \left\lceil B\left(1-\alpha\right)\right\rceil \right\rangle }^{\dagger}$.
We reject $\mathrm{H}_{0}^{\mathsf{b}}$ if $\sup_{\tau\in\mathcal{T}}\left|\widehat{\gamma}\left(\tau\right)\right|>s_{\gamma,\left\langle \left\lceil B\left(1-\alpha\right)\right\rceil \right\rangle }^{\dagger}$.

We can also use a one-sided UCB to test the hypothesis that the conditional
ITE distribution given $\tilde{X}\in A_{0}$ stochastically dominates
the conditional distribution given $\tilde{X}\in A_{1}$, which can
be formulated as testing ``$\mathrm{H}_{0}^{\mathsf{c}}$: $\delta\left(\tau\right)\leq0$,
for all $\tau\in\left[\underline{\tau},\overline{\tau}\right]$''
against the alternative hypothesis ``$\mathrm{H}_{1}^{\mathsf{c}}$:
$\delta\left(\tau\right)>0$, for some unknown $\tau\in\left[\underline{\tau},\overline{\tau}\right]$''.
Let $\dot{s}_{\delta,1-\alpha}^{\mathsf{unif}}$ denote the $\left(1-\alpha\right)$-th
quantile of the resampling distribution of $\sup_{\tau\in\left[\underline{\tau},\overline{\tau}\right]}\left\{ \widehat{\delta}^{\dagger}\left(\tau\right)-\widehat{\delta}\left(\tau\right)\right\} $.
A one-sided bootstrap UCB is given by $\left\{ \left[\widehat{\delta}\left(\tau\right)-\dot{s}_{\delta,1-\alpha}^{\mathsf{unif}},\infty\right):\tau\in\left[\underline{\tau},\overline{\tau}\right]\right\} $.
We accept $\mathrm{H}_{0}^{\mathsf{c}}$ if the constant zero function
is covered by the UCB (i.e., the lower bound of the UCB is smaller
than zero for all $\tau$). We can show that under $\mathrm{H}_{0}^{\mathsf{c}}$,
\[
\Pr\left[\underset{\tau\in\left[\underline{\tau},\overline{\tau}\right]}{\sup}\widehat{\delta}\left(\tau\right)>\dot{s}_{\delta,1-\alpha}^{\mathsf{unif}}\right]\leq\Pr\left[\underset{\tau\in\left[\underline{\tau},\overline{\tau}\right]}{\sup}\left\{ \widehat{\delta}\left(\tau\right)-\delta\left(\tau\right)\right\} >\dot{s}_{\delta,1-\alpha}^{\mathsf{unif}}\right],
\]
and the right hand side of the inequality converges to $\alpha$ as
$n_{0},n_{1}\uparrow\infty$. This result shows that the proposed
test is asymptotically valid. We can easily adapt Algorithm \ref{alg:difference quantile functions}
for practical computation of the critical value $\dot{s}_{\delta,1-\alpha}^{\mathsf{unif}}$.
Steps 1-4 are the same as those in Algorithm \ref{alg:difference quantile functions}.
Then, we order $\left\{ \max_{\tau\in\mathcal{T}}\left\{ \widehat{\delta}^{\dagger\left(r\right)}\left(\tau\right)-\widehat{\delta}\left(\tau\right)\right\} \right\} _{r=1}^{B}$
to get the corresponding order statistics $\dot{s}_{\delta,\left\langle 1\right\rangle }^{\dagger}\leq\cdots\leq\dot{s}_{\delta,\left\langle B\right\rangle }^{\dagger}$.
The critical value is given by $\dot{s}_{\delta,\left\langle \left\lceil B\left(1-\alpha\right)\right\rceil \right\rangle }^{\dagger}$.
We reject $\mathrm{H}_{0}^{\mathsf{c}}$ if $\sup_{\tau\in\mathcal{T}}\widehat{\delta}\left(\tau\right)>\dot{s}_{\delta,\left\langle \left\lceil B\left(1-\alpha\right)\right\rceil \right\rangle }^{\dagger}$.

\section{Monte Carlo simulations\label{sec:Monte-Carlo-simulations}}

Section \ref{subsec:Validity-of-theory} examines the quality of the
Gaussian approximation to the finite sample distributions of the estimators
proposed in Section \ref{subsec:Empirical-CDF-and-Q}. The Gaussian
approximation is justified by the asymptotic results in Sections \ref{subsec:Asymptotic-normality-CDF}
and \ref{subsec:Asymptotic-normality-quantiles}. Section \ref{subsec:Finite-sample-performance}
provides simulation results to assess the finite sample performances
of the inference methods proposed in Section \ref{sec:Bootstrap-inference}.

We consider the same DGP as in the simulation section of \citetalias{feng2019estimation}.
The same DGP is also used in the simulations in \citetalias{Ma2023}.
The outcome and treatment status are generated by $Y=\left(\epsilon+1\right)^{2+D}$
and $D=\mathbbm{1}\left(-0.5+0.5\cdot Z+\eta\geq0\right)$, where
$\left(\epsilon,\eta\right)=\left(\varPhi\left(U\right),\varPhi\left(V\right)\right)$,
$\left(U,V\right)$ follow a mean-zero bivariate normal distribution
with $\mathrm{Var}\left[U\right]=\mathrm{Var}\left[V\right]=1$ and
$\mathrm{Cov}\left[U,V\right]=0.3$. Here, $\varPhi$ denotes the
CDF of $\mathrm{N}\left(0,1\right)$. The IV is generated by $Z=\mathbbm{1}\left(N>0\right)$,
where $N\sim\mathrm{N}\left(0,1\right)$ is independent of $\left(\epsilon,\eta\right)$.
It is straightforward to check that the ITE is given by $\varDelta=\epsilon\left(\epsilon+1\right)^{2}$,
where $\epsilon=\varPhi\left(U\right)$ follows a uniform distribution
on $\left[0,1\right]$. Therefore, the support of $\varDelta$ is
$\left[0,4\right]$. Throughout the simulations, the number of Monte
Carlo replications is set to $1,000$, and the number of bootstrap
replications is set to 500. Let $n$ denote the sample size in each
of the Monte Carlo replications.

\subsection{Validity of the asymptotic theory\label{subsec:Validity-of-theory}}

\begin{figure}[H]
\caption{Simulated finite sample distributions of $\widehat{Q}_{\varDelta}\left(\tau\right)$
superimposed by the large sample (Gaussian) density: histogram = simulated
distribution of $\widehat{Q}_{\varDelta}(\tau)$ based on $1,000$
replications; solid curve = density of $\mathrm{N}\left(Q_{\varDelta}(\tau),V_{Q}(\tau)/n\right)$}
\label{fig:quatile_dist} \centering %
\begin{tabular}{cc}
\includegraphics[scale=0.04]{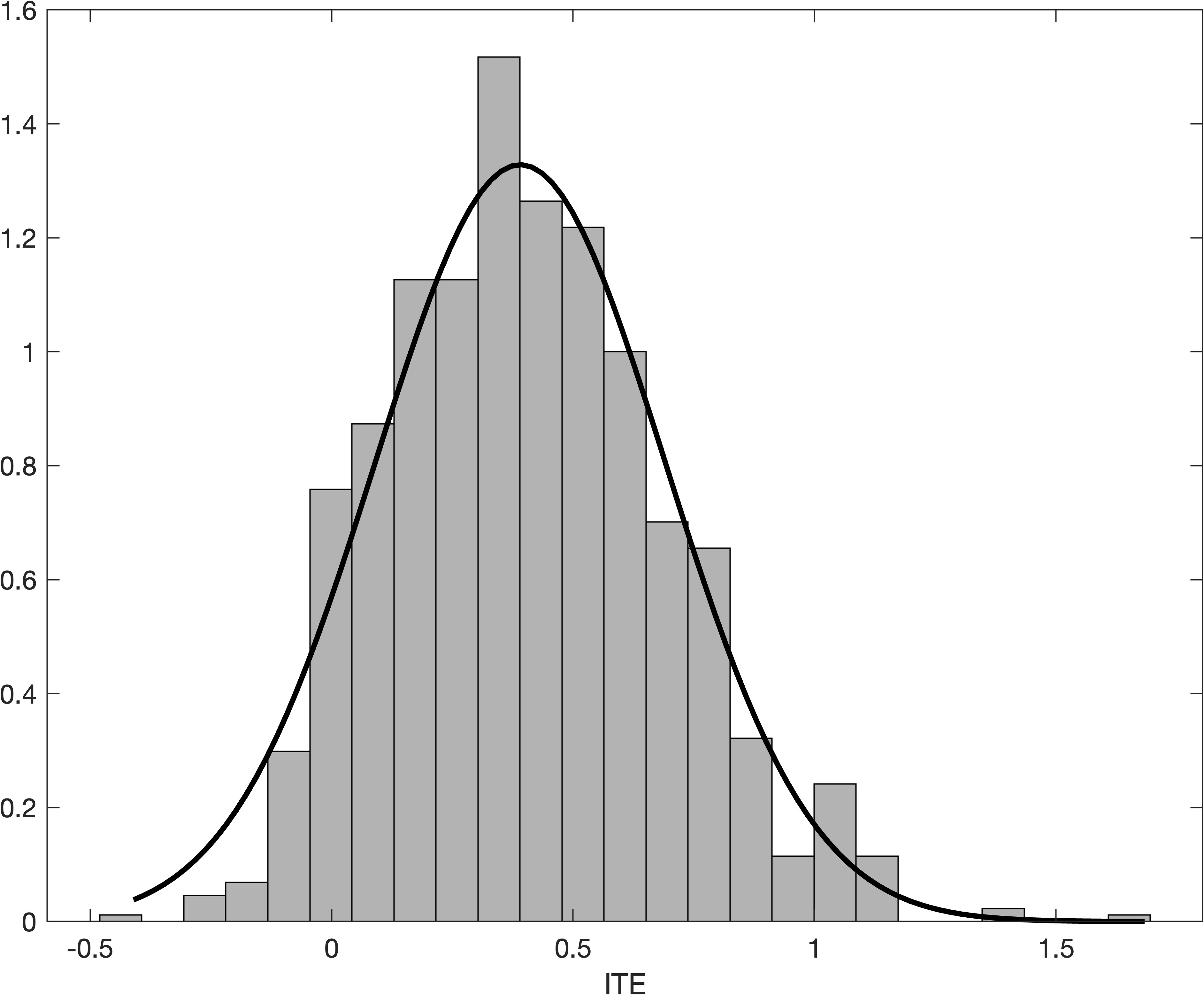} & \includegraphics[scale=0.04]{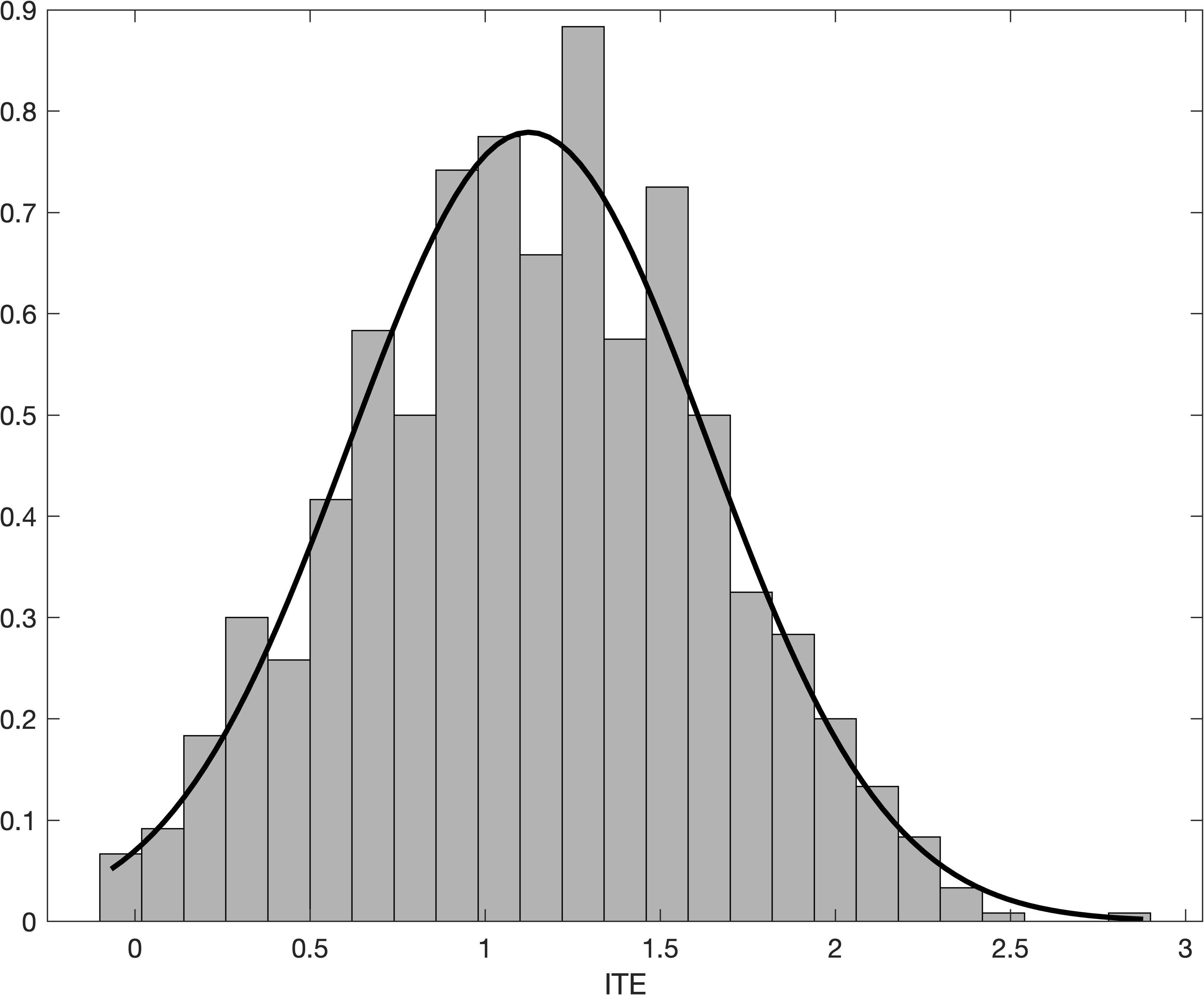}\tabularnewline
(a) $\tau=0.25,n=250$ & (b) $\tau=0.50,n=250$\tabularnewline
\includegraphics[scale=0.04]{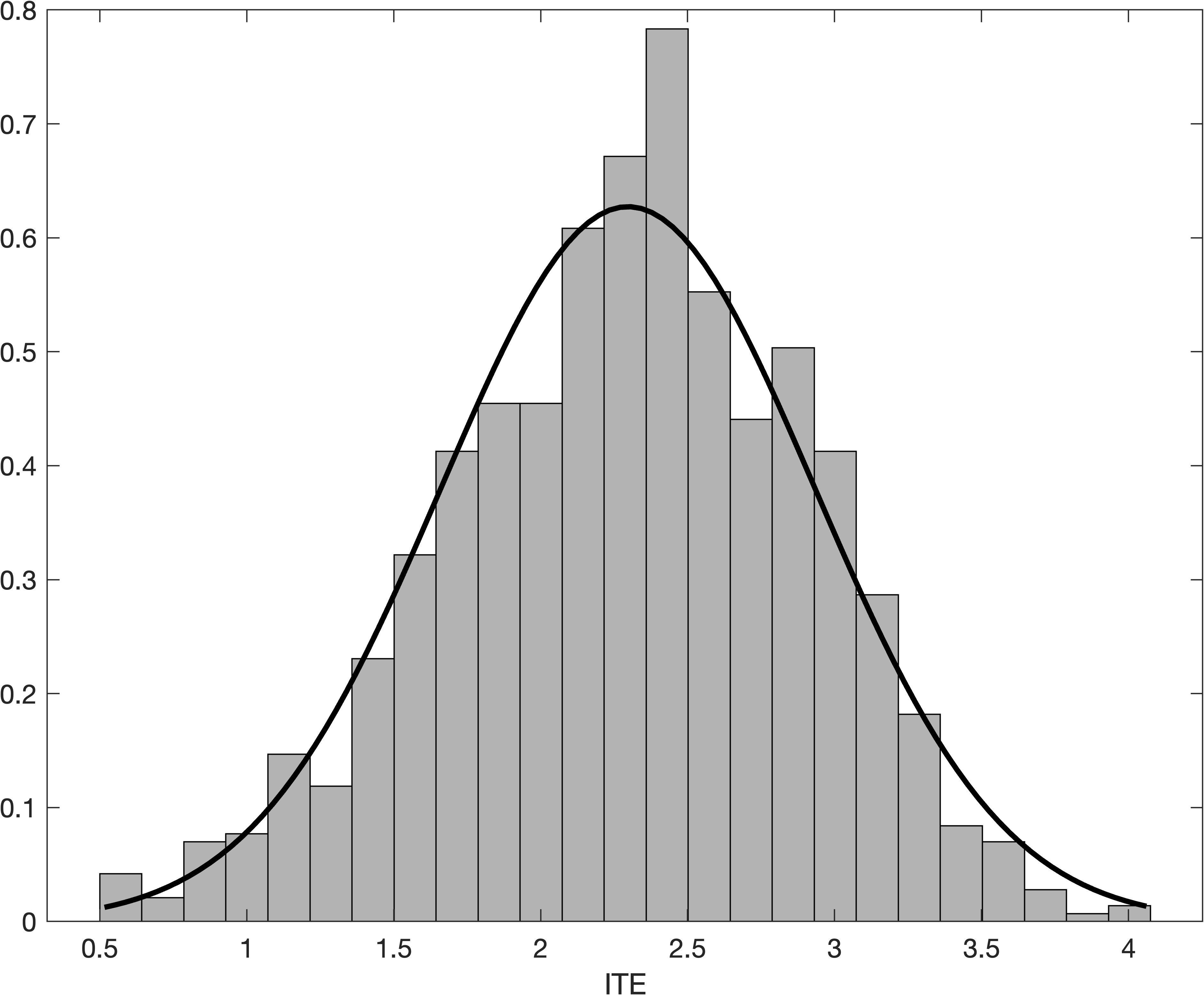} & \includegraphics[scale=0.04]{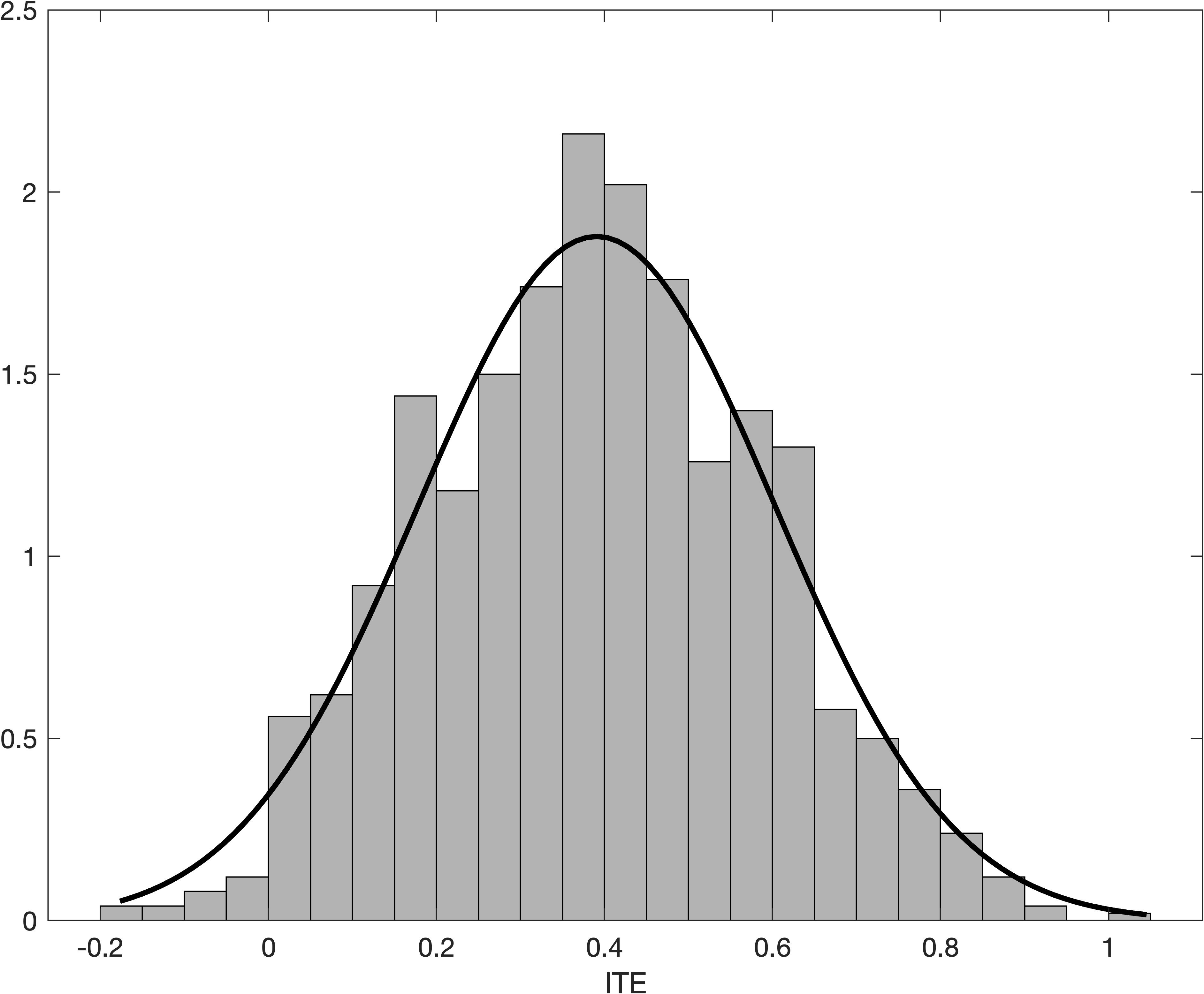}\tabularnewline
(c) $\tau=0.75,n=250$ & (d) $\tau=0.25,n=500$\tabularnewline
\includegraphics[scale=0.04]{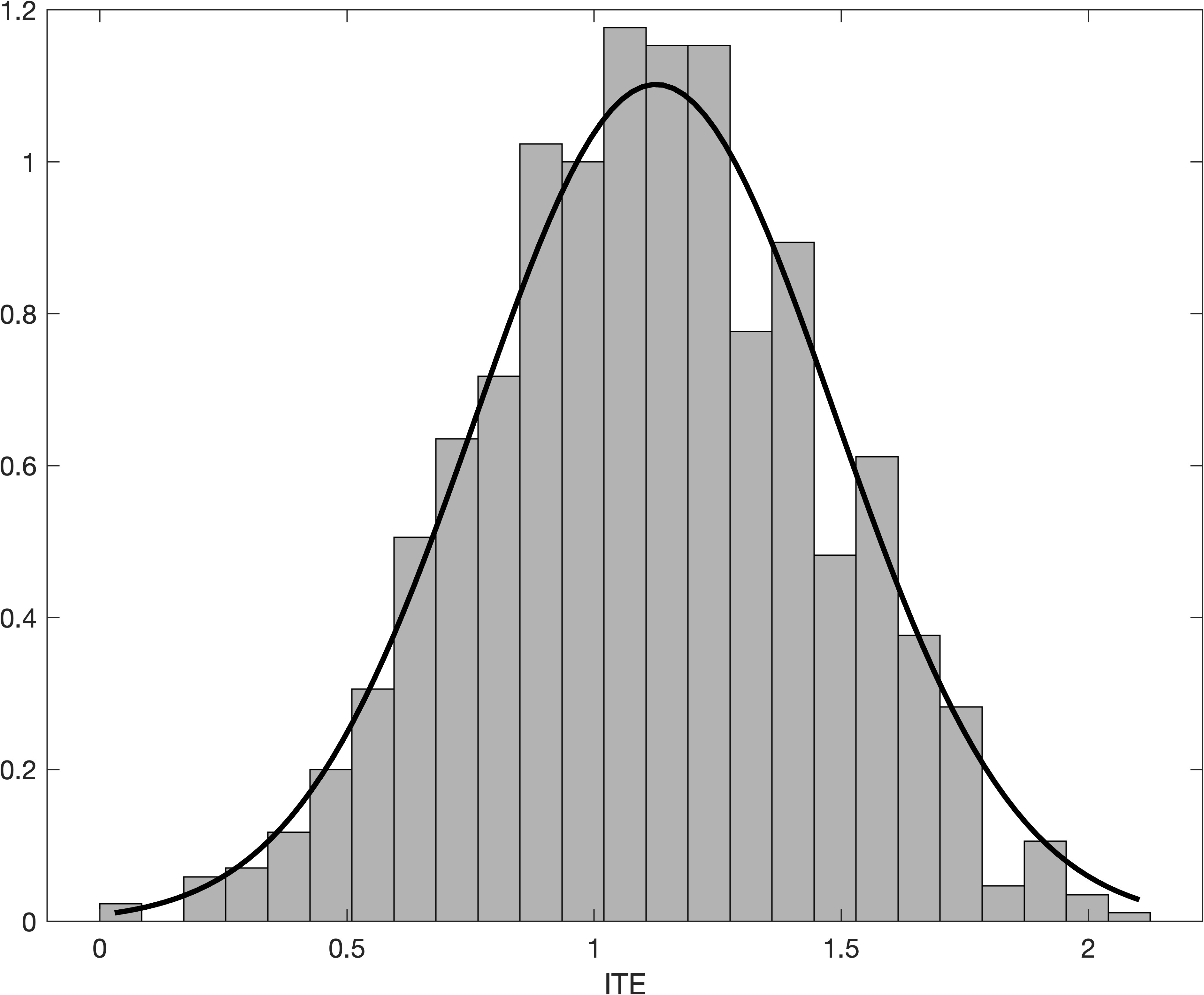} & \includegraphics[scale=0.04]{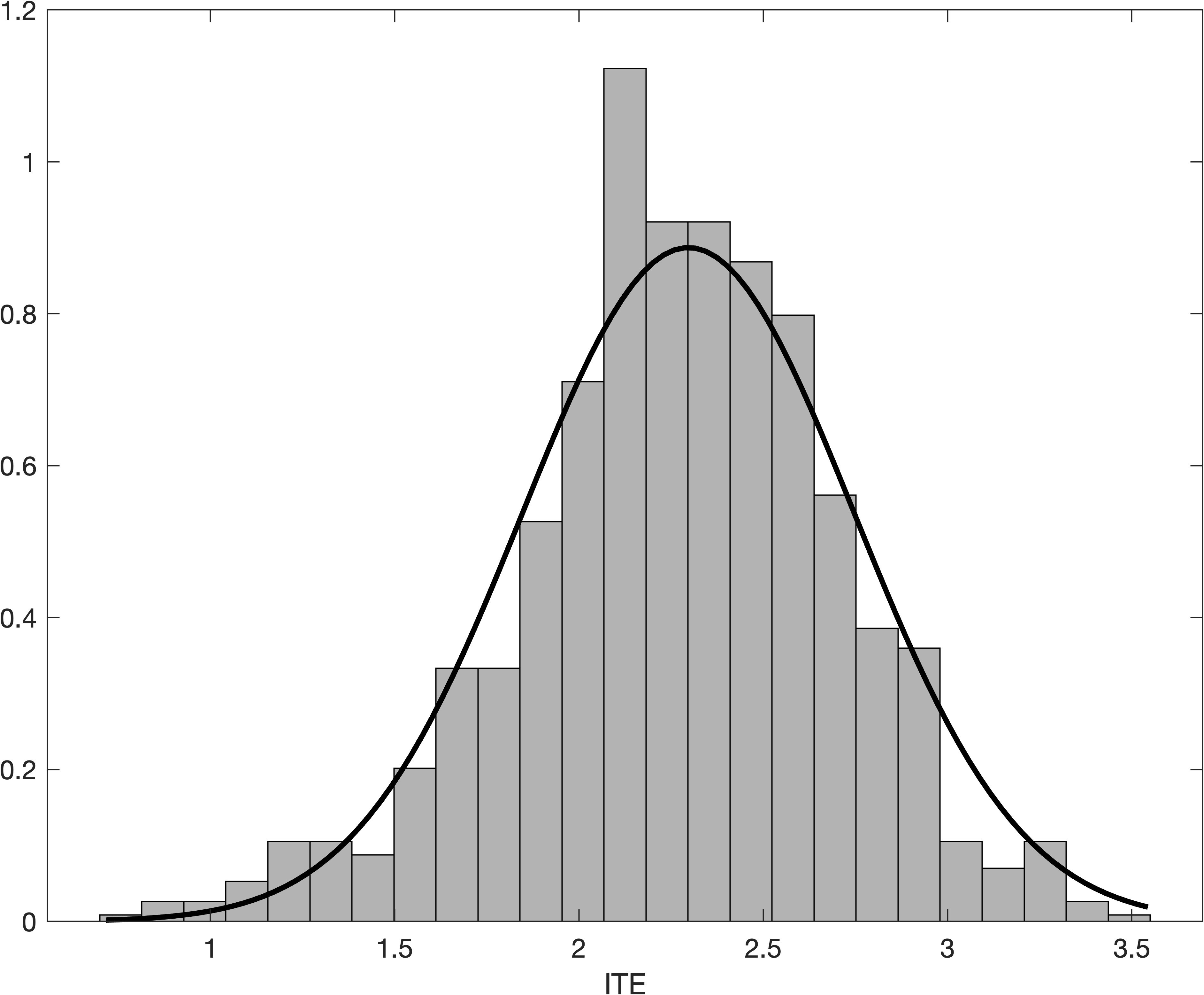}\tabularnewline
(e) $\tau=0.50,n=500$ & (f) $\tau=0.75,n=500$\tabularnewline
\end{tabular}
\end{figure}

\begin{figure}[H]
\caption{Simulated finite sample distributions of $\widehat{Q}_{\varDelta}\left(\tau\right)$
superimposed by the large sample (Gaussian) density: histogram = simulated
distribution of $\widehat{Q}_{\varDelta}(\tau)$ based on $1,000$
replications; solid curve = density of $\mathrm{N}\left(Q_{\varDelta}(\tau),V_{Q}(\tau)/n\right)$}
\label{fig:extreme_quatile_dist} \centering %
\begin{tabular}{cc}
\includegraphics[scale=0.04]{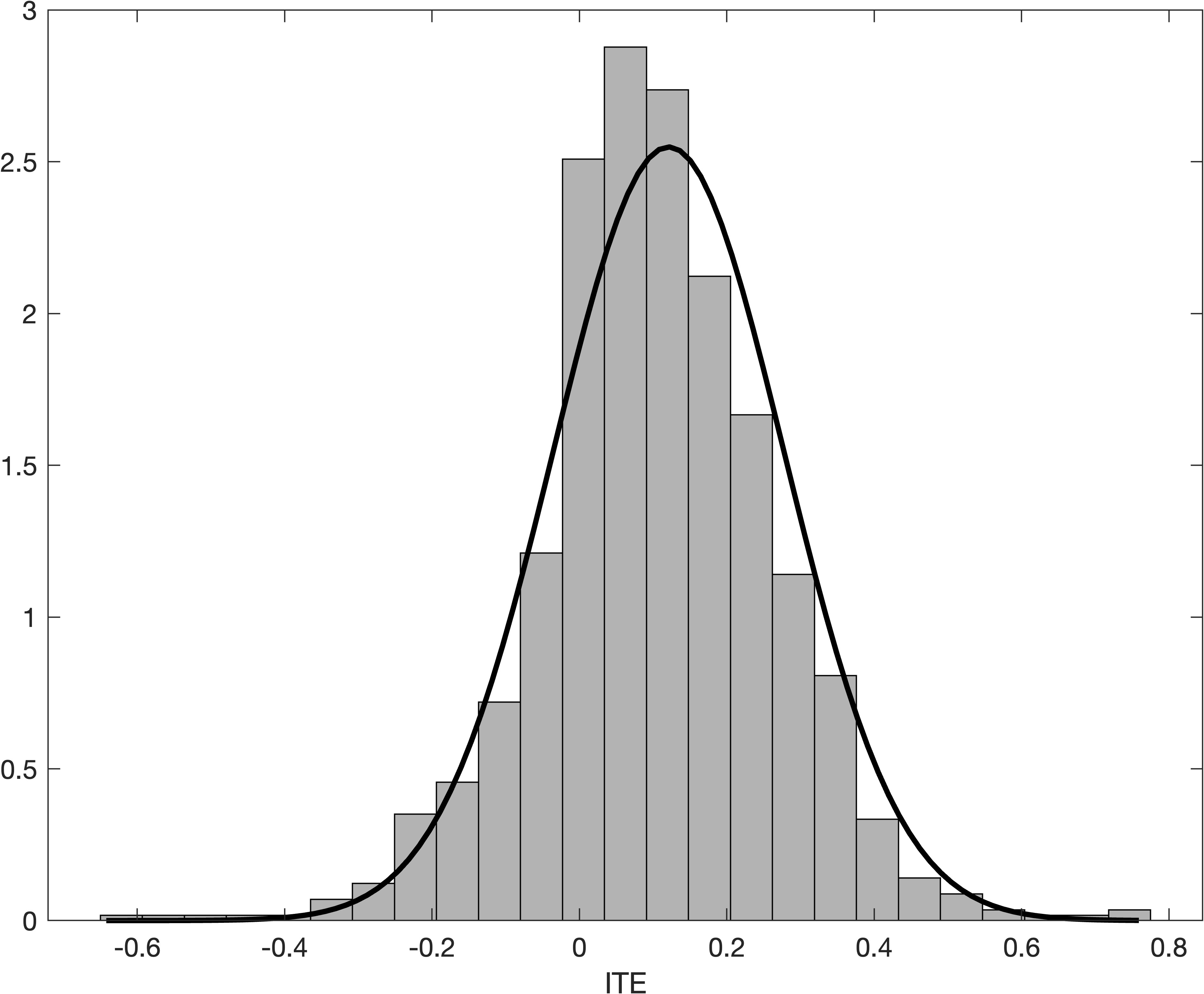} & \includegraphics[scale=0.04]{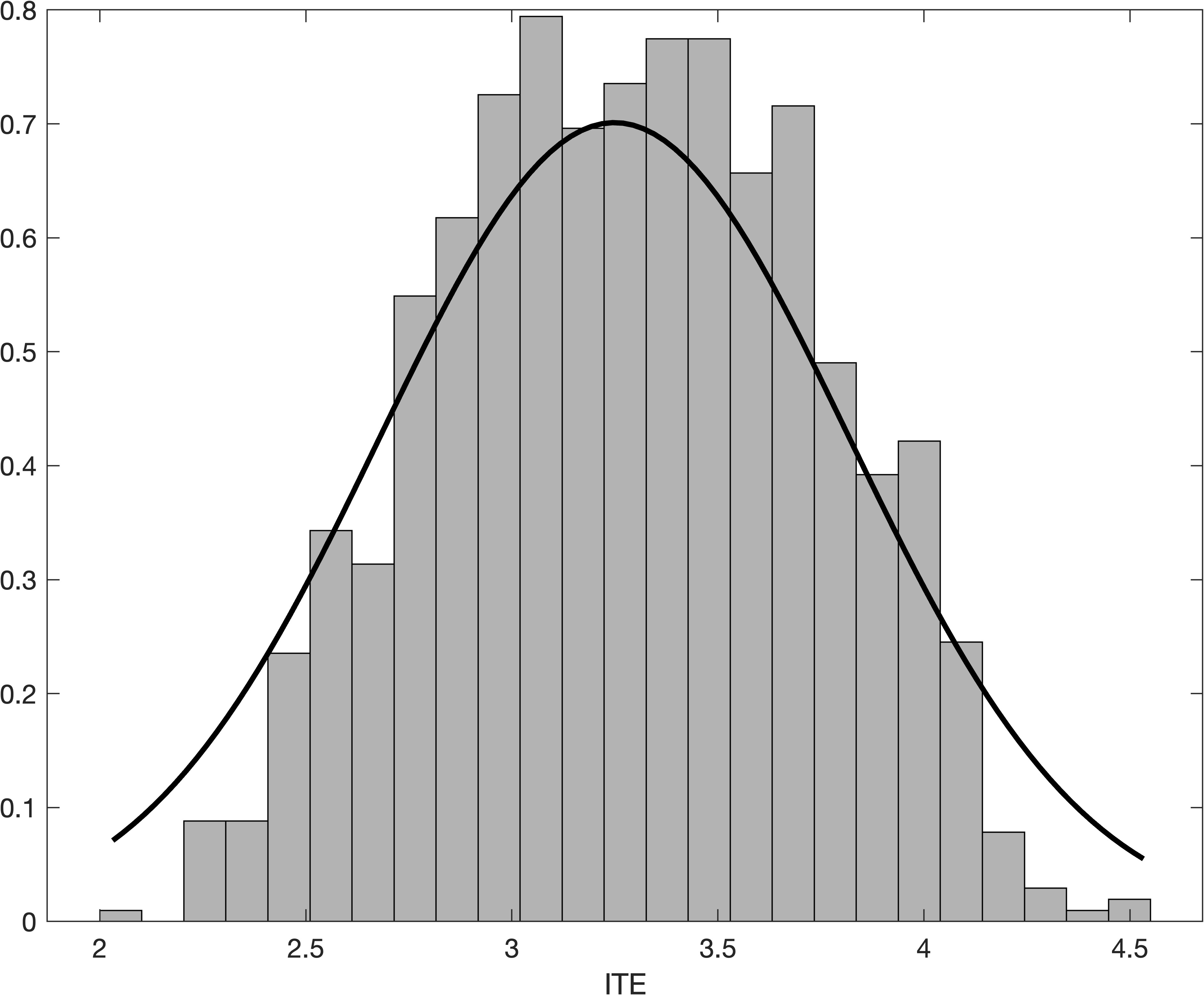}\tabularnewline
(a) $\tau=0.10,n=250$ & (b) $\tau=0.90,n=250$\tabularnewline
\includegraphics[scale=0.04]{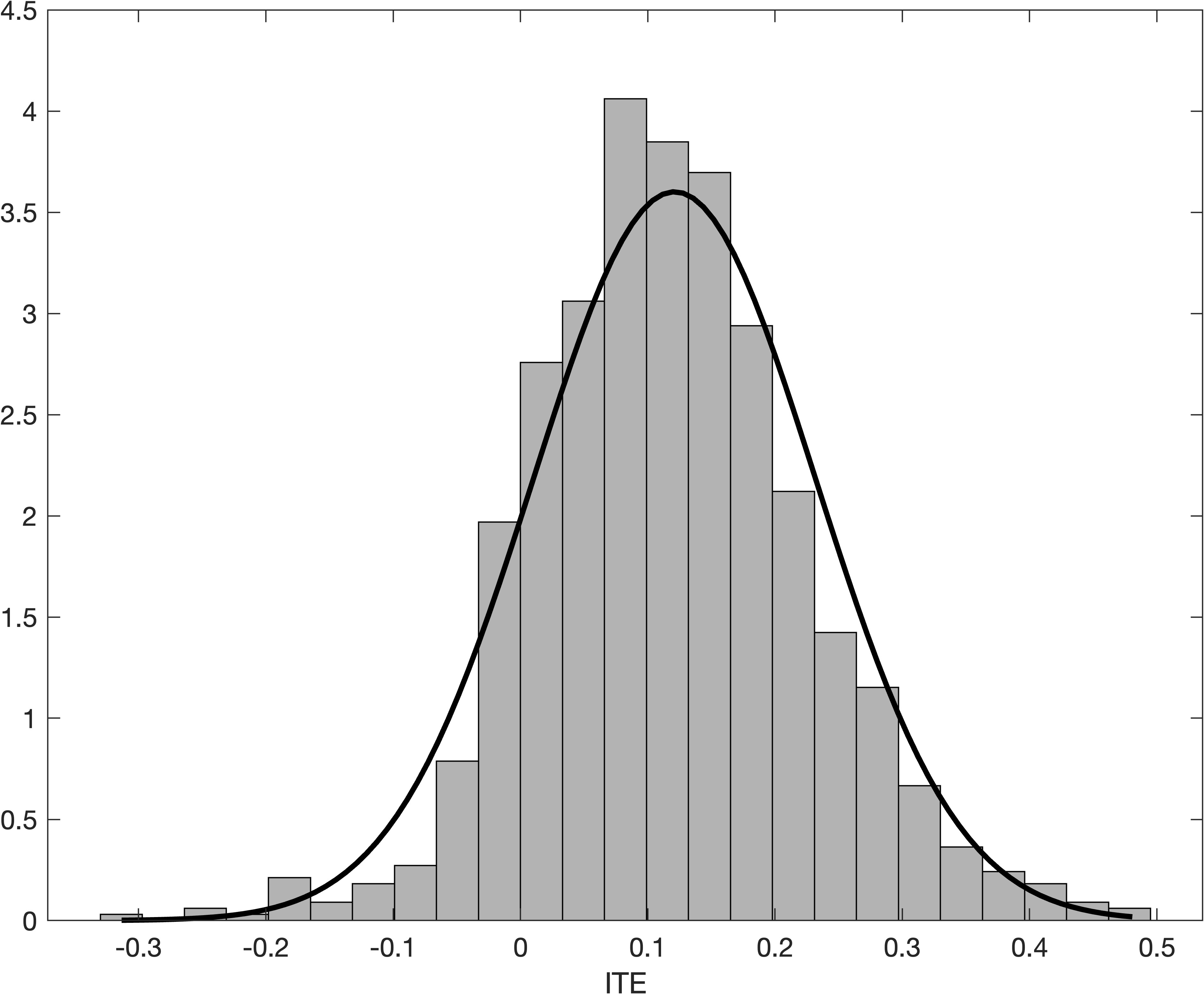} & \includegraphics[scale=0.04]{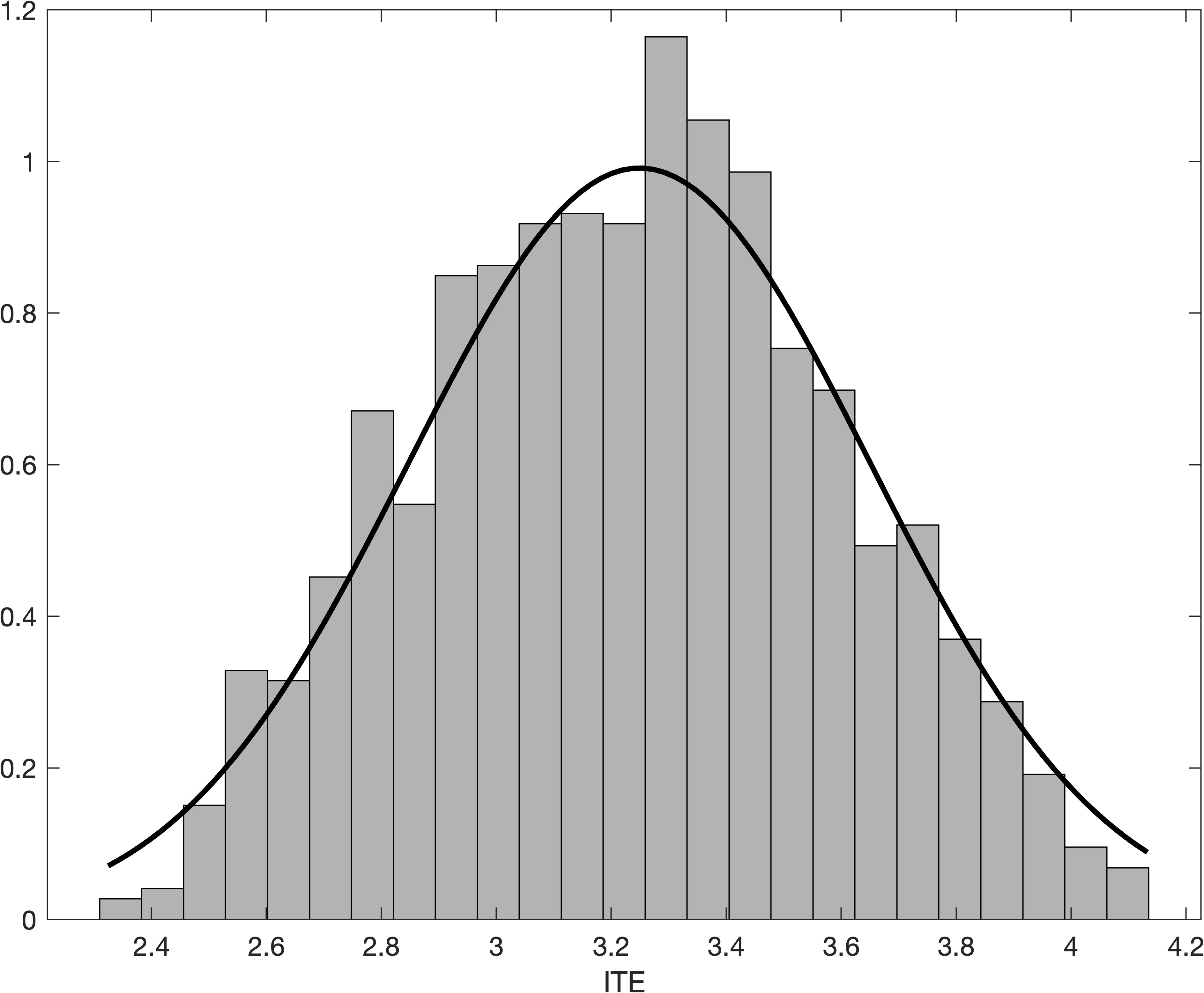}\tabularnewline
(c) $\tau=0.10,n=500$ & (d) $\tau=0.90,n=500$\tabularnewline
\end{tabular}
\end{figure}

To avoid redundancy, we focus on estimating the $\tau$-th quantile
$Q_{\varDelta}\left(\tau\right)$ using the empirical quantiles of
pseudo ITEs and omit the results that assess the quality of the estimator
of the cumulative probabilities. In Figure \ref{fig:quatile_dist},
each histogram displays realizations of $\widehat{Q}_{\varDelta}\left(\tau\right)$,
the $\tau$-th empirical quantile of pseudo ITEs, computed over $1,000$
simulation replications. The solid curve in each panel represents
the large sample density of $\widehat{Q}_{\varDelta}\left(\tau\right)$,
i.e., the Gaussian density with mean $Q_{\varDelta}\left(\tau\right)$
and variance $V_{Q}\left(\tau\right)/n$, as characterized by Corollary
\ref{cor:quantile}(ii), for $\tau\in\{0.25,0.5,0.75\}$ and for $n=250$
and $500$. Figure \ref{fig:extreme_quatile_dist} displays analogous
results for more extreme quantiles, with $\tau\in\{0.1,0.9\}$. Both
figures demonstrate close agreement between the simulated distributions
of $\widehat{Q}_{\varDelta}\left(\tau\right)$ and the corresponding
large sample Gaussian distributions for moderate sample sizes across
a range of probability levels, including relatively extreme levels
such as $0.1$ and $0.9$.

\subsection{Finite sample performances of the inference methods\label{subsec:Finite-sample-performance}}

This section evaluates the finite sample performances of the confidence
intervals and UCBs proposed in Algorithms \ref{alg:confidence interval for CDF}
to \ref{alg:variable width quantile}. We consider the same DGP as
in the preceding subsection and examine the inference methods for
four target parameters: (i) bootstrap percentile confidence intervals
for the cumulative probabilities $F_{\varDelta}\left(v\right)$ for
fixed values of $v$; (ii) bootstrap UCBs for the CDF (values $F_{\varDelta}\left(v\right)$
of the CDF over a range of $v$'s); (iii) bootstrap percentile confidence
intervals for the ITE quantiles $Q_{\varDelta}\left(\tau\right)$
for fixed values of $\tau$; (iv) bootstrap UCBs for the quantile
function (the values $Q_{\varDelta}\left(\tau\right)$ of the quantile
function over a range of $\tau$'s). The sample sizes considered are
$n=250$, $500$ and $1,000$.

Table \ref{table: pointwise_CDF} reports the pointwise coverage probabilities
and the expected lengths of the bootstrap percentile confidence interval
(denoted as \textbf{BP}) proposed in Algorithm \ref{alg:confidence interval for CDF}
for the cumulative probabilities $F_{\varDelta}\left(v\right)$, at
$v\in\{0.5,1,2,3,3.5\}$. For comparison, the table also includes
a ``naive'' confidence interval (\textbf{NAI}), which is constructed
using the standard error $\sqrt{\widehat{F}_{\varDelta}\left(v\right)\left(1-\widehat{F}_{\varDelta}\left(v\right)\right)/n}$
and accounts only for the component $V_{1}\left(v\right)$ of the
asymptotic variance given in Theorem \ref{thm:weak_convergence}(ii),
ignoring the ITE estimation error. The results show that the bootstrap
percentile confidence interval for $F_{\varDelta}\left(v\right)$
described in Algorithm \ref{alg:confidence interval for CDF} provides
coverage probabilities close to the nominal level across all values
of $v$ and sample sizes. In contrast, the ``naive'' confidence
intervals severely undercover, highlighting the importance of accounting
for the estimation error captured by $V_{2}\left(v\right)$, which
may contribute more to the asymptotic variance than the canonical
sampling variation $V_{1}\left(v\right)$. 

Table \ref{table: uniform_CDF} reports the simultaneous coverage
probabilities of the \textbf{constant-width UCB} from Algorithm \ref{alg:confidence bands for CDF}
and the \textbf{variable-width UCB}, constructed analogously to Algorithm
\ref{alg:variable width quantile}, for the CDF $F_{\varDelta}$ over
equally spaced grid points in the intervals $[0.04,3.96]$ and $[0.10,3.90]$
respectively with the step size $0.01$. For comparison, we include\textbf{
Interpolated BP} which constructs a band by interpolating the pointwise
bootstrap percentile confidence intervals in Algorithm \ref{alg:confidence interval for CDF}.
Table \ref{table: uniform_CDF} shows that the UCBs lead to good simultaneous
coverage. Although the interpolated BP intervals perform well pointwise
(as in Table \ref{table: pointwise_CDF}), they perform poorly for
uniform coverage. We also calculate the average expected widths of
the two confidence bands and show the results in Table \ref{table: uniform_CDF}.\footnote{The average expected width is computed by first averaging the widths
in all simulation replications at each grid point and then averaging
over all grid points in the given range.}

Table \ref{table: pointwise_quantile} presents results showing the
finite sample performance of the bootstrap percentile confidence intervals
(Algorithm \ref{alg:confidence interval for quantile}) for the $\tau$-th
quantile of the ITEs, with $\tau\in\left\{ 0.1,0.25,0.5,0.75,0.9\right\} $,
and the interquartile range (IQR). Table \ref{table: uniform_quantile}
examines the UCB for the quantile function $Q_{\varDelta}$ over equally
spaced grid points in the intervals $[0.05,0.95]$ and $[0.2,0.8]$,
with the step size $0.01$. Similar to the results discussed in the
preceding paragraph, Table \ref{table: pointwise_quantile} confirms
that the bootstrap percentile confidence intervals for ITE quantiles
and the IQR achieve good pointwise coverage, while Table \ref{table: uniform_quantile}
shows that the UCBs for the quantile function, both the constant-width
UCB from Algorithm \ref{alg:confidence band quantile} and the variable-width
UCB from Algorithm \ref{alg:variable width quantile}, provide reliable
simultaneous coverage. It is worth noting that all of the bootstrap
percentile confidence intervals and UCBs exhibit good coverage accuracy,
even in relatively small samples ($n=250$). When the sample size
$n=500$ or $1000$, the variable-width UCB appears narrower than
the constant-width counterpart.

\begin{table}[H]
\caption{Coverage probability (CP) and the average length (CIL) of the $(1-\alpha)\times100\%$
pointwise confidence intervals for the CDF $F_{\Delta}\left(v\right)$
of ITE. BP = bootstrap percentile confidence interval, NAI = a \textquotedblleft naive\textquotedblright{}
confidence interval. The nominal coverage levels are $1-\alpha=0.90,0.95,0.99$.}
\label{table: pointwise_CDF}
\centering{}%
\begin{tabular}{cclcccccc}
\toprule 
$v$ & $n$ & Methods & \multicolumn{3}{c}{CP} & \multicolumn{3}{c}{CIL}\tabularnewline
\midrule 
 &  &  & 0.90 & 0.95 & 0.99 & 0.90 & 0.95 & 0.99\tabularnewline
\midrule 
$0.5$ & 250 & BP & 0.904 & 0.943 & 0.989 & 0.372 & 0.433 & 0.536\tabularnewline
 &  & NAI & 0.301 & 0.358 & 0.421 & 0.092 & 0.110 & 0.144\tabularnewline
 & 500 & BP & 0.898 & 0.959 & 0.992 & 0.301 & 0.355 & 0.451\tabularnewline
 &  & NAI & 0.285 & 0.334 & 0.428 & 0.066 & 0.079 & 0.104\tabularnewline
 & 1000 & BP & 0.895 & 0.950 & 0.990 & 0.219 & 0.260 & 0.338\tabularnewline
 &  & NAI & 0.277 & 0.320 & 0.426 & 0.047 & 0.056 & 0.074\tabularnewline
\midrule 
$1$ & 250 & BP & 0.880 & 0.945 & 0.984 & 0.356 & 0.414 & 0.513\tabularnewline
 &  & NAI & 0.292 & 0.348 & 0.458 & 0.100 & 0.119 & 0.157\tabularnewline
 & 500 & BP & 0.904 & 0.957 & 0.990 & 0.289 & 0.339 & 0.429\tabularnewline
 &  & NAI & 0.326 & 0.358 & 0.464 & 0.072 & 0.086 & 0.113\tabularnewline
 & 1000 & BP & 0.902 & 0.944 & 0.987 & 0.218 & 0.257 & 0.332\tabularnewline
 &  & NAI & 0.286 & 0.352 & 0.435 & 0.051 & 0.061 & 0.081\tabularnewline
\midrule 
$2$ & 250 & BP & 0.886 & 0.936 & 0.983 & 0.294 & 0.343 & 0.427\tabularnewline
 &  & NAI & 0.325 & 0.383 & 0.460 & 0.094 & 0.111 & 0.146\tabularnewline
 & 500 & BP & 0.906 & 0.953 & 0.987 & 0.237 & 0.276 & 0.345\tabularnewline
 &  & NAI & 0.351 & 0.419 & 0.507 & 0.066 & 0.079 & 0.104\tabularnewline
 & 1000 & BP & 0.910 & 0.951 & 0.992 & 0.183 & 0.216 & 0.275\tabularnewline
 &  & NAI & 0.312 & 0.365 & 0.474 & 0.047 & 0.057 & 0.074\tabularnewline
\midrule 
$3$ & 250 & BP & 0.883 & 0.945 & 0.984 & 0.213 & 0.251 & 0.323\tabularnewline
 &  & NAI & 0.314 & 0.402 & 0.502 & 0.073 & 0.086 & 0.114\tabularnewline
 & 500 & BP & 0.904 & 0.946 & 0.988 & 0.162 & 0.190 & 0.244\tabularnewline
 &  & NAI & 0.313 & 0.362 & 0.470 & 0.050 & 0.060 & 0.078\tabularnewline
 & 1000 & BP & 0.915 & 0.962 & 0.991 & 0.126 & 0.149 & 0.190\tabularnewline
 &  & NAI & 0.315 & 0.365 & 0.489 & 0.036 & 0.042 & 0.056\tabularnewline
\midrule 
$3.5$ & 250 & BP & 0.880 & 0.943 & 0.990 & 0.167 & 0.199 & 0.261\tabularnewline
 &  & NAI & 0.425 & 0.459 & 0.627 & 0.056 & 0.066 & 0.087\tabularnewline
 & 500 & BP & 0.889 & 0.944 & 0.992 & 0.120 & 0.144 & 0.188\tabularnewline
 &  & NAI & 0.355 & 0.418 & 0.521 & 0.038 & 0.045 & 0.059\tabularnewline
 & 1000 & BP & 0.904 & 0.956 & 0.987 & 0.092 & 0.109 & 0.141\tabularnewline
 &  & NAI & 0.323 & 0.389 & 0.494 & 0.026 & 0.031 & 0.041\tabularnewline
\bottomrule
\end{tabular}
\end{table}

\begin{table}[H]
\centering \caption{Simultaneous coverage probability (Simultaneous CP) and the average
expected width (CBW) of the $(1-\alpha)\times100\%$ UCBs with constant
or variable width, and the confidence band constructed by interpolating
the pointwise bootstrap percentile confidence intervals (Interpolated
BP) for $F_{\varDelta}$. The nominal coverage levels are $1-\alpha=0.90,0.95,0.99$.}
\label{table: uniform_CDF} %
\begin{tabular}{cclcccccc}
\toprule 
Range & $n$ & Methods & \multicolumn{3}{c}{Simultaneous CP} & \multicolumn{3}{c}{CBW}\tabularnewline
\midrule 
 &  &  & 0.90 & 0.95 & 0.99 & 0.90 & 0.95 & 0.99\tabularnewline
\midrule 
$[0.04,3.96]$ & 250 & Constant-width UCB & 0.927 & 0.961 & 0.989 & 0.536 & 0.588 & 0.682\tabularnewline
 &  & Variable-width UCB & 0.879 & 0.941 & 0.991 & 0.586 & 0.662 & 0.773\tabularnewline
 &  & Interpolated BP & 0.429 & 0.589 & 0.862 & 0.277 & 0.324 & 0.407\tabularnewline
 & 500 & Constant-width UCB & 0.962 & 0.980 & 0.993 & 0.448 & 0.496 & 0.588\tabularnewline
 &  & Variable-width UCB & 0.884 & 0.967 & 0.996 & 0.512 & 0.592 & 0.720\tabularnewline
 &  & Interpolated BP & 0.414 & 0.622 & 0.861 & 0.218 & 0.256 & 0.327\tabularnewline
 & 1000 & Constant-width UCB & 0.974 & 0.989 & 1.000 & 0.355 & 0.394 & 0.474\tabularnewline
 &  & Variable-width UCB & 0.901 & 0.970 & 0.995 & 0.428 & 0.508 & 0.648\tabularnewline
 &  & Interpolated BP & 0.389 & 0.607 & 0.860 & 0.165 & 0.195 & 0.252\tabularnewline
\midrule 
$[0.10,3.90]$ & 250 & Constant-width UCB & 0.929 & 0.961 & 0.991 & 0.535 & 0.586 & 0.678\tabularnewline
 &  & Variable-width UCB & 0.860 & 0.930 & 0.990 & 0.568 & 0.642 & 0.754\tabularnewline
 &  & Interpolated BP & 0.516 & 0.658 & 0.899 & 0.279 & 0.326 & 0.410\tabularnewline
 & 500 & Constant-width UCB & 0.960 & 0.977 & 0.994 & 0.448 & 0.494 & 0.584\tabularnewline
 &  & Variable-width UCB & 0.879 & 0.956 & 0.994 & 0.494 & 0.571 & 0.695\tabularnewline
 &  & Interpolated BP & 0.496 & 0.680 & 0.886 & 0.220 & 0.259 & 0.329\tabularnewline
 & 1000 & Constant-width UCB & 0.971 & 0.987 & 0.997 & 0.354 & 0.393 & 0.470\tabularnewline
 &  & Variable-width UCB & 0.885 & 0.960 & 0.992 & 0.406 & 0.479 & 0.608\tabularnewline
 &  & Interpolated BP & 0.447 & 0.659 & 0.880 & 0.167 & 0.197 & 0.254\tabularnewline
\bottomrule
\end{tabular}
\end{table}

\begin{table}[H]
\caption{Coverage probability (CP) and the expected length (CIL) of the $(1-\alpha)\times100\%$
bootstrap percentile confidence intervals for $Q_{\varDelta}\left(\tau\right)$
and the interquartile range (IQR). The nominal coverage levels are
$1-\alpha=0.90,0.95,0.99$.}
\label{table: pointwise_quantile}
\centering{}%
\begin{tabular}{ccccccccc}
\toprule 
$\tau$ & $n$ &  & \multicolumn{3}{c}{CP} & \multicolumn{3}{c}{CIL}\tabularnewline
\midrule 
 &  &  & 0.90 & 0.95 & 0.99 & 0.90 & 0.95 & 0.99\tabularnewline
\midrule 
0.10 & 250 &  & 0.881 & 0.936 & 0.990 & 0.607 & 0.749 & 1.059\tabularnewline
 & 500 &  & 0.907 & 0.948 & 0.985 & 0.375 & 0.459 & 0.636\tabularnewline
 & 1000 &  & 0.905 & 0.945 & 0.981 & 0.245 & 0.294 & 0.396\tabularnewline
\midrule 
0.25 & 250 &  & 0.902 & 0.951 & 0.991 & 0.896 & 1.076 & 1.436\tabularnewline
 & 500 &  & 0.913 & 0.956 & 0.990 & 0.641 & 0.761 & 0.994\tabularnewline
 & 1000 &  & 0.900 & 0.940 & 0.989 & 0.468 & 0.555 & 0.717\tabularnewline
\midrule 
0.50 & 250 &  & 0.884 & 0.942 & 0.983 & 1.485 & 1.751 & 2.239\tabularnewline
 & 500 &  & 0.906 & 0.957 & 0.993 & 1.119 & 1.323 & 1.706\tabularnewline
 & 1000 &  & 0.902 & 0.946 & 0.985 & 0.818 & 0.973 & 1.269\tabularnewline
\midrule 
0.75 & 250 &  & 0.888 & 0.935 & 0.982 & 1.741 & 2.049 & 2.606\tabularnewline
 & 500 &  & 0.908 & 0.957 & 0.989 & 1.374 & 1.628 & 2.099\tabularnewline
 & 1000 &  & 0.916 & 0.956 & 0.993 & 1.011 & 1.202 & 1.571\tabularnewline
\midrule 
0.90 & 250 &  & 0.878 & 0.941 & 0.987 & 1.254 & 1.482 & 1.909\tabularnewline
 & 500 &  & 0.893 & 0.954 & 0.990 & 1.021 & 1.198 & 1.525\tabularnewline
 & 1000 &  & 0.921 & 0.961 & 0.989 & 0.814 & 0.959 & 1.220\tabularnewline
\midrule
\midrule 
IQR & 250 &  & 0.913 & 0.953 & 0.986 & 1.578 & 1.857 & 2.352\tabularnewline
 & 500 &  & 0.913 & 0.957 & 0.992 & 1.272 & 1.505 & 1.941\tabularnewline
 & 1000 &  & 0.923 & 0.969 & 0.995 & 0.953 & 1.133 & 1.481\tabularnewline
\bottomrule
\end{tabular}
\end{table}

\begin{table}[H]
\caption{Simultaneous coverage probability (Simultaneous CP) and the average
expected width (CBW) for the $(1-\alpha)\times100\%$ UCB of $Q_{\varDelta}$.
The nominal coverage levels are $1-\alpha=0.90,0.95,0.99$.}
\label{table: uniform_quantile} \centering{}%
\begin{tabular}{cclcccccc}
\toprule 
Range & $n$ & Methods & \multicolumn{3}{c}{Simultaneous CP} & \multicolumn{3}{c}{CBW}\tabularnewline
\midrule 
 &  &  & 0.90 & 0.95 & 0.99 & 0.90 & 0.95 & 0.99\tabularnewline
\midrule 
$[0.05,\ 0.95]$ & 250 & Constant-width & 0.911 & 0.950 & 0.986 & 2.580 & 2.908 & 3.533\tabularnewline
 &  & Variable-width & 0.881 & 0.939 & 0.987 & 2.567 & 3.026 & 4.148\tabularnewline
 & 500 & Constant-width & 0.934 & 0.974 & 0.991 & 2.003 & 2.258 & 2.751\tabularnewline
 &  & Variable-width & 0.875 & 0.944 & 0.990 & 1.834 & 2.125 & 2.836\tabularnewline
 & 1000 & Constant-width & 0.941 & 0.974 & 0.996 & 1.499 & 1.688 & 2.060\tabularnewline
 &  & Variable-width & 0.866 & 0.930 & 0.982 & 1.310 & 1.488 & 1.888\tabularnewline
\midrule 
$[0.20,\ 0.80]$ & 250 & Constant-width & 0.919 & 0.952 & 0.987 & 2.495 & 2.832 & 3.471\tabularnewline
 &  & Variable-width & 0.854 & 0.923 & 0.979 & 2.330 & 2.704 & 3.496\tabularnewline
 & 500 & Constant-width & 0.943 & 0.975 & 0.991 & 1.929 & 2.188 & 2.691\tabularnewline
 &  & Variable-width & 0.877 & 0.938 & 0.984 & 1.719 & 1.971 & 2.510\tabularnewline
 & 1000 & Constant-width & 0.952 & 0.979 & 0.996 & 1.426 & 1.620 & 1.998\tabularnewline
 &  & Variable-width & 0.893 & 0.944 & 0.989 & 1.265 & 1.438 & 1.789\tabularnewline
\bottomrule
\end{tabular}
\end{table}

\section{Empirical application: 401(k) program and savings\label{sec:Empirical-application}}

We revisit the empirical application of \citetalias{feng2019estimation}
and conduct inference on the distribution of ITEs of participating
in 401(k) retirement programs on personal savings. Following \citetalias{feng2019estimation},
the outcome variable is family net financial assets; the treatment
indicator reflects participation in 401(k) programs; the IV is eligibility
for 401(k); and the covariates include categorical variables for income
and age (each grouped into four categories based on distributional
quartiles), an indicator for marital status, and a dummy for family
size less than 3. We show that many of the qualitative statements
in the empirical application sections of \citetalias{feng2019estimation}
can be confirmed by using the inference methods proposed in this paper.
At the same time, our CDF-based approach allows one to directly target
important distributional characteristics, such as the proportion of
individuals with positive ITEs, and conduct valid inference.

Table \ref{table: 401k_summary} reports the 95\% confidence intervals
for three features of the ITE distribution: the proportion of positive
ITEs ($\Pr[\varDelta>0]$), the median, and the interquartile range
(IQR). For the full sample, the confidence interval for the proportion
of positive ITEs is $\left[0.851,0.919\right]$, indicating that while
most households benefited, a non-negligible fraction experienced negative
effects. Note that the \citetalias{feng2019estimation}  estimate
for the same feature is 0.917, which is near the right boundary of
our 95\% confidence interval. Thus, our result suggests that the proportion
of individuals with negative ITEs may be larger than that reported
in \citetalias{feng2019estimation}. In particular, at the 5\% significance
level, we cannot reject the null hypothesis that 14.9\% of individuals
experience a negative ITE. The median ITE has a confidence interval
of $\left[6.96,9.74\right]$ (in thousands of dollars), confirming
a significantly positive center of the treatment effect distribution.
The IQR, with a confidence interval of $\left[16.68,23.38\right]$,
reveals considerable variation in treatment effects across households.

Subsample analysis based on covariate categories reveals notable patterns.
The proportion of individuals with positive ITEs tends to increase
with income and age, but remains relatively stable across groups defined
by marital status and family size. Regarding the median impact of
the program, even in the two subgroups that benefit the least-- the
lowest income group and the youngest age group--the median ITE remains
significantly positive. In terms of dispersion, the IQR of the ITE
distribution increases substantially with income and age. Married
individuals also exhibit greater dispersion in their ITE distribution
than unmarried individuals. These findings suggest that treatment
effect heterogeneity is more pronounced among higher-income, older,
and married subpopulations.

Our subsample analysis also suggests that a larger proportion of young
individuals may have negative ITEs than that reported in \citetalias{feng2019estimation}.
According to their estimates, 15.93\% of young individuals (with age
in the first quartile) have negative effects. However, our 95\% confidence
interval suggests that 29.4\% of young individuals may experience
negative ITEs.

\begin{figure}[H]
\caption{Comparison of ITE distributions (quantile functions) between groups
$A_{1}$ and $A_{0}$ based on each covariate. Solid line = estimated
quantile function, shaded area = $95\%$ variable-width UCB.}
\label{fig:401k_q_comp} \centering %
\begin{tabular}{cc}
\includegraphics[scale=0.05]{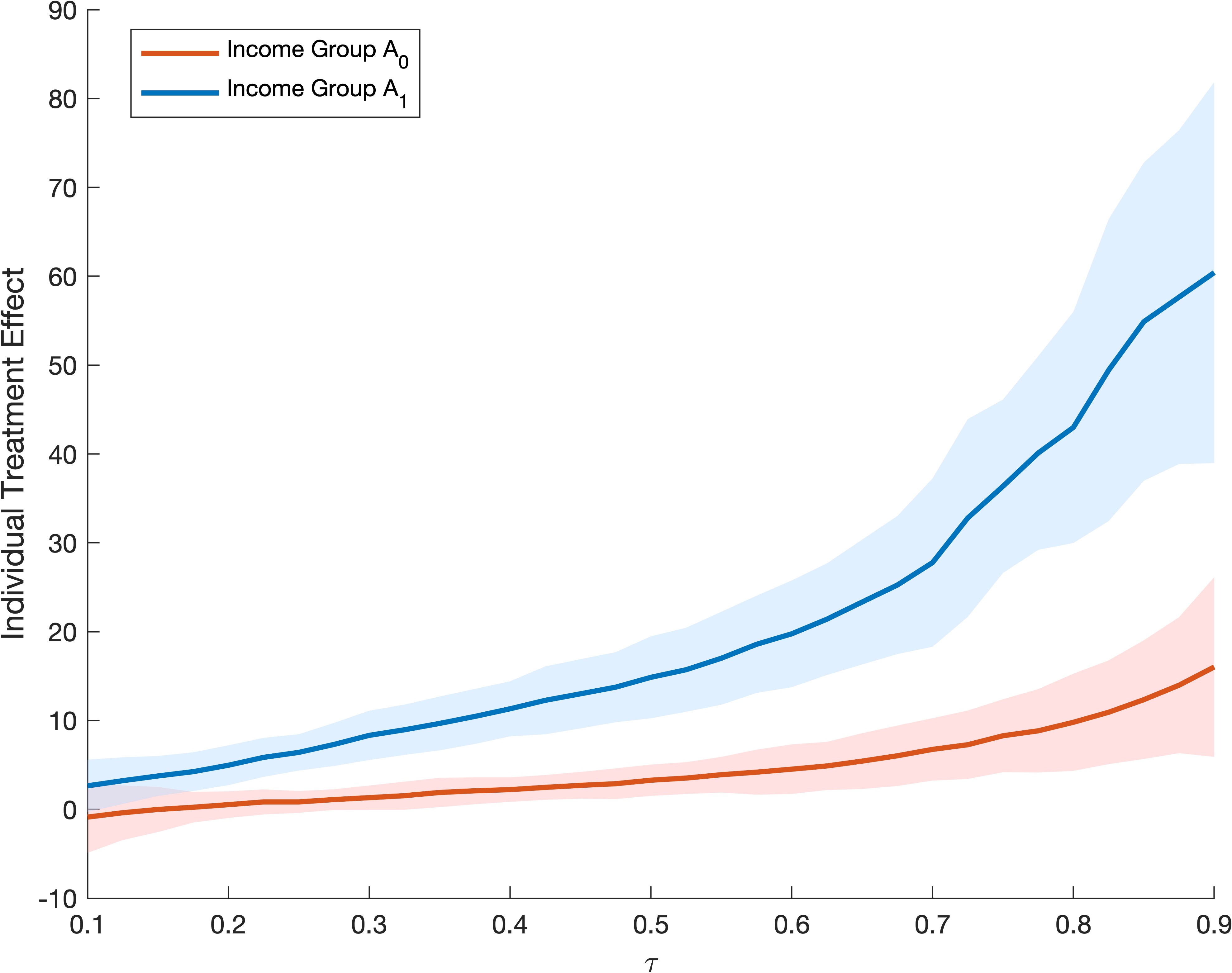} & \includegraphics[scale=0.05]{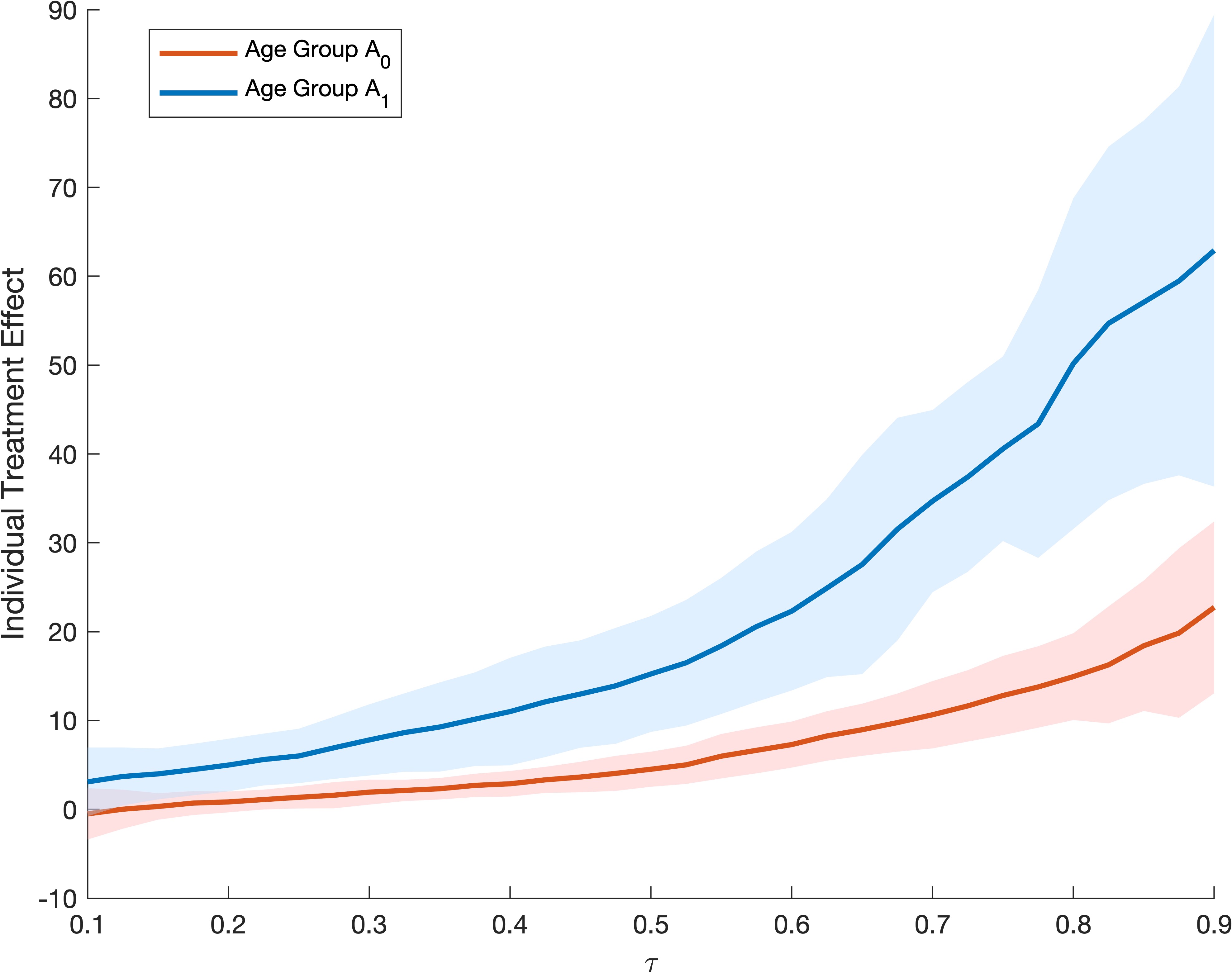}\tabularnewline
(a) $A_{1}:$ Income $>$ median, $A_{0}:$ Income $\leq$ median & (b) $A_{1}:$ Age $>$ median, $A_{0}:$ Age $\leq$ median\tabularnewline
\includegraphics[scale=0.05]{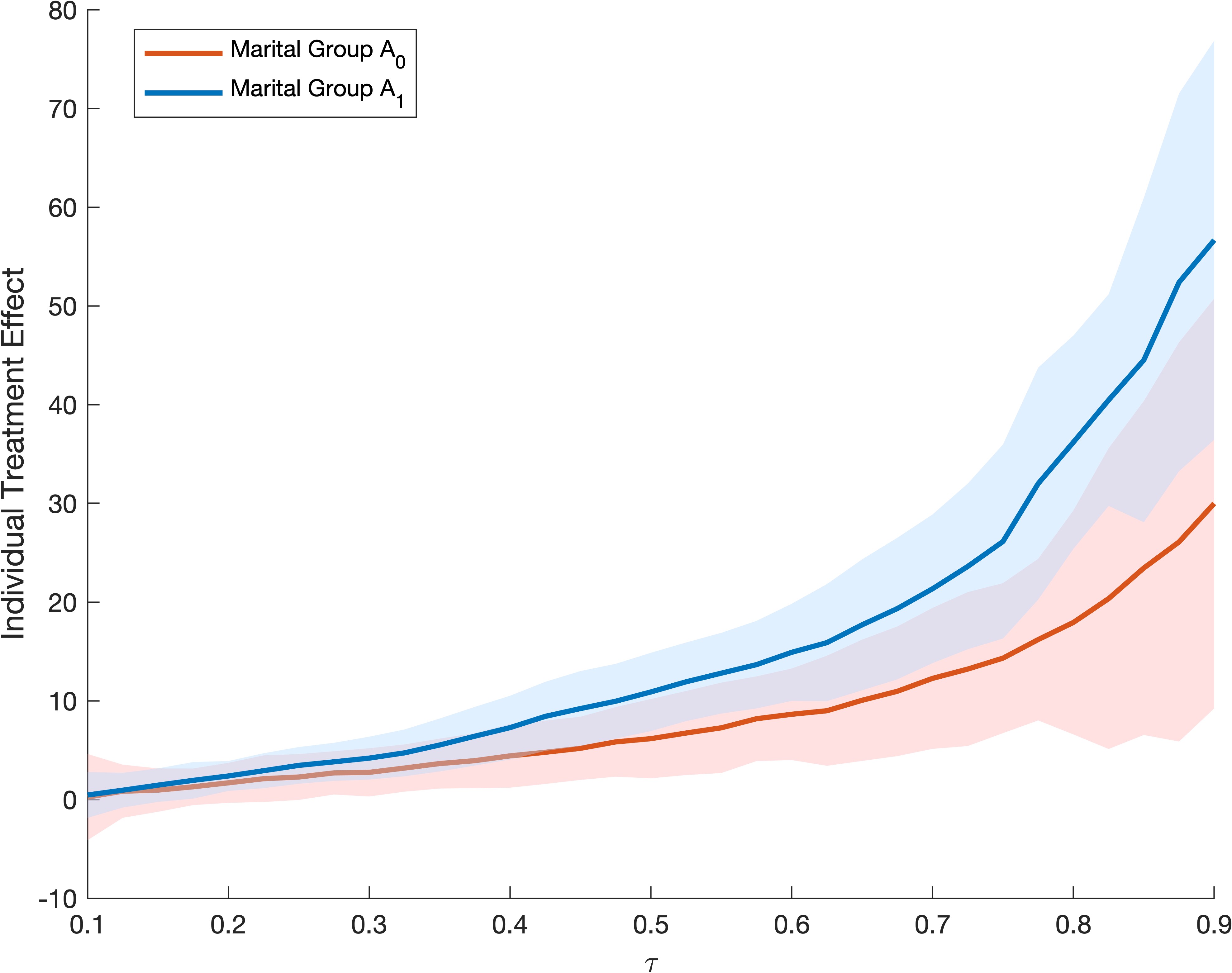} & \includegraphics[scale=0.05]{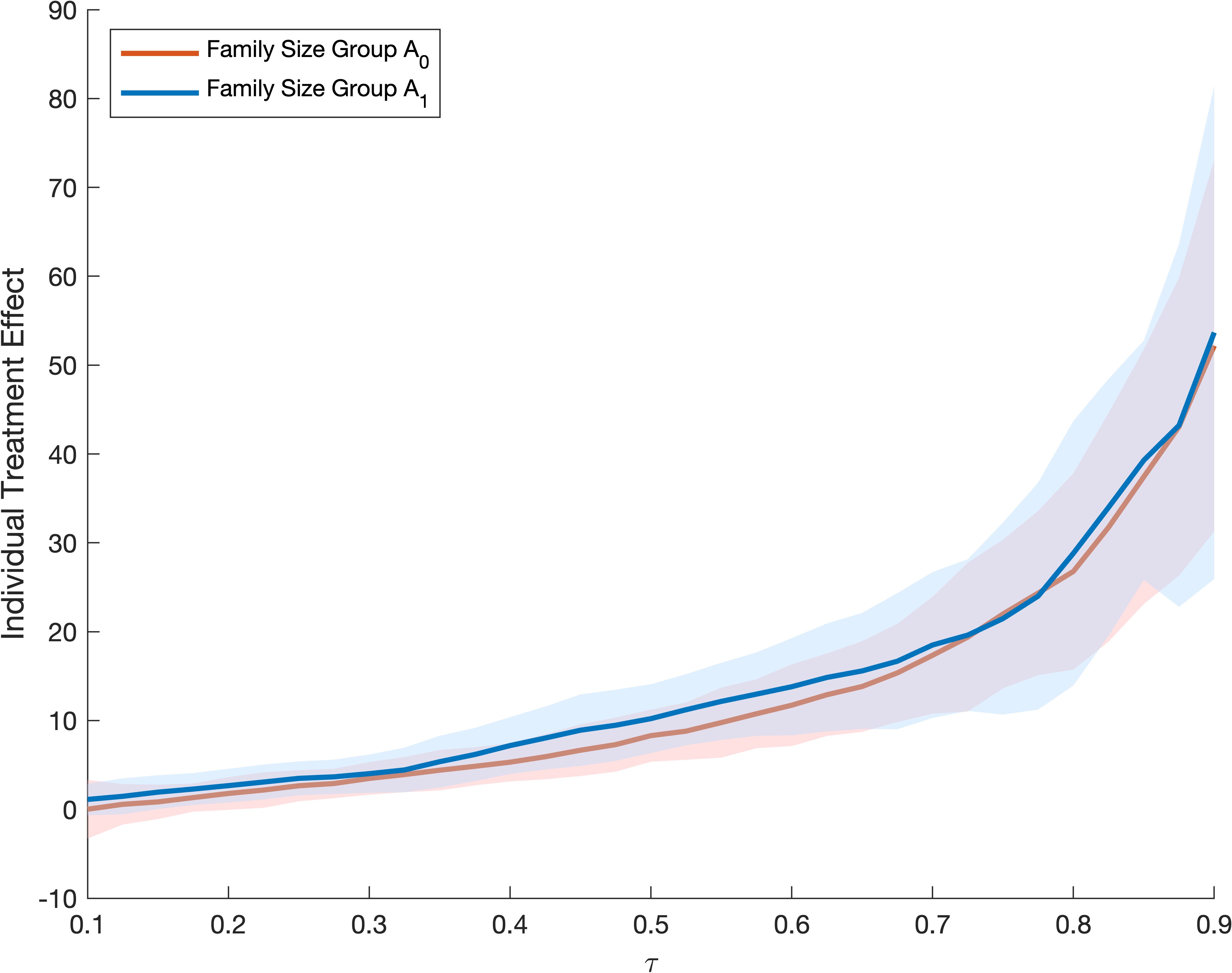}\tabularnewline
(c) $A_{1}:$ Married, $A_{0}:$ Unmarried & (d) $A_{1}:$ Family size $\geq3$, $A_{0}:$ Family size $<3$\tabularnewline
\end{tabular}
\end{figure}

Table \ref{table: 401k_quantile_diff} summarizes how the ITE distribution
varies with each of the four covariates by reporting confidence intervals
for differences in three representative quantiles ($\tau=0.25,0.5,0.75$)
and for the difference in the IQR of the ITE distribution between
groups $A_{1}$ and $A_{0}$, computed using Algorithm \ref{alg:difference quantiles}.
Parallel to Figures 4-7 of \citetalias{feng2019estimation}, Figure
\ref{fig:401k_q_comp} visualizes the quantile functions $Q_{\Delta\mid\tilde{X}}\left(\cdot\mid A_{0}\right)$
and $Q_{\Delta\mid\tilde{X}}\left(\cdot\mid A_{1}\right)$ together
with their $95\%$ variable-width UCBs (Algorithm \ref{alg:variable width quantile}
with range $\left[\underline{\tau},\overline{\tau}\right]$ =$\left[0.1,0.9\right]$).
Panels (a) and (b) of Figures \ref{fig:401k_q_comp} indicate that
the ITE distribution shifts to the right and becomes more dispersed
as income and age increase. A similar but weaker pattern is observed
in Panel (c), where marital status changes from unmarried to married.
By contrast, family size shows little influence on the ITE distribution
as Panel (d) shows.

\begin{figure}[H]
\caption{Comparison of ITE distributions (quantile function) between groups
$A_{1}$ and $A_{0}$ based on each covariate. Solid line = estimate
of the quantile difference function $Q_{\Delta\mid\tilde{X}}\left(\cdot\mid A_{1}\right)-Q_{\Delta\mid\tilde{X}}\left(\cdot\mid A_{0}\right)$,
shaded area = $95\%$ variable-width UCB.}
\label{fig:401k_q_diff} \centering %
\begin{tabular}{cc}
\includegraphics[scale=0.05]{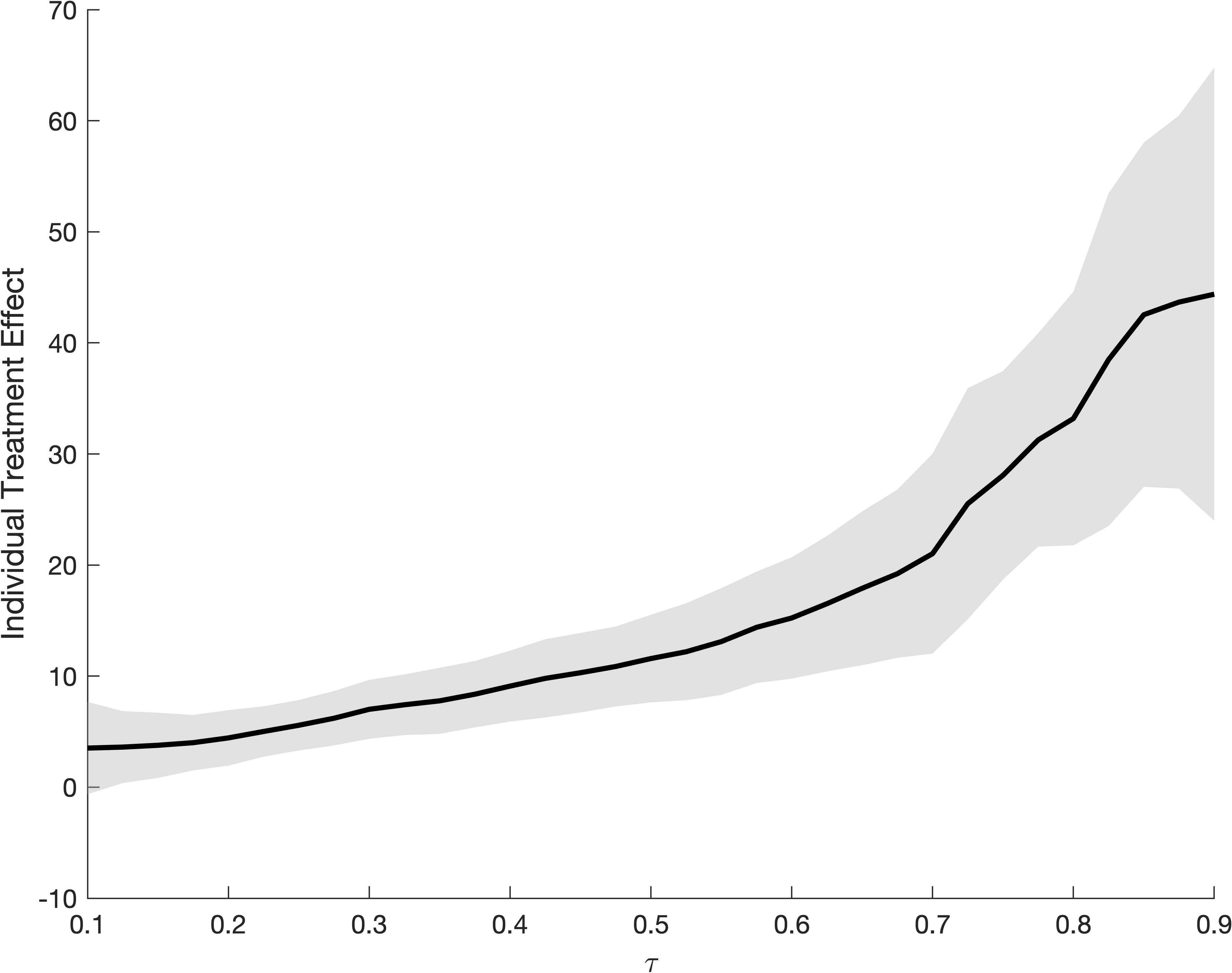} & \includegraphics[scale=0.05]{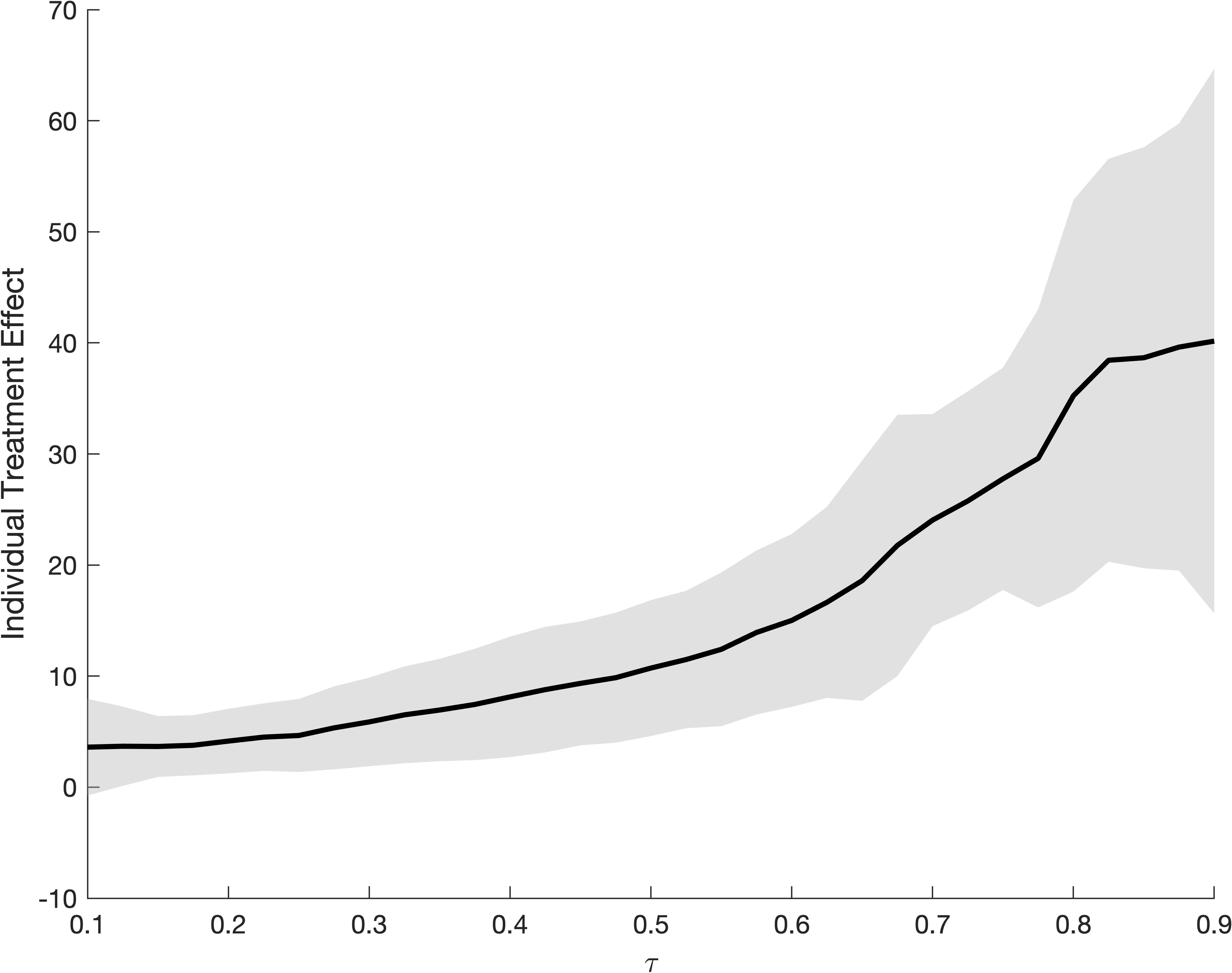}\tabularnewline
(a) $A_{1}:$ Income $>$ median, $A_{0}:$ Income $\leq$ median & (b) $A_{1}:$ Age $>$ median, $A_{0}:$ Age $\leq$ median\tabularnewline
\includegraphics[scale=0.05]{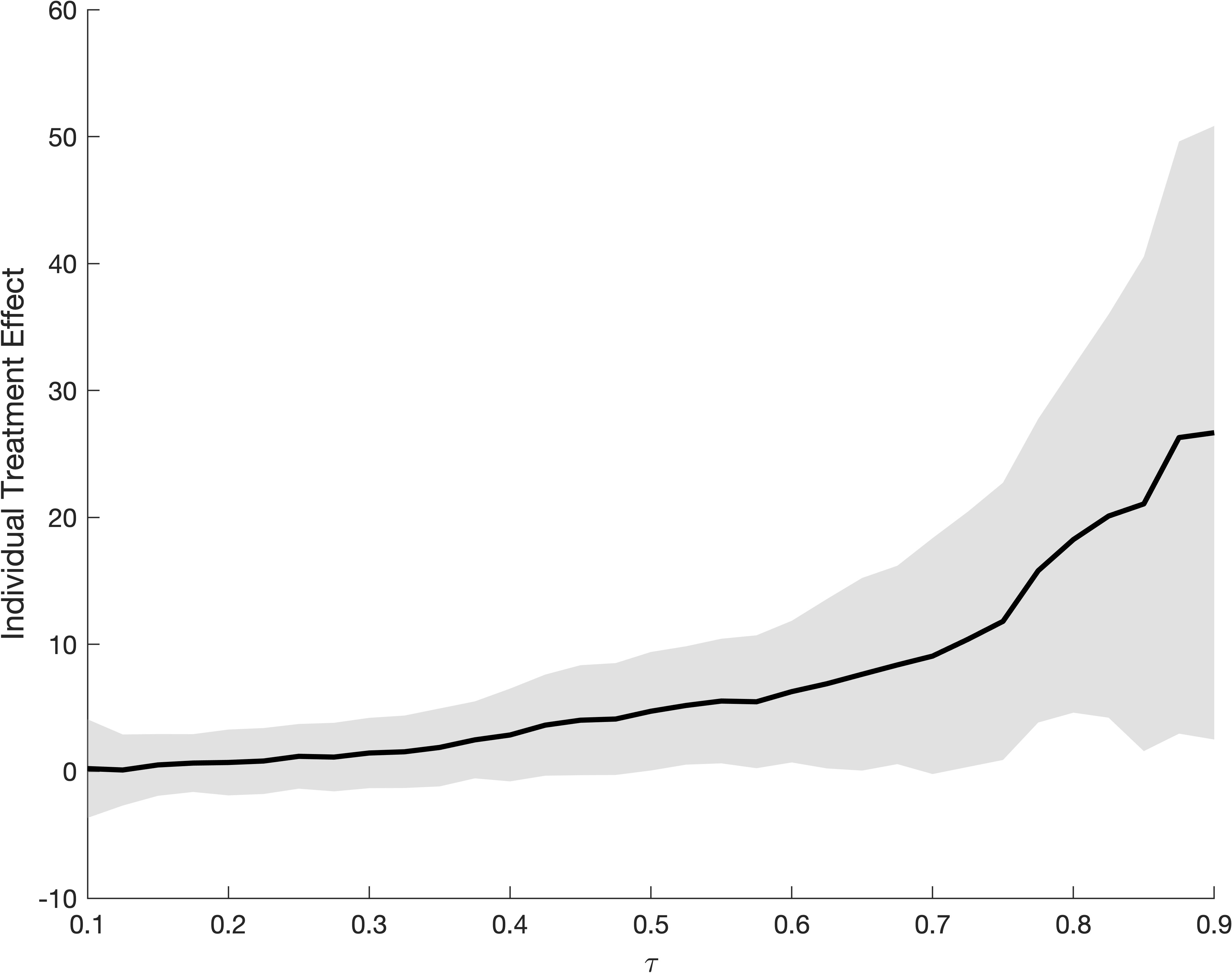} & \includegraphics[scale=0.05]{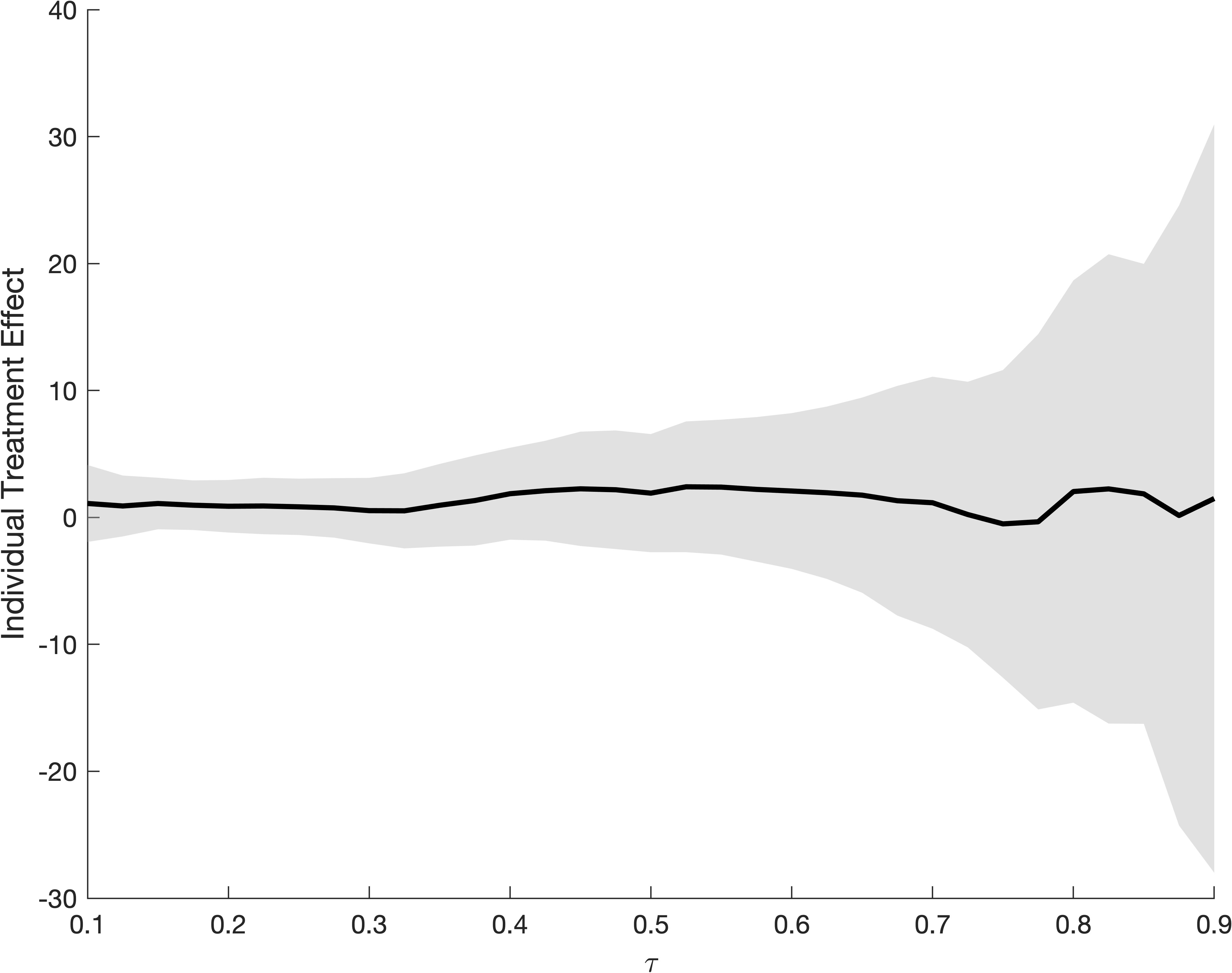}\tabularnewline
(c) $A_{1}:$ Married, $A_{0}:$ Unmarried & (d) $A_{1}:$ Family size $\geq3$, $A_{0}:$ Family size $<3$\tabularnewline
\end{tabular}
\end{figure}

Figure \ref{fig:401k_q_diff} depicts the estimator of the quantile
difference function $Q_{\Delta\mid\tilde{X}}\left(\cdot\mid A_{1}\right)-Q_{\Delta\mid\tilde{X}}\left(\cdot\mid A_{0}\right)$
on $\left[\underline{\tau},\overline{\tau}\right]$ =$\left[0.1,0.9\right]$
and its $95\%$ UCB (Algorithm \ref{alg:variable width quantile difference}
with $\left[\underline{\tau},\overline{\tau}\right]=\left[0.1,0.9\right]$).
Panel (a) of Figure \ref{fig:401k_q_diff} suggests that the ITE distribution
for individuals with above the median income stochastically dominates
that for individuals with below the median income. Similarly, Panel
(b) of Figure \ref{fig:401k_q_diff} suggests that the ITE distribution
for older individuals (age above the median) stochastically dominates
that for younger individuals (age below the median). Furthermore,
Panel (c) suggests that the ITE distribution for married individuals
may stochastically dominate that for unmarried individuals, with particularly
clear dominance in the upper tail. On the other hand, Panel (d) shows
that we cannot reject the null hypothesis of equality in the ITE distributions
between individuals with larger and smaller family sizes (family size
above or below three).

\begin{table}[H]
\caption{$95\%$ bootstrap percentile confidence intervals for distributional
features of ITEs of participation in the 401(k) retirement program
on personal savings (in thousands of dollars): proportion of positive
ITEs ($\Pr[\varDelta>0]$), median, and interquartile range (IQR).}
\label{table: 401k_summary}
\centering{}%
\begin{tabular}{lrccc}
\toprule 
 & $n$ & $\Pr[\varDelta>0]$ & Median & IQR\tabularnewline
\midrule 
Full sample & 8,702 & {[}0.851, 0.919{]} & {[}6.96, 9.74{]} & {[}16.68, 23.38{]}\tabularnewline
\midrule 
\multicolumn{5}{l}{Subsample conditional on:}\tabularnewline
\midrule 
Income $\leq$ 1st quartile & 777 & {[}0.528, 0.923{]} & {[}0.08, 2.39{]} & {[}1.84, 6.48{]}\tabularnewline
Income 1st to 2nd quartile & 2,637 & {[}0.765, 0.916{]} & {[}2.79, 5.46{]} & {[}6.52, 12.51{]}\tabularnewline
Income 2nd to 3rd quartile & 2,672 & {[}0.827, 0.938{]} & {[}5.86, 10.02{]} & {[}11.15, 18.66{]}\tabularnewline
Income $>$ 3rd quartile & 2,616 & {[}0.944, 0.987{]} & {[}20.10, 33.92{]} & {[}31.29, 53.79{]}\tabularnewline
\midrule 
Age $\leq$ 1st quartile & 2,504 & {[}0.706, 0.884{]} & {[}2.09, 4.26{]} & {[}6.42, 11.21{]}\tabularnewline
Age 1st to 2nd quartile & 2,072 & {[}0.840, 0.957{]} & {[}5.36, 9.89{]} & {[}9.44, 18.92{]}\tabularnewline
Age 2nd to 3rd quartile & 1,892 & {[}0.904, 0.985{]} & {[}10.69, 18.32{]} & {[}19.18, 34.97{]}\tabularnewline
Age $>$ 3rd quartile & 2,234 & {[}0.845, 0.961{]} & {[}12.03, 24.32{]} & {[}32.91, 57.99{]}\tabularnewline
\midrule 
Married & 2,955 & {[}0.811, 0.943{]} & {[}4.18, 7.77{]} & {[}9.88, 17.39{]}\tabularnewline
Unmarried & 5,747 & {[}0.846, 0.923{]} & {[}8.52, 12.69{]} & {[}20.17, 30.30{]}\tabularnewline
\midrule 
Family size $<3$ & 5,744 & {[}0.826, 0.914{]} & {[}6.16, 9.62{]} & {[}16.24, 25.87{]}\tabularnewline
Family size $\geq3$ & 2,958 & {[}0.880, 0.964{]} & {[}7.18, 11.90{]} & {[}14.83, 25.56{]}\tabularnewline
\bottomrule
\end{tabular}
\end{table}

\begin{table}[H]
\caption{$95\%$ bootstrap percentile confidence intervals for the quantile
differences $\delta\left(\tau\right)\protect\coloneqq Q_{\Delta\mid\tilde{X}}\left(\tau\mid A_{1}\right)-Q_{\Delta\mid\tilde{X}}\left(\tau\mid A_{0}\right)$
and the IQR difference $\delta\left(0.75\right)-\delta\left(0.25\right)$
in the ITE distribution between groups $A_{1}$ and $A_{0}$, where
$A_{1}$ and $A_{0}$ are determined by each covariate.}
\label{table: 401k_quantile_diff}
\centering{}%
\begin{tabular}{llcccc}
\toprule 
Group $A_{1}$ & Group $A_{0}$ & $\delta\left(0.25\right)$ & $\delta\left(0.5\right)$ & $\delta\left(0.75\right)$ & $\delta\left(0.75\right)-\delta\left(0.25\right)$\tabularnewline
\midrule 
Income $>$ median & Income $\leq$ median & {[}3.61, 6.54{]} & {[}8.93, 14.09{]} & {[}21.27, 34.59{]} & {[}16.64, 29.20{]}\tabularnewline
\midrule 
Age $>$ median & Age $\leq$ median & {[}2.77, 6.20{]} & {[}7.56, 14.63{]} & {[}21.26, 36.01{]} & {[}17.20, 31.31{]}\tabularnewline
\midrule 
Married & Unmarried & {[}-0.25, 2.64{]} & {[}1.57, 7.01{]} & {[}6.06, 19.84{]} & {[}5.17, 17.87{]}\tabularnewline
\midrule 
Family size $\geq3$ & Family size $<3$ & {[}-0.45, 2.34{]} & {[}-1.58, 4.47{]} & {[}-7.12, 8.09{]} & {[}-7.63, 7.05{]}\tabularnewline
\bottomrule
\end{tabular}
\end{table}

\subsubsection*{Declaration of generative AI and AI-assisted technologies in the
manuscript preparation process}

During the preparation of this work, the authors used AI-assisted
technologies for language refinement and readability improvements.
After using these tools, the authors reviewed and edited the content
as needed and take full responsibility for the content of the published
article.

\bibliographystyle{chicago}
\bibliography{ITE_CDF}

\end{document}